\documentclass[a4paper,11pt]{article}
\pdfoutput=1
\usepackage{./auxi/jheppub}
\usepackage[utf8]{inputenc}

\usepackage{slashed} % for Feynman slashed notation
\usepackage{mathtools} % for subequation numbering
\usepackage{tikz}
\usepackage{dsfont} % for displaying the blackboard 1 properly
\usetikzlibrary{positioning,arrows,calc}
\usetikzlibrary{arrows.meta}
\usetikzlibrary{decorations.pathmorphing}
\usetikzlibrary{decorations.markings}
\usetikzlibrary{shapes.geometric}
\usetikzlibrary{patterns}
\usepackage{xparse}
\usepackage{diagbox} % for category cells in tables
\usepackage{braket}
\usepackage{subcaption} % to use subfigure environment (pics side by side)
\usepackage[export]{adjustbox} %to align pictures in eqs
\usepackage{amsmath}

\definecolor{navyblue}{rgb}{0.0, 0.0, 0.5}

\newcommand{\phih}{{\hat{\phi}}}

\newcommand{\Psih}{{\hat{\Psi}}}
\newcommand{\Psib}{{\bar{\Psi}}}
\newcommand{\Psihb}{{\hat{\Psib}}}

\newcommand{\Disp}{\mathbb{D}}
\newcommand{\muh}{\hat{\mu}}
\newcommand{\nuh}{\hat{\nu}}
\newcommand{\rhoh}{\hat{\rho}}
\newcommand{\sigmah}{\hat{\sigma}}
\newcommand{\Dispb}[1]{\Disp_{#1}}

\newcommand{\NormPhi}{\Nm_{\phi}}
\newcommand{\NormPsi}{\Nm_{\Psi}}
\newcommand{\NormPhiHat}{\Nm_{\phih}}
\newcommand{\NormPhiHatIndex}[1]{\Nm_{\phih^{#1}}}
\newcommand{\NormTilt}{\Nm_{t}}
\newcommand{\NormDisp}{\Nm_{\Disp}}
\newcommand{\NormPsiHat}{\Nm_{\Psih}}
\newcommand{\Norms}{\Nm_{s_{\pm}}}
\newcommand{\NormPhiSquared}{\Nm_{\phi^2}}
\newcommand{\NormBilinear}{\Nm_{\Psib \Psi}}

\tikzset{
  vtx/.style={
    circle,
    draw=blue,
    fill=blue,
    inner sep=1pt
  },
  wcirc/.style={
    circle,
    draw=white,
    fill=white,
    inner sep=2pt
  },
  bcirc/.style={
    circle,
    draw=black,
    fill=black,
    inner sep=1pt
  },
  dcirc/.style={
    circle,
    draw=blue,
    fill=blue,
    inner sep=1pt
  },
  rcirc/.style={
    circle,
    draw=red,
    fill=red,
    inner sep=1pt
  },
  phi/.style={
    thick
  },
  sigma/.style={
    thick,
    dashed
  },
  vl1/.style={
    thick,
    blue
  },
  vl2/.style={
    thick,
    dashed,
    blue
  },
  valign/.style={
    baseline={([yshift=-.55ex]current bounding box.center)}
  }
}
\usetikzlibrary{ decorations.markings}

\newcommand{\Op}{\mathcal{O}}

\newcommand{\vvev}[1]{\langle\!\langle\, #1 \, \rangle\!\rangle}
\newcommand{\vev}[1]{\langle\, #1 \, \rangle}

\newif\ifstartcompletesineup
\newif\ifendcompletesineup
\pgfkeys{
    /pgf/decoration/.cd,
    start up/.is if=startcompletesineup,
    start up=true,
    start up/.default=true,
    start down/.style={/pgf/decoration/start up=false},
    end up/.is if=endcompletesineup,
    end up=true,
    end up/.default=true,
    end down/.style={/pgf/decoration/end up=false}
}
\pgfdeclaredecoration{complete sines}{initial}
{
    \state{initial}[
        width=+0pt,
        next state=upsine,
        persistent precomputation={
            \ifstartcompletesineup
                \pgfkeys{/pgf/decoration automaton/next state=upsine}
                \ifendcompletesineup
                    \pgfmathsetmacro\matchinglength{
                        0.5*\pgfdecoratedinputsegmentlength / (ceil(0.5* \pgfdecoratedinputsegmentlength / \pgfdecorationsegmentlength) )
                    }
                \else
                    \pgfmathsetmacro\matchinglength{
                        0.5 * \pgfdecoratedinputsegmentlength / (ceil(0.5 * \pgfdecoratedinputsegmentlength / \pgfdecorationsegmentlength ) - 0.499)
                    }
                \fi
            \else
                \pgfkeys{/pgf/decoration automaton/next state=downsine}
                \ifendcompletesineup
                    \pgfmathsetmacro\matchinglength{
                        0.5* \pgfdecoratedinputsegmentlength / (ceil(0.5 * \pgfdecoratedinputsegmentlength / \pgfdecorationsegmentlength ) - 0.4999)
                    }
                \else
                    \pgfmathsetmacro\matchinglength{
                        0.5 * \pgfdecoratedinputsegmentlength / (ceil(0.5 * \pgfdecoratedinputsegmentlength / \pgfdecorationsegmentlength ) )
                    }
                \fi
            \fi
            \setlength{\pgfdecorationsegmentlength}{\matchinglength pt}
        }] {}
    \state{downsine}[width=\pgfdecorationsegmentlength,next state=upsine]{
        \pgfpathsine{\pgfpoint{0.5\pgfdecorationsegmentlength}{0.5\pgfdecorationsegmentamplitude}}
        \pgfpathcosine{\pgfpoint{0.5\pgfdecorationsegmentlength}{-0.5\pgfdecorationsegmentamplitude}}
    }
    \state{upsine}[width=\pgfdecorationsegmentlength,next state=downsine]{
        \pgfpathsine{\pgfpoint{0.5\pgfdecorationsegmentlength}{-0.5\pgfdecorationsegmentamplitude}}
        \pgfpathcosine{\pgfpoint{0.5\pgfdecorationsegmentlength}{0.5\pgfdecorationsegmentamplitude}}
}
    \state{final}{}
}

\definecolor{bgbox}{RGB}{255,254,230}
\definecolor{setupplane}{RGB}{230,230,230}
\definecolor{gluoncolor}{RGB}{207,54,108}
\definecolor{vertexcolor}{RGB}{53,152,219}
\definecolor{SEcolor}{RGB}{176,156,255}
\definecolor{blobcolor}{RGB}{190,180,230}
\tikzset{
corner/.style={line width=1pt,dashed,draw=black,dash pattern=on 6pt off 4pt},
scalar/.style={line width=1pt,draw=black},
gluon/.style={line width=1pt,decorate, draw=gluoncolor,
    decoration={complete sines,aspect=0,amplitude=1.25mm,segment length=1.5mm,start up,end up}},
ghost/.style={line width=1pt,loosely dotted,draw=black},
wilson/.style={line width=2pt,draw=black},
 }
\NewDocumentCommand\semiloop{O{black}mmmO{}O{above}}
{%
\draw[#1] let \p1 = ($(#3)-(#2)$) in (#3) arc (#4:({#4+180}):({0.5*veclen(\x1,\y1)})node[midway, #6] {#5};)
}

\usepackage{bbm}

\def\veps{\varepsilon}
\newcommand{\PropagatorScalar}{
\begin{tikzpicture}[baseline,valign]
  \draw[dashed] (0, 0) -- (0.8, 0);
 \end{tikzpicture}
 }
 \newcommand{\PropagatorFermion}{
 \begin{tikzpicture}[baseline,valign]
  \draw[] (0, 0) -- (0.8, 0);
  \draw[-Latex] (.45,0)--(.5,0);
 \end{tikzpicture} 
 }
 \newcommand{\VertexFourScalars}{
 \begin{tikzpicture}[baseline,valign]
  \draw[dashed] ( 0.3, 0.3) -- (-0.3, -0.3);
  \draw[dashed] (-0.3, 0.3) -- ( 0.3, -0.3);
  \node at (0,0) [bcirc] {};
 \end{tikzpicture} 
 }
  \newcommand{\VertexYukawa}{
 \begin{tikzpicture}[baseline,valign]
  \draw[] ( 0.35, 0.35) -- (0,0);
  \draw[] (-0.35, 0.35) -- ( 0,0);
  \draw[dashed] ( 0, 0) -- (0, -.3);
  \draw[-Latex] (.2,.2)--(.1,.1);
  \draw[-Latex] (-.15,.15)--(-.25,.25);
  \node at (0,0) [rcirc] {};
 \end{tikzpicture} 
 }
 
\newcommand{\VertexDefect}{
\begin{tikzpicture}[baseline,valign]
  \draw[thick] (-0.1, 0) -- (0.7, 0);
  \draw[dashed](0.3, 0) -- (0.3, 0.6);
  \node at (0.3, 0) [dcirc] {};
 \end{tikzpicture}
 }

\newcommand{\OnePointOne}{
\begin{tikzpicture}[baseline,valign]
  \draw[thick] (0, 0) -- (2, 0);
  \draw[dashed] (1, 0) to (1, 2);
  \node at (1,0) [dcirc] {};
  \draw[fill=white] (1,2) circle (.075);
\end{tikzpicture}
}
\newcommand{\OnePointTwo}{
\begin{tikzpicture}[baseline,valign]
  \draw[thick] (-0.2, 0) -- (1.8, 0);
  \draw[dashed] (0.3, 0) to(0.8, 1);
  \draw[dashed] (0.8, 0) to(0.8, 1);
  \draw[dashed] (1.3,0) to (0.8,1);
  \draw[dashed] (0.8,1) to (0.8,2);
  \node at (0.8, 1) [bcirc] {};
  \node at (0.3,0) [dcirc] {};
  \node at (0.8,0) [dcirc] {};
  \node at (1.3,0) [dcirc] {};
  \draw[fill=white] (.8,2) circle (.075);
\end{tikzpicture}
}
\newcommand{\OnePointThree}{
\begin{tikzpicture}[baseline,valign,decoration={markings, 
    mark= at position 0.52 with {\arrow{stealth}},
     mark= at position 1 with {\arrow{stealth}}}]
  \draw[thick] (0, 0) -- (2, 0);
  \draw[dashed] (1, 0) to (1, 0.5);
  \draw[dashed] (1,1.5) to (1,2);
  \draw[postaction=decorate] (1,1) circle (0.5);
  \node at (1,0) [dcirc] {};
  \node at (1,0.5) [rcirc] {};
  \node at (1,1.5) [rcirc] {};
  \draw[fill=white] (1,2) circle (.075);
\end{tikzpicture}
}
\newcommand{\OnePointFour}{
\begin{tikzpicture}[baseline,valign]
  \draw[thick] (0, 0) -- (2, 0);
  \draw[dashed] (1, 0) to (1, 2);
  \draw[dashed] (1,1) circle (0.5);
  \node at (1,0) [dcirc] {};
  \node at (1,0.5) [bcirc] {};
  \node at (1,1.5) [bcirc] {};
  \draw[fill=white] (1,2) circle (.075);
\end{tikzpicture}
}
\newcommand{\OnePointFive}{
\begin{tikzpicture}[baseline,valign]
  \draw[thick] (0, 0) -- (2, 0);
  \draw[dashed] (0.5, 0) to (0.5, 1);
  \draw[dashed] (1,0) to (1.25,1);
  \draw[dashed] (1.5,0) to (1.25,1);
  \draw[dashed] (0.5,1) to[in=90,out=90] (1.25,1);
  \draw[dashed] (0.5,1) to (0.2,2);
  \draw[dashed] (0.5,1) to (1.25,1);
  \node at (1,0) [dcirc] {};
  \node at (0.5,0) [dcirc] {};
  \node at (1.5,0) [dcirc] {};
  \node at (0.5,1) [bcirc] {};
  \node at (1.25,1) [bcirc] {};
  \draw[fill=white] (.2,2) circle (.075);
\end{tikzpicture}
}
\newcommand{\OnePointSix}{
\begin{tikzpicture}[baseline,valign]
  \draw[thick] (0, 0) -- (2, 0);
  \draw[dashed] (0.5, 0) to(1, 1);
  \draw[dashed] (1, 0) to(1, 1);
  \draw[dashed] (1.5,0) to (1,1);
  \draw[dashed] (1,1) to (1,2);
  \draw[dashed] (1,1.5) to (0.2,0);
  \draw[dashed] (1,1.5) to (1.8,0);
  \node at (1, 1) [bcirc] {};
  \node at (0.5,0) [dcirc] {};
  \node at (1,0) [dcirc] {};
  \node at (1.5,0) [dcirc] {};
  \node at (1,1.5) [bcirc] {};
  \node at (0.2,0) [dcirc] {};
  \node at (1.8,0) [dcirc] {};
  \draw[fill=white] (1,2) circle (.075);
\end{tikzpicture}
}
\newcommand{\OnePointSeven}{
\begin{tikzpicture}[baseline,valign,decoration={markings, 
    mark= at position 0.52 with {\arrow{stealth}},
     mark= at position 1 with {\arrow{stealth}}}]
  \draw[thick] (0, 0) -- (2, 0);
  \draw[dashed] (0.5, 0) to(1, 0.75);
  \draw[dashed] (1, 0) to(1, 0.75);
  \draw[dashed] (1.5,0) to (1,0.75);
  \draw[dashed] (1,0.75) to (1,1);
  \draw[dashed] (1,1.5) to (1,2);
  \draw[postaction=decorate] (1,1.25) circle (0.25);
  \node at (1, 0.75) [bcirc] {};
  \node at (0.5,0) [dcirc] {};
  \node at (1,0) [dcirc] {};
  \node at (1.5,0) [dcirc] {};
  \node at (1,1.5) [rcirc] {};
  \node at (1,1) [rcirc] {};
  \draw[white] (0,3) -- (2,3);
  \draw[fill=white] (1,2) circle (.075);
\end{tikzpicture}
}
\newcommand{\OnePointEight}{
 \begin{tikzpicture}[baseline,valign,decoration={markings, 
    mark= at position 0.45 with {\arrow{stealth}},
     mark= at position 0.9 with {\arrow{stealth}}}]
  \draw[thick] (0, 0) -- (2, 0);
  \draw[dashed] (0.4, 0) to(0.55, 0.20);
  \draw[dashed] (0.85, 0.75) to(1, 1);
  \draw[dashed] (1, 0) to(1, 1);
  \draw[dashed] (1.5,0) to (1,1);
  \draw[dashed] (1,1) to (1,2);
  \draw[postaction=decorate] (0.65,0.45) circle (0.25);
  \node at (1, 1) [bcirc] {};
  \node at (0.4,0) [dcirc] {};
  \node at (1,0) [dcirc] {};
  \node at (1.5,0) [dcirc] {};
  \node at (0.8,0.7) [rcirc] {};
  \node at (0.5,0.2) [rcirc] {};
  \draw[white] (0,3) -- (2,3);
  \draw[fill=white] (1,2) circle (.075);
\end{tikzpicture}
}
\newcommand{\OnePointNine}{
\begin{tikzpicture}[baseline,valign,decoration={markings, 
    mark= at position 0.52 with {\arrow{stealth}},
     mark= at position 1 with {\arrow{stealth}}}]
  \draw[thick] (0, 0) -- (2, 0);
  \draw[dashed] (1, 0) to (1, 0.25);
  \draw[dashed] (1,1.5) to (1,2);
  \draw[postaction=decorate] (1,1.25) circle (0.25);
  \draw[postaction=decorate] (1,0.5) circle (0.25);
  \draw[dashed] (1,0.75) to (1,1);
  \node at (1,0) [dcirc] {};
  \node at (1,0.75) [rcirc] {};
  \node at (1,1) [rcirc] {};
  \node at (1,0.25) [rcirc] {};
  \node at (1,1.5) [rcirc] {};
  \draw[white] (0,3) -- (2,3);
  \draw[fill=white] (1,2) circle (.075);
\end{tikzpicture}
}
\newcommand{\OnePointTen}{
\begin{tikzpicture}[baseline,valign,decoration={markings, 
    mark= at position 0.52 with {\arrow{stealth}}}]
  \draw[thick] (0, 0) -- (2, 0);
  \draw[dashed] (0.5, 0) to(0.5, 1);
  \draw[dashed] (1, 0) to(1, 0.5);
  \draw[dashed] (1.5,0) to (1.5,1);
  \draw[dashed] (1,1.5) to (1,2);
  \draw[postaction=decorate] (0.5,1) to (1,1.5);
  \draw[postaction=decorate] (1,0.5) to (0.5,1);
  \draw[postaction=decorate] (1.5,1) to (1,0.5);
  \draw[postaction=decorate] (1,1.5) to (1.5,1);
  \node at (0.5,0) [dcirc] {};
  \node at (1,0) [dcirc] {};
  \node at (1.5,0) [dcirc] {};
  \node at (0.5,1) [rcirc] {};
  \node at (1,0.5) [rcirc] {};
  \node at (1,1.5) [rcirc] {};
  \node at (1.5,1) [rcirc] {};
  \draw[white] (0,3) -- (2,3);
  \draw[fill=white] (1,2) circle (.075);
\end{tikzpicture}
}
\newcommand{\OnePointEleven}{
\begin{tikzpicture}[baseline,valign,decoration={markings, 
    mark= at position 0.15 with {\arrow{stealth}},
     mark= at position 0.4 with {\arrow{stealth}},
     mark= at position 0.65 with {\arrow{stealth}},
     mark= at position 0.9 with {\arrow{stealth}}}]
  \draw[thick] (0, 0) -- (2, 0);
  \draw[dashed] (1, 0) to (1, 0.5);
  \draw[dashed] (0.5,1) to (1.5,1);
  \draw[dashed] (1,1.5) to (1,2);
  \draw[postaction=decorate] (1,1) circle (0.5);
  \node at (1,0) [dcirc] {};
  \node at (1,0.5) [rcirc] {};
  \node at (1,1.5) [rcirc] {};
  \node at (0.5,1) [rcirc] {};
  \node at (1.5,1) [rcirc] {};
  \draw[white] (0,3) -- (2,3);
  \draw[fill=white] (1,2) circle (.075);
\end{tikzpicture}
}

\newcommand{\TwoPointScalarsOne}{
\begin{tikzpicture}[baseline,valign]
  \draw[thick] (0, 0) -- (2, 0);
  \draw[dashed] (0.3, 0) to[out=90,in=90] (1.7, 0);
  \draw[white] (0.5,0.9) -- (1.5,0.9);
  \draw[fill=white] (.3,0) circle (.075);
  \draw[fill=white] (1.7,0) circle (.075);
\end{tikzpicture}
}
\newcommand{\TwoPointScalarsTwo}{
\begin{tikzpicture}[baseline,valign,decoration={markings, 
    mark= at position 0.3 with {\arrow{stealth}},
     mark= at position 0.78 with {\arrow{stealth}}}]
  \draw[thick] (0, 0) -- (2, 0);
  \draw[dashed] (0.2, 0) to[out=90,in=50] (0.7, 0.5);
  \draw[dashed] (1.3, 0.5) to[out=140,in=90] (1.8, 0);
  \draw[postaction=decorate] (1,0.5) circle (0.3);
  \node at (0.7,0.5) [rcirc] {};
  \node at (1.3,0.5) [rcirc] {};
  \draw[white] (0.5,0.9) -- (1.5,0.9);
  \draw[fill=white] (.2,0) circle (.075);
  \draw[fill=white] (1.8,0) circle (.075);
\end{tikzpicture}
}
\newcommand{\TwoPointScalarsThree}{
\begin{tikzpicture}[baseline,valign]
  \draw[thick] (0, 0) -- (2, 0);
  \draw[dashed] ( 0.4, 0) -- (1, 0.8) -- (1.6, 0);
  \draw[dashed] ( 0.75, 0) -- (1, 0.8) -- (1.25, 0);
  \node at (1.0, 0.8) [bcirc] {};
  \node at (0.75, 0) [dcirc] {};
  \node at (1.25, 0) [dcirc] {};
  \draw[white] (0.5,0.9) -- (1.5,0.9);
  \draw[fill=white] (.4,0) circle (.075);
  \draw[fill=white] (1.6,0) circle (.075);
\end{tikzpicture}
}

\newcommand{\TwoPointFermionsOne}{
\begin{tikzpicture}[baseline,valign,decoration={markings, 
    mark= at position 0.55 with {\arrow{stealth}}}]
  \draw[thick] (0, 0) -- (2, 0);
  \draw[postaction={decorate}] (1.7, 0) to[out=90,in=90] (0.3, 0);
  \draw[white] (0,0.65) -- (2,0.65);
  \draw[fill=white] (.3,0) circle (.075);
  \draw[fill=white] (1.7,0) circle (.075);
\end{tikzpicture}
}
\newcommand{\TwoPointFermionsTwo}{
\begin{tikzpicture}[baseline,valign,decoration={markings, 
    mark= at position 0.3 with {\arrow{stealth}},
    mark= at position 0.8 with {\arrow{stealth}}}]
  \draw[thick] (0, 0) -- (2, 0);
  \draw[postaction=decorate]  (1.4, 0) to[out=90,in=90] (0.3, 0);
  \draw[dashed] (0.85,0.3) to[out=90, in=90] (1.7,0);
  \node at (0.85,0.3) [rcirc] {};
  \node at (1.7,0) [dcirc] {};
  \draw[white] (0,0.65) -- (2,0.65);
  \draw[fill=white] (.3,0) circle (.075);
  \draw[fill=white] (1.4,0) circle (.075);
\end{tikzpicture}
}
\newcommand{\TwoPointFermionsThree}{
\begin{tikzpicture}[baseline,valign,decoration={markings, 
    mark= at position 0.55 with {\arrow{stealth}},
    mark= at position 0.2 with {\arrow{stealth}},
    mark= at position 0.9 with {\arrow{stealth}}}]
  \draw[thick] (0, 0) -- (2, 0);
  \draw[postaction={decorate}] (1.7, 0) to[out=90,in=90] (0.3, 0);
  \draw[dashed] (1.35,0.35) to[out=90, in=90] (0.6,0.35);
  \node at (0.6,0.35) [rcirc] {};
  \node at (1.35,0.35) [rcirc] {};
  \draw[white] (0,0.65) -- (2,0.65);
  \draw[fill=white] (.3,0) circle (.075);
  \draw[fill=white] (1.7,0) circle (.075);
\end{tikzpicture}
}
\newcommand{\TwoPointFermionsFour}{
\begin{tikzpicture}[baseline,valign,decoration={markings, 
    mark= at position 0.55 with {\arrow{stealth}},
    mark= at position 0.2 with {\arrow{stealth}},
    mark= at position 0.9 with {\arrow{stealth}}}]
  \draw[thick] (0, 0) -- (2, 0);
  \draw[postaction={decorate}] (1.7, 0) to[out=90,in=90] (0.3, 0);
  \draw[dashed] (0.6,0.35) -- (0.6,0);
  \draw[dashed] (1.35,0.35) -- (1.35,0);
  \node at (0.6,0.35) [rcirc] {};
  \node at (1.35,0.35) [rcirc] {};
  \node at (0.6,0) [dcirc] {};
  \node at (1.35,0) [dcirc] {};
  \draw[white] (0,0.65) -- (2,0.65);
  \draw[fill=white] (.3,0) circle (.075);
  \draw[fill=white] (1.7,0) circle (.075);
\end{tikzpicture}
}

\newcommand{\ThreePointScalars}{
\begin{tikzpicture}[baseline,valign]
  \draw[thick] (-0.4, 0) -- (1.4, 0);
  \draw[dashed] (-0.2, 0) -- (0.5, 0.7) -- (1.2, 0);
  \draw[dashed] ( 0.3, 0) -- (0.5, 0.7) -- (0.7, 0);
  \node at (0.5, 0.7) [bcirc] {};
  \node at (0.3,0) [dcirc] {};
  \draw[fill=white] (-.2,0) circle (.075);
  \draw[fill=white] (1.2,0) circle (.075);
  \draw[fill=white] (.7,0) circle (.075);
\end{tikzpicture}
}

\newcommand{\ThreePointScalarsSquareOne}{
\begin{tikzpicture}[baseline,valign]
  \draw[thick] (-0.4, 0) -- (1.4, 0);
  \draw[dashed] (-0.2, 0) to[out=90,in=90] (1.2, 0);
  \draw[dashed] (0.2, 0) to[out=90,in=90] (1.2, 0);
  \draw[fill=white] (-.2,0) circle (.075);
  \draw[fill=white] (1.2,0) circle (.075);
  \draw[fill=white] (.2,0) circle (.075);
\end{tikzpicture}
}
\newcommand{\ThreePointScalarsSquareTwo}{
\begin{tikzpicture}[baseline,valign]
  \draw[thick] (-0.4, 0) -- (1.4, 0);
  \draw[dashed] (-0.2, 0) to[out=90,in=90] (1.2, 0);
  \draw[dashed] ( 0.2, 0) -- (0.5, 0.4) -- (1.2, 0);
  \node at (0.5, 0.4) [bcirc] {};
  \draw[fill=white] (-.2,0) circle (.075);
  \draw[fill=white] (1.2,0) circle (.075);
  \draw[fill=white] (.2,0) circle (.075);
\end{tikzpicture} 
}
\newcommand{\ThreePointScalarsSquareThree}{
\begin{tikzpicture}[baseline,valign]
  \draw[thick] (-0.4, 0) -- (1.4, 0);
  \draw[dashed] ( 0.2, 0) to[out=90,in=90] (1.2, 0);
  \draw[dashed] (-0.2, 0) to[out=90,in=90] (1.2, 0);
  \draw[dashed] (0.5, 0) -- (0.7, 0.25);
  \draw[dashed] (0.9, 0) -- (0.7, 0.25);
  \node at (0.7, 0.28) [bcirc] {};
  \node at (0.5, 0.00) [dcirc] {};
  \node at (0.9, 0.00) [dcirc] {};
  \draw[fill=white] (-.2,0) circle (.075);
  \draw[fill=white] (1.2,0) circle (.075);
  \draw[fill=white] (.2,0) circle (.075);
\end{tikzpicture}
}
\newcommand{\ThreePointScalarsSquareFour}{
\begin{tikzpicture}[baseline,valign,decoration={markings, 
    mark= at position 0.37 with {\arrow{stealth}},
    mark= at position 0.82 with {\arrow{stealth}}}]
  \draw[thick] (-0.4, 0) -- (1.4, 0);
  \draw[dashed] (-0.2, 0) to[out=90,in=50] (0.02,0.32);
  \draw[dashed] (0.38,0.45) to[out=20, in=90] (1.2, 0);
  \draw[dashed] (0.2, 0) to[out=90,in=90] (1.2, 0);
  \draw[postaction=decorate] (0.18,0.4) circle (0.18);
  \node at (0.38,0.45) [rcirc] {};
  \node at (0.02,0.32) [rcirc] {};
  \draw[fill=white] (-.2,0) circle (.075);
  \draw[fill=white] (1.2,0) circle (.075);
  \draw[fill=white] (.2,0) circle (.075);
\end{tikzpicture}
}

\newcommand{\ThreePointScalarsTwoFermions}{
\begin{tikzpicture}[baseline,valign,decoration={markings, 
    mark= at position 0.3 with {\arrow{stealth}},   
    mark= at position 0.8 with {\arrow{stealth}}}]
  \draw[thick] (0, 0) -- (2, 0);
  \draw[postaction=decorate]  (1.4, 0) to[out=90,in=90] (0.3, 0);
  \draw[dashed] (0.85,0.3) to[out=90, in=90] (1.7,0);
  \node at (0.85,0.3) [rcirc] {};
  \draw[white] (0,-0.2) -- (2,-0.2);
  \draw[fill=white] (1.4,0) circle (.075);
  \draw[fill=white] (.3,0) circle (.075);
  \draw[fill=white] (1.7,0) circle (.075);
\end{tikzpicture}
}

\newcommand{\FourPointScalarsOne}{
\begin{tikzpicture}[baseline,valign]
  \draw[thick] (-0.4, 0) -- (1.4, 0);
  \draw[dashed] (-0.2, 0) to[out=90,in=90] (0.4, 0);
  \draw[dashed] (0.6, 0) to[out=90,in=90] (1.2, 0);
  \draw[white] (0,0.8) -- (1,0.8);
  \draw[fill=white] (-.2,0) circle (.075);
  \draw[fill=white] (.4,0) circle (.075);
  \draw[fill=white] (.6,0) circle (.075);
  \draw[fill=white] (1.2,0) circle (.075);
\end{tikzpicture}
}
\newcommand{\FourPointScalarsTwo}{
\begin{tikzpicture}[baseline,valign]
  \draw[thick] (-0.4, 0) -- (1.6, 0);
  \draw[dashed] (-0.2, 0) to[out=90,in=90] (0.4, 0);
  \draw[dashed] ( 0.6, 0) -- (1, 0.7) -- (1.4, 0);
  \draw[dashed] ( 0.85, 0) -- (1, 0.7) -- (1.15, 0);
  \node at (1.0, 0.70) [bcirc] {};
  \node at (0.85, 0.00) [dcirc] {};
  \node at (1.15, 0.00) [dcirc] {};
  \draw[white] (0,0.8) -- (1,0.8);
  \draw[fill=white] (-.2,0) circle (.075);
  \draw[fill=white] (.4,0) circle (.075);
  \draw[fill=white] (.6,0) circle (.075);
  \draw[fill=white] (1.4,0) circle (.075);
\end{tikzpicture}
}
\newcommand{\FourPointScalarsThree}{
\begin{tikzpicture}[baseline,valign,decoration={markings, 
    mark= at position 0.3 with {\arrow{stealth}},
     mark= at position 0.78 with {\arrow{stealth}}}]
  \draw[thick] (-0.3, 0) -- (1.5, 0);
  \draw[dashed] (-0.2, 0) to[out=90,in=90] (0.4, 0);
  \draw[dashed] (0.6, 0) to[out=90,in=50] (0.82, 0.3);
  \draw[dashed] (1.18, 0.3) to[out=50,in=90] (1.4, 0);
  \draw[postaction=decorate] (1,0.3) circle (0.18);
  \node at (0.82,0.3) [rcirc] {};
  \node at (1.18,0.3) [rcirc] {};
  \draw[white] (0,0.8) -- (1,0.8);
  \draw[fill=white] (-.2,0) circle (.075);
  \draw[fill=white] (.4,0) circle (.075);
  \draw[fill=white] (.6,0) circle (.075);
  \draw[fill=white] (1.4,0) circle (.075);
\end{tikzpicture}
}
\newcommand{\FourPointScalarsFour}{
\begin{tikzpicture}[baseline,valign]
  \draw[thick] (-0.2, 0) -- (1.2, 0);
  \draw[dashed] (0, 0) -- (0.5, 0.7) -- (1, 0);
  \draw[dashed] ( 0.3, 0) -- (0.5, 0.7) -- (0.7, 0);
  \node at (0.5, 0.7) [bcirc] {};
  \draw[white] (0,0.8) -- (1,0.8);
  \draw[fill=white] (0,0) circle (.075);
  \draw[fill=white] (.3,0) circle (.075);
  \draw[fill=white] (.7,0) circle (.075);
  \draw[fill=white] (1,0) circle (.075);
\end{tikzpicture}
}

\newcommand{\FourPointFermionsOne}{
\begin{tikzpicture}[baseline,valign,decoration={markings, 
    mark= at position 0.55 with {\arrow{stealth}}}]
  \draw[thick] (-0.5, 0) -- (1.5, 0);
  \draw[postaction={decorate}] (1.3, 0) to[out=90,in=90] (0.6, 0);
  \draw[postaction={decorate}] (0.4, 0) to[out=90,in=90] (-0.3, 0);
  \draw[white] (0,0.65) -- (1,0.65);
  \draw[fill=white] (-.3,0) circle (.075);
  \draw[fill=white] (.4,0) circle (.075);
  \draw[fill=white] (.6,0) circle (.075);
  \draw[fill=white] (1.3,0) circle (.075);
\end{tikzpicture}
}
\newcommand{\FourPointFermionsTwo}{
\begin{tikzpicture}[baseline,valign,decoration={markings, 
    mark= at position 0.55 with {\arrow{stealth}}}]
  \draw[thick] (-0.5, 0) -- (1.5, 0);
  \draw[postaction={decorate}] (0.4, 0) to[out=90,in=90] (-0.3, 0);
  \draw[postaction={decorate}] (1.3, 0) to[out=90,in=90] (0.6, 0);
  \draw[dashed] (1.2,0.15) to[out=90, in=90] (0.7,0.15);
  \node at (0.7,0.15) [rcirc] {};
  \node at (1.2,0.15) [rcirc] {};
  \draw[white] (0,0.65) -- (1,0.65);
  \draw[fill=white] (-.3,0) circle (.075);
  \draw[fill=white] (.4,0) circle (.075);
  \draw[fill=white] (.6,0) circle (.075);
  \draw[fill=white] (1.3,0) circle (.075);
\end{tikzpicture} 
}
\newcommand{\FourPointFermionsThree}{
\begin{tikzpicture}[baseline,valign,decoration={markings, 
    mark= at position 0.55 with {\arrow{stealth}}}]
  \draw[thick] (-0.5, 0) -- (1.5, 0);
  \draw[postaction={decorate}] (0.4, 0) to[out=90,in=90] (-0.3, 0);
  \draw[postaction={decorate}] (1.3, 0) to[out=90,in=90] (0.6, 0);
  \draw[dashed] (1.1,0.18) -- (1.1,0);
  \draw[dashed] (0.8,0.18) -- (0.8,0);
  \node at (0.8,0.18) [rcirc] {};
  \node at (1.1,0.18) [rcirc] {};
  \node at (0.8,0) [dcirc] {};
  \node at (1.1,0) [dcirc] {};
  \draw[white] (0,0.65) -- (1,0.65);
  \draw[fill=white] (-.3,0) circle (.075);
  \draw[fill=white] (.4,0) circle (.075);
  \draw[fill=white] (.6,0) circle (.075);
  \draw[fill=white] (1.3,0) circle (.075);
\end{tikzpicture} 
}
\newcommand{\FourPointFermionsFour}{
\begin{tikzpicture}[baseline,valign,decoration={markings, 
    mark= at position 0.5 with {\arrow{stealth}}}]
  \draw[thick] (-0.5, 0) -- (1.5, 0);
  \draw[postaction=decorate] (0.2,0) -- (0.2,0.4);
  \draw[postaction=decorate] (0.8,0) -- (0.8,0.4);
  \draw[postaction=decorate] (0.2, 0.4) to[out=90,in=90] (-0.2, 0);
  \draw[postaction=decorate] (0.8, 0.4) to[out=90,in=90] (1.2, 0);
  \draw[dashed] (0.2, 0.4) to (0.8, 0.4);
  \node at (0.2,0.4) [rcirc] {};
  \node at (0.8,0.4) [rcirc] {};
  \draw[fill=white] (-.2,0) circle (.075);
  \draw[fill=white] (.2,0) circle (.075);
  \draw[fill=white] (.8,0) circle (.075);
  \draw[fill=white] (1.2,0) circle (.075);
\end{tikzpicture} 
}
\newcommand{\FourPointFermionsFive}{
\begin{tikzpicture}[baseline,valign,decoration={markings, 
    mark= at position 0.6 with {\arrow{stealth}}}]
  \draw[thick] (-0.5, 0) -- (1.5, 0);
  \draw[postaction=decorate] (0.2,0) -- (0.2,0.4);
  \draw[postaction=decorate] (0.8,0) -- (0.8,0.4);
  \draw (0.2,0.4) -- (0.8,0.6);
  \draw (0.8,0.4) -- (0.2,0.6);
  \draw[postaction=decorate] (0.8, 0.6) to[out=20,in=90] (1.2, 0);
  \draw[postaction=decorate] (0.2, 0.6) to[out=160,in=90] (-0.2, 0);
  \draw[dashed] (0.2, 0.4) to (0.8, 0.4);
  \node at (0.2,0.4) [rcirc] {};
  \node at (0.8,0.4) [rcirc] {};
  \draw[fill=white] (-.2,0) circle (.075);
  \draw[fill=white] (.2,0) circle (.075);
  \draw[fill=white] (.8,0) circle (.075);
  \draw[fill=white] (1.2,0) circle (.075);
\end{tikzpicture}  
}

\newcommand{\FourPointFermionsScalarsOne}{
\begin{tikzpicture}[baseline,valign,decoration={markings, 
    mark= at position 0.6 with {\arrow{stealth}}}]
  \draw[thick] (0, 0) -- (2, 0);
  \draw[postaction=decorate] (0.2,0.4) -- (0.2,0);
  \draw[postaction=decorate] (0.8,0) -- (0.8,0.4);
  \draw[dashed] (0.2,0.4) -- (0.5,0.6);
  \draw[dashed] (0.8,0.4) -- (1.2,0.6);
  \draw[dashed] (0.5, 0.6) to[out=20,in=90] (1.2, 0);
  \draw[dashed] (1.2, 0.6) to[out=20,in=90] (1.5, 0);
  \draw[postaction=decorate] (0.8, 0.4) to (0.2, 0.4);
  \node at (0.2,0.4) [rcirc] {};
  \node at (0.8,0.4) [rcirc] {};
  \draw[fill=white] (.2,0) circle (.075);
  \draw[fill=white] (.8,0) circle (.075);
  \draw[fill=white] (1.2,0) circle (.075);
  \draw[fill=white] (1.5,0) circle (.075);
\end{tikzpicture}
}
\newcommand{\FourPointFermionsScalarsTwo}{
\begin{tikzpicture}[baseline,valign,decoration={markings, 
    mark= at position 0.5 with {\arrow{stealth}}}]
  \draw[thick] (0, 0) -- (2, 0);
  \draw[postaction=decorate] (0.2,0.4) -- (0.2,0);
  \draw[postaction=decorate] (0.8,0) -- (0.8,0.4);
  \draw[dashed] (0.2, 0.4) to[out=90,in=90] (1.5, 0);
  \draw[dashed] (0.8, 0.4) to[out=20,in=90] (1.2, 0);
  \draw[postaction=decorate] (0.8, 0.4) to (0.2, 0.4);
  \node at (0.2,0.4) [rcirc] {};
  \node at (0.8,0.4) [rcirc] {};
  \draw[fill=white] (.2,0) circle (.075);
  \draw[fill=white] (.8,0) circle (.075);
  \draw[fill=white] (1.2,0) circle (.075);
  \draw[fill=white] (1.5,0) circle (.075);
\end{tikzpicture}
}

\newcommand{\OnePointSquaredScalarOne}{
\begin{tikzpicture}[baseline,valign]
  \draw[thick] (0.2, 0) -- (1.8, 0);
  \draw[dashed] (0.5, 0) to (1, 1.5);
  \draw[dashed] (1.5, 0) to (1, 1.5);
  \node at (0.5,0) [dcirc] {};
  \node at (1.5,0) [dcirc] {};
  \draw[fill=white] (1,1.5) circle (.075);
\end{tikzpicture}
}
\newcommand{\OnePointSquaredScalarTwo}{
\begin{tikzpicture}[baseline,valign]
  \draw[thick] (0.2, 0) -- (1.8, 0);
  \draw[dashed] (0.5, 0) to (1, 1.5);
  \draw[dashed] (1.5, 0) to (1, 1.5);
  \draw[dashed] (0.9,0) to (1.2,0.8);
  \draw[dashed] (1.2,0) to (1.2,0.8);
  \node at (1.5,0) [dcirc] {};
  \node at (0.5,0) [dcirc] {};
  \node at (0.9,0) [dcirc] {};
  \node at (1.2,0) [dcirc] {};
  \node at (1.2,0.8) [bcirc] {};
  \draw[fill=white] (1,1.5) circle (.075);
\end{tikzpicture}
}
\newcommand{\OnePointSquaredScalarThree}{
\begin{tikzpicture}[baseline,valign]
  \draw[thick] (0.2, 0) -- (1.8, 0);
  \draw[dashed] (0.5, 0) to (1, 0.8);
  \draw[dashed] (1.5, 0) to (1, 0.8);
  \draw[dashed] (1,0.8) to[in=0,out=0] (1,1.5);
  \draw[dashed] (1,0.8) to[in=180,out=180] (1,1.5);
  \node at (0.5,0) [dcirc] {};
  \node at (1.5,0) [dcirc] {};
  \node at (1,0.8) [bcirc] {};
  \draw[fill=white] (1,1.5) circle (.075);
\end{tikzpicture}
}
\newcommand{\OnePointSquaredScalarFour}{
\begin{tikzpicture}[baseline,valign,decoration={markings, 
    mark= at position 0.5 with {\arrow{stealth}},
     mark= at position 1 with {\arrow{stealth}}}]
  \draw[thick] (0.2, 0) -- (1.8, 0);
  \draw[dashed] (0.4, 0) to (0.55, 0.45);
  \draw[dashed] (0.75, 0.95) to (1, 1.5);
  \draw[postaction=decorate] (0.65,0.7) circle (0.25);
  \draw[dashed] (1.5, 0) to (1, 1.5);
  \node at (0.4,0) [dcirc] {};
  \node at (1.5,0) [dcirc] {};
  \node at (0.75,0.95) [rcirc] {};
  \node at (0.55,0.45) [rcirc] {};
  \draw[fill=white] (1,1.5) circle (.075);
\end{tikzpicture}
}

\newcommand{\OnePointBilinear}{
\begin{tikzpicture}[baseline,valign,decoration={markings, 
    mark= at position 0.55 with {\arrow{stealth}}}]
  \draw[thick] (0.2,0) -- (1.8,0);
  \draw[dashed] (1,0) -- (1,0.7);
  \draw[postaction=decorate] (1,0.7) to[in=0,out=0] (1,1.5);
  \draw[postaction=decorate] (1,1.5) to[in=180,out=180] (1,0.7);
  \node at (1,0.7) [rcirc] {};
  \node at (1,0) [dcirc] {};
  \draw[fill=white] (1,1.5) circle (.075);
 \end{tikzpicture}
 }
 
\newcommand{\BulkToDefectTwoPointOne}{
\begin{tikzpicture}[baseline,valign]
  \draw[thick] (0.2, 0) -- (1.8, 0);
  \draw[dashed] (1, 0) to (1, 1.5);
  \draw[fill=white] (1,1.5) circle (.075);
\end{tikzpicture} 
}
\newcommand{\BulkToDefectTwoPointTwo}{
\begin{tikzpicture}[baseline,valign]
  \draw[thick] (0.2, 0) -- (1.8, 0);
  \draw[dashed] (0.5, 0) to(1, 0.7);
  \draw[dashed] (1, 0) to(1, 0.7);
  \draw[dashed] (1.5,0) to (1,0.7);
  \draw[dashed] (1,0.7) to (1,1.5);
  \node at (1, 0.7) [bcirc] {};
  \node at (0.5,0) [dcirc] {};
  \node at (1.5,0) [dcirc] {};
  \draw[fill=white] (1,1.5) circle (.075);
\end{tikzpicture}
}
\newcommand{\BulkToDefectTwoPointThree}{
\begin{tikzpicture}[baseline,valign,decoration={markings, 
    mark= at position 0.52 with {\arrow{stealth}},
    mark= at position 1 with {\arrow{stealth}}}]
  \draw[thick] (0.2, 0) -- (1.8, 0);
  \draw[dashed] (1, 0) to (1, 0.5);
  \draw[dashed] (1,1) to (1,1.5);
  \draw[postaction=decorate] (1,0.75) circle (0.25);
  \node at (1,0.5) [rcirc] {};
  \node at (1,1) [rcirc] {};
  \draw[fill=white] (1,1.5) circle (.075);
\end{tikzpicture}
}

\newcommand{\TwoPointBulkScalarsOne}{
\begin{tikzpicture}[baseline,valign]
  \draw[thick] (0,0) -- (3,0);
  \draw[dashed] (0.5,1) -- (2.5,1);
  \draw[white] (0,2) -- (3,2);
  \draw[fill=white] (.5,1) circle (.075);
  \draw[fill=white] (2.5,1) circle (.075);
 \end{tikzpicture}
 }
 \newcommand{\TwoPointBulkScalarsTwo}{
 \begin{tikzpicture}[baseline,valign,decoration={markings,
 mark= at position 0.27 with {\arrow{stealth}},
  mark= at position 0.77 with {\arrow{stealth}}}]
  \draw[thick] (0,0) -- (3,0);
  \draw[dashed] (0.5,1) -- (1,1);
  \draw[postaction=decorate] (1.5,1) circle (0.5);
  \draw[dashed] (2,1) -- (2.5,1);
  \node at (1,1) [rcirc] {};
  \node at (2,1) [rcirc] {};
  \draw[white] (0,2) -- (3,2);
  \draw[fill=white] (.5,1) circle (.075);
  \draw[fill=white] (2.5,1) circle (.075);
 \end{tikzpicture}
 }
 \newcommand{\TwoPointBulkScalarsThree}{
  \begin{tikzpicture}[baseline,valign]
  \draw[thick] (0, 0) -- (3, 0);
  \draw[dashed] (1, 0) to (1, 2);
  \draw[dashed] (2, 0) to (2, 2);
  \node at (1,0) [dcirc] {};
  \node at (2,0) [dcirc] {};
  \draw[fill=white] (1,2) circle (.075);
  \draw[fill=white] (2,2) circle (.075);
\end{tikzpicture}
 }
 \newcommand{\TwoPointBulkScalarsFour}{
 \begin{tikzpicture}[baseline,valign]
  \draw[thick] (0, 0) -- (3, 0);
  \draw[dashed] (0.3, 0) to(0.8, 1);
  \draw[dashed] (0.8, 0) to(0.8, 1);
  \draw[dashed] (1.3,0) to (0.8,1);
  \draw[dashed] (0.8,1) to (0.8,2);
  \draw[dashed] (2,0) to (2,2);
  \node at (0.8, 1) [bcirc] {};
  \node at (0.3,0) [dcirc] {};
  \node at (0.8,0) [dcirc] {};
  \node at (1.3,0) [dcirc] {};
  \node at (2,0) [dcirc] {};
  \draw[white] (0,3) -- (3,3);
  \draw[fill=white] (.8,2) circle (.075);
  \draw[fill=white] (2,2) circle (.075);
\end{tikzpicture} 
}
\newcommand{\TwoPointBulkScalarsFive}{
\begin{tikzpicture}[baseline,valign,decoration={markings,
 mark= at position 0.53 with {\arrow{stealth}},
  mark= at position 1 with {\arrow{stealth}}}]
  \draw[thick] (0, 0) -- (3, 0);
  \draw[dashed] (1, 0) to (1, 0.5);
  \draw[dashed] (1,1.5) to (1,2);
  \draw[dashed] (2,0) to (2,2);
  \draw[postaction=decorate] (1,1) circle (0.5);
  \node at (1,0) [dcirc] {};
  \node at (1,0.5) [rcirc] {};
  \node at (1,1.5) [rcirc] {};
  \node at (2,0) [dcirc] {};
  \draw[white] (0,3) -- (3,3);
  \draw[fill=white] (1,2) circle (.075);
  \draw[fill=white] (2,2) circle (.075);
\end{tikzpicture}
}
\newcommand{\TwoPointBulkScalarsSix}{
\begin{tikzpicture}[baseline,valign]
  \draw[thick] (0,0) -- (3,0); 
  \draw[dashed] (1,2) -- (2,0);
  \draw[dashed] (1,0) -- (2,2);
  \node at (1.5,1) [bcirc] {};
  \node at (1,0) [dcirc] {};
  \node at (2,0) [dcirc] {};
  \draw[white] (0,3) -- (3,3);
  \draw[fill=white] (1,2) circle (.075);
  \draw[fill=white] (2,2) circle (.075);
\end{tikzpicture}
}

\newcommand{\TwoPointBulkFermionsOne}{
\begin{tikzpicture}[baseline,valign,decoration={markings,
  mark= at position 0.55 with {\arrow{stealth}}}]
  \draw[thick] (0,0) -- (2,0);
  \draw[postaction=decorate] (1.8,1) -- (0.2,1);
  \draw[white] (0,1.95) -- (2,1.95);
  \draw[fill=white] (.2,1) circle (.075);
  \draw[fill=white] (1.8,1) circle (.075);
 \end{tikzpicture}
 }
 \newcommand{\TwoPointBulkFermionsTwo}{
 \begin{tikzpicture}[baseline,valign,decoration={markings,
  mark= at position 0.55 with {\arrow{stealth}}}]
  \draw[thick] (0,0) -- (2,0);
  \draw[dashed] (1,0) -- (1,1);
  \draw[postaction=decorate] (1,1) -- (0.5,2);
  \draw[postaction=decorate] (1.5,2) -- (1,1);
  \node at (1,1) [rcirc] {};
  \node at (1,0) [dcirc] {};
  \draw[fill=white] (.5,2) circle (.075);
  \draw[fill=white] (1.5,2) circle (.075);
 \end{tikzpicture}
 }
 \newcommand{\TwoPointBulkFermionsThree}{
  \begin{tikzpicture}[baseline,valign,decoration={markings,
  mark= at position 0.55 with {\arrow{stealth}}}]
  \draw[thick] (0, 0) -- (2, 0);
  \draw[postaction=decorate] (1.8, 1) to (0.2, 1);
  \draw[dashed] (0.7,1) to[in=90,out=90] (1.3,1);
  \node at (0.7,1) [rcirc] {};
  \node at (1.3,1) [rcirc] {};
  \draw[white] (0,1.95) -- (2,1.95);
  \draw[fill=white] (.2,1) circle (.075);
  \draw[fill=white] (1.8,1) circle (.075);
\end{tikzpicture}
}
\newcommand{\TwoPointBulkFermionsFour}{
\begin{tikzpicture}[baseline,valign,decoration={markings,
  mark= at position 0.55 with {\arrow{stealth}}}]
  \draw[thick] (0, 0) -- (2, 0);
  \draw[dashed] (0.5, 0) to(0.5, 1);
  \draw[dashed] (1.5, 0) to(1.5, 1);
  \draw[postaction=decorate] (1.5,1) to (0.5,1);
  \draw[postaction=decorate] (0.5,1) to (0.5,2);
  \draw[postaction=decorate] (1.5,2) to (1.5,1);
  \node at (0.5, 1) [rcirc] {};
  \node at (1.5,1) [rcirc] {};
  \node at (0.5,0) [dcirc] {};
  \node at (1.5,0) [dcirc] {};
  \draw[fill=white] (.5,2) circle (.075);
  \draw[fill=white] (1.5,2) circle (.075);
\end{tikzpicture} 
}
\usepackage{dsfont}
\usepackage{float}
\usepackage{pgfplots}
\pgfplotsset{compat=1.14}

\let\oldbfseries=\bfseries
\let\oldmdseries=\mdseries
\let\oldnormalfont=\normalfont
\renewcommand{\bfseries}{\oldbfseries\boldmath}
\renewcommand{\mdseries}{\oldmdseries\unboldmath}
\renewcommand{\normalfont}{\oldnormalfont\unboldmath}

\makeatletter
\newlength{\apb@width}
\newcommand{\autoparbox}[2][c]{\settowidth{\apb@width}{#2}\parbox[#1]{\apb@width}{#2}}

\makeatother

\DeclareMathOperator{\tr}{tr}

\def\Fm{{\mathcal{F}}}

\def\Km{{\mathcal{K}}}

\def\Nm{{\mathcal{N}}}
\def\Om{{\mathcal{O}}}

\def\eps{\epsilon}
\def\veps{\varepsilon}

\def\pd{\partial}

\newcommand{\beq}{\begin{equation}}
\newcommand{\eeq}{\end{equation}}

\newmuskip\pFqskip
\mathchardef\pFcomma=\mathcode`,

\makeatletter
\renewcommand*\env@matrix[1][\arraystretch]{%
  \edef\arraystretch{#1}%
  \hskip -\arraycolsep
  \let\@ifnextchar\new@ifnextchar
  \array{*\c@MaxMatrixCols c}}
\makeatother

\graphicspath{{./figures/}}

\begin{document}  
%%%%%%%%%%%%%%%%%%%%%%%%%%%%%%
	
	\begin{titlepage}
	\noindent HU-EP-23/10-RTG \\ 
	  DESY-23-055 \bigskip
		
		\medskip
		\begin{center} 
			{\LARGE \bf Line defect correlators in fermionic CFTs}
			
			\bigskip
			\bigskip
			\bigskip
			
		{Julien Barrat$^{1}$, Pedro Liendo$^{2}$, Philine van Vliet$^{2}$\\ }
		\bigskip
		\bigskip
		\textit{${}^{1}$
		Institut f\"ur Physik und IRIS Adlershof, Humboldt-Universit{\"a}t zu Berlin, Zum Gro{\ss}en Windkanal 2, 12489 Berlin, Germany}
	
		\textit{		${}^{2}$
		Deutsches Elektronen-Synchrotron DESY, Notkestr. 85, 22607 Hamburg, Germany}
		\vskip 5mm

		\texttt{barratju@hu-berlin.de,~pedro.liendo@desy.de,	philine.vanvliet@desy.de} \\
		\end{center}

		\bigskip
		\bigskip

		\begin{abstract}
		
            \noindent Scalar-fermion models, such as the Gross--Neveu--Yukawa model, admit natural $1d$ defects given by the exponential of a scalar field integrated along a straight line.
            In $4-\varepsilon$ dimensions the defect coupling is weakly relevant and the setup defines a non-trivial interacting defect CFT.
            In this work we study correlation functions on these defect CFTs to order $\varepsilon$.
            We focus on $1d$ correlators constrained to the line, which include canonical operators like the displacement and the one-dimensional analog of the spin field.
            These results give access to perturbative CFT data that can be used as input in the numerical bootstrap.
            We also consider local operators outside the line, in particular two-point functions of scalars whose dynamics are non-trivial due to the presence of the defect.
		
		\end{abstract}
		
		\noindent 
		
	\end{titlepage}
	
%%%%%%%%%%%%%%%%%%%%%%%%%%%%%%%%%%%%%
	
	\setcounter{tocdepth}{2}	
	\tableofcontents
	\newpage
	
%%%%%%%%%%%%%%%%%%%%%%%%%%%%%%%%%%%%%
%%%%%%%%%%%%%%%%%%%%%%%%%%%%%%%%%%%%%

\section{Introduction}
\label{sec:intro}

Extended objects or defects are important observables that probe new dynamics inaccessible to local operators. In the context of conformal field theories (CFTs), they have gotten significant attention in recent years as part of the defect conformal bootstrap program \cite{Liendo:2012hy,Billo:2016cpy,Lauria:2017wav,Lemos:2017vnx,Lauria:2018klo}. There have been many recent developments which include 
studies of conformal boundaries \cite{Gliozzi:2015qsa,Rastelli:2017ecj,deLeeuw:2017dkd,Bissi:2018mcq,Kaviraj:2018tfd,Mazac:2018biw,Dey:2020jlc,Dey:2020lwp,Behan:2020nsf,Barrat:2020vch, Antunes:2021qpy,Barrat:2021tpn, Bissi:2022bgu,Gimenez-Grau:2020jvf,Behan:2021tcn,Nishioka:2022odm,Barrat:2022eim, Herzog:2023dop}, monodromy defects \cite{Billo:2013jda,Gaiotto:2013nva,Yamaguchi:2016pbj,Soderberg:2017oaa,Giombi:2021uae,Gimenez-Grau:2021wiv,Bianchi:2021snj,SoderbergRousu:2023pbe}, and line and surface defects \cite{Bianchi:2018zpb,Liendo:2018ukf,Gimenez-Grau:2019hez,Lauria:2020emq,Barrat:2020vch,Drukker:2020atp,Penati:2021tfj,Gimenez-Grau:2022czc,Drukker:2022txy}.
In this work we consider line defects in fermionic models. 
The general setup is a scalar-fermion theory in $4-\veps$ dimensions with Yukawa interactions and a quartic potential for the scalars. Although some parts of our analysis are quite general, we mostly focus on three interesting models, characterized by the number of scalar fields in the $4-\veps$ description.

The \textit{Gross--Neveu--Yukawa} (GNY) model contains a single real scalar and $N_f$ Dirac fermions. It has a perturbative fixed point in $4-\veps$ and is expected to flow to the same three-dimensional universality class as the classic Gross--Neveu (GN) model. The GN model is a fermionic CFT  with a four-fermion interaction originally formulated in $d$ dimensions displaying asymptotic freedom in the large-$N_f$ limit \cite{Gross:1974jv}, and is believed to have a non-trivial interacting fixed point in $2+\veps$ dimensions.
In $d>2$ dimensions, the GN model is renormalizable in the large-$N_f$ limit \cite{Gross:1975vu}, but not for finite $N_f$. 
The GNY model can therfore be considered a UV-completion of the GN model \cite{Hasenfratz:1991it,Zinn-Justin:1991ksq}.\footnote{For a specific number of fermions, $N_f = 1/4$, the interacting fixed point in $d = 3$ exhibits emergent supersymmetry (SUSY) \cite{Lee:2006if,Grover:2013rc}.}

If we consider a complex scalar and $N_f$ Dirac fermions we obtain the \textit{Nambu--Jona-Lasinio--Yukawa} (NJLY) model. Similarly to the GNY model, the NJLY model can be thought of as a UV completion of the Nambu--Jona-Lasinio (NJL) model \cite{Nambu:1961tp}, a purely fermionic model which exhibits asymptotic freedom in the large-$N_f$ limit and has the same symmetries as QCD. 
Similarly to the discussion above, in $d=3$ both the NJL and NJLY models are expected to describe the same universality class.\footnote{This model shows emergent SUSY as well, now for $N_f = 1/2$, leading to a single Majorana fermion in $d=3$ \cite{Fei:2016sgs}.}

With three real scalars we have the \textit{chiral Heisenberg}  (cH) model, which has an $\mathrm{O}(3)$ symmetry in addition to the $U(N_f)$ symmetry of the fermions. This model has been studied less than the GNY and NJLY models in the literature, it is however expected to describe the antiferromagnetic critical point of graphene \cite{Janssen:2014gea}. The cH model also has a $d=2+\veps$ description known as the $SU(2)$ Gross--Neveu model, where the fermion bilinear is contracted with a Pauli matrix (see for example \cite{Gracey:2018qba}).

It was recently pointed out that all these models admit line defects that can be studied perturbatively \cite{Giombi:2022vnz,Pannell:2023pwz}. In the $4 - \veps$ description, the defect is given by an exponential of a scalar field integrated along a line. In $d=4$ a free scalar has dimension $\Delta_{\phi} =1$, which makes the defect coupling marginal, and is therefore a good candidate for describing a non-trivial defect CFT in $d=4-\veps$ dimensions. In \cite{Giombi:2022vnz} (see also \cite{Pannell:2023pwz}) it was shown that this is indeed the case. 

It was also pointed out in \cite{Giombi:2022vnz}, that the GN model in $2+\veps$ dimensions admits a natural line defect defined as the exponential of a fermion bilinear. In $d=2$ a free fermion has dimension $\Delta_\psi = \frac{1}{2}$ and the defect coupling is again marginal. In $2+\veps$ dimensions one can find a non-trival defect CFT which is expected to be in the same universality class as the defect CFT in $4-\veps$ dimensions described above. Most likely, this $d=2 + \veps$ picture of the defect can also be generalized to the NJL and the cH models discussed above. For the chiral Heisenberg model, the defect is given by the exponential of a fermion bilinear, similar to the GN model description. In the NJL model one can construct two fermion bilinears: $\sim \bar{\psi} \psi$, $\bar{\psi} \gamma_5 \psi$, and the defect is given by an exponential of both these terms, neatly matching the $4-\veps$ analysis of \cite{Pannell:2023pwz}. In this work however, we focus exclusively on the $4-\veps$ expansion.

The line defect considered here is closely related to the localized magnetic field or pinning line defect for the $\mathrm{O}(N)$ models studied in \cite{Allais:2014fqa,allais2014magnetic,Parisen_Toldin_2017,Cuomo:2021kfm,Gimenez-Grau:2022czc}. Such a defect models impurities localized in space, which can be implemented in lattice formulations by turning on a background field. These types of magnetic defects are therefore very natural from an experimental point of view, and indeed are expected to be observable in nature (see \cite{Cuomo:2021kfm} and references therein). 

We focus on what can be considered the two canonical configurations in defect CFT: four-point functions on the defect, and two-point functions of bulk operators outside the defect. Both these configurations have gotten significant attention in recent years \cite{deLeeuw:2017dkd,Barrat:2020vch,Dey:2020lwp,Barrat:2021yvp,Cavaglia:2022yvv}, as they are natural correlators to be studied using modern bootstrap techniques, both numerical \cite{Gimenez-Grau:2019hez,Paulos:2019fkw,Cavaglia:2021bnz,Padayasi:2021sik,Gimenez-Grau:2022czc,Cavaglia:2022qpg} and analytical \cite{Bissi:2018mcq,Mazac:2018mdx,Liendo:2019jpu,Bianchi:2020hsz,Gimenez-Grau:2021wiv,Ghosh:2021ruh,Barrat:2022psm,Bianchi:2022ppi,Li:2023whn}. For the magnetic line defect in the $\mathrm{O}(N)$ model, perturbative correlators were calculated in \cite{Gimenez-Grau:2022czc,Gimenez-Grau:2022ebb,Bianchi:2022sbz,Nishioka:2022qmj}. The results of this paper generalize the analysis of $\mathrm{O}(N)$ models to include fermions.

Notice that in our setup the defect remains one-dimensional, while the bulk is allowed to change dimension. 
It is also possible to keep the codimension fixed and to allow the defect to change dimension, as is the case for monodromy defects \cite{Billo:2013jda,Gaiotto:2013nva}. We do not consider monodromy defects here, for interesting recent progress see for example \cite{Bianchi:2021snj}. Interpolating between different dimensions and/or codimensions poses several challenges, as it is not clear how to represent correlators across dimensions. This problem was recently tackled in the context of BCFT \cite{Herzog:2022jlx} (see also \cite{Carmi:2018qzm,Giombi:2021cnr}). However, for higher codimension defects the analysis is more involved. We do not study fermions across dimensions in this work, but we discuss possible future directions in the conclusions.

The outline of the paper is as follows.
In Section~\ref{sec:setup} we discuss the fixed point of the line defect in generalized Yukawa CFTs, and compute the two-loop $\beta$-function of the defect scalar. In Section~\ref{sec:defcorr} we focus on operators on the defect and compute two-, three-, and four-point correlators of scalars and fermions.
We check that our results are consistent with an expansion of the four-point function in conformal blocks.
In Section~\ref{sec:bulkcorr}, we move to bulk operators in the presence of the defect, and study one- and two-point functions for  the scalars. In this section we also sketch the diagrams that contribute to two-point functions of fermions in the presence of the line.  
We conclude in Section~\ref{sec:conclusion} and give an outlook on further research.

\section{Yukawa CFTs with a line defect}
\label{sec:setup}

We are studying a general class of Yukawa models with $\mathrm{O}(N)$ flavor symmetry.
These theories are described by the following action in $d$-dimensional Euclidean space, with $2<d<4$:
\begin{equation}
    S
    =
    \int d^d x\,
    \left( \frac{1}{2} \partial_\mu \phi^a \partial_\mu \phi^a
    +
    i \bar{\Psi} \slashed{\partial} \Psi
    +
    g\, \bar{\Psi} \Sigma^a \phi^a \Psi
    +
    \frac{\lambda}{4!} \phi^a \phi^a \phi^b \phi^b \right)\,,
    \label{eq:GNY-lag}
\end{equation}
with $\mu = 0\,,\ldots\,,d-1$, $x^0 = \tau$ the Euclidean time direction, and $a=1\,, \ldots\,, N$ the index of the $\mathrm{O}(N)$ symmetry.
A choice of the matrix $\Sigma^a$ for a given $N$ corresponds to a choice of model, and in this work we focus on the ones mentioned in Section \ref{sec:intro} and listed in Appendix \ref{app:betafcn}.
For the GNY ($N=1$) and NLJY ($N=2$) models, $\Sigma^a$ is a matrix acting on the fermion flavor space ($i = 1, \ldots, N_f$) and on the spinor space, defining how the field $\phi^a$ interacts with fermions:
\begin{align}
    \text{GNY:} \qquad & \Psi = (\psi^1\,, \ldots \psi^{N_f})\,, \quad \slashed{\pd} = \mathds{1}_{N_f} \gamma_\mu \pd_\mu\,, \quad \Sigma = \mathds{1}_{N_f} \mathds{1}_{4}\,, \label{eq:GNY} \\
    \text{NJLY:}  \qquad & \Psi = (\psi^1\,, \ldots \psi^{N_f})\,, \quad \slashed{\pd} = \mathds{1}_{N_f} \gamma_\mu \pd_\mu\,, \quad \Sigma^1 = \mathds{1}_{N_f} \mathds{1}_4\,, \Sigma^2 = \mathds{1}_{N_f}\, i \gamma^5\,. \label{eq:NJLY}
\end{align}
Here, $\psi^i$ denotes Dirac fermions, and the $\gamma$-matrices are in the four-dimensional representation detailed in Appendix \ref{app:spinors}.
Note that the GNY model consists of a single scalar field, while the NJLY model contains one scalar field and one pseudoscalar field.
For the cH model ($N=3$), we use a $2N_f \times 2N_f$ representation of the $\gamma$-matrices to express the action via
\begin{equation}
    \text{cH:} \qquad \Psi = (\Psi_+, \Psi_{-})\,, \Psi_{\pm} = (\psi_{\pm}^1\,, \ldots\,, \psi_{\pm}^{N_f}) \quad \slashed{\pd} = (\mathds{1}_2 \otimes \gamma_\mu) \pd_\mu\,, \quad \Sigma^a = \sigma^a \otimes \mathds{1}_{2N_f}\,.
\end{equation}
In this case, the fermions are Dirac fermions only when $N_f = 2$ and all the fields $\phi^a$ are scalars.
Note that the Yukawa vertex is \textit{not} flavor-preserving in this case.

In order to perform calculations for all these models at once, we use the fact that the following identities hold:\footnote{For the case of the cH model, one should replace $\gamma_\mu$ by $\mathds{1}_2 \otimes \gamma_\mu$ on the left-hand sides, and $\gamma^\mu \gamma^\nu \gamma^\rho$ by $\mathds{1}_2 \otimes (\gamma^\mu \gamma^\nu \gamma^\rho)$ on the right-hand side of \eqref{eq:Sigma-identity2}.
Moreover, the Pauli matrices $\sigma^a$ are normalized such that $\sigma^a \sigma^a = \mathds{1}$.}
\begin{align}
    \tr\, \Sigma^a \gamma^\mu \Sigma^b \gamma^\nu
    &=
    4 N_f \delta^{ab} \delta^{\mu\nu}\,, \label{eq:Sigma-identity1} \\
    \gamma^\mu \Sigma^a \gamma^\nu \Sigma^a \gamma^\rho
    &=
    N \gamma^\mu \gamma^\nu \gamma^\rho\,. \label{eq:Sigma-identity2}
\end{align}

The $\beta-$functions of the couplings in Eq.~\eqref{eq:GNY-lag} are known to several loop orders for each model \cite{Zerf:2017zqi}.
For general Yukawa and scalar couplings, they can be found up to two loops in (the appendix of) \cite{Fei:2016sgs}.
We use their conventions in the rest of this paper.
For the purpose of writing our results for the three Yukawa models presented in Appendix \ref{app:betafcn} in a compact way, we write the $\beta-$functions in terms of the number of scalars $N = 1,2,3$. 
In this parametrization, setting $N_f \to 0$ gives results that can be compared with the $\mathrm{O}(N)$ model.
However, the exact $\beta-$function depends on the chosen Yukawa couplings that appear in the Lagrangian in Eq.~\eqref{eq:GNY-lag}, and the parametrization of $N$ should be considered with care and not be extended to $N >3$.
Below, we only list the expressions up to $\Om(\veps)$.

The $\beta-$functions are given by \cite{Karkkainen:1993ef,Rosenstein:1993zf,PhysRevB.80.075432} 
\begin{align}
    \beta_{\lambda}
    &=
    -\veps \lambda +\frac{1}{(4 \pi )^2} \left(8 g^2 \lambda N_f -48 g^4 N_f + \frac{N+8}{3} \lambda^2 \right) + \Om(\lambda^3, g^6, \lambda^2 g^2, \lambda g^4)\:, \\
    \beta_{g}
    &=
    -\veps \frac{g}{2}+\frac{\kappa_1 g^3}{(4 \pi )^2} + \Om(g^5)\:,
\end{align}
where $1 \leq N \leq 3$.
The Wilson-Fisher-Yukawa (WFY) fixed point can be reached for the following values of the couplings at one loop in $\veps:=4-d$:
\begin{align}
 \frac{\lambda_{\star}}{(4 \pi )^2} =  \frac{3 \kappa_{2} \veps }{2 \kappa_{1}(N+8)} + \Om(\veps^2) \:, \quad \frac{g^{2}_\star}{(4 \pi)^2} &=  \frac{\veps }{2 \kappa_{1}} + \Om(\veps^2)
\,, 
\label{eq:WFY-FP}
\end{align}
where we see that $g \sim \Op(\sqrt{\veps})$, while $\lambda \sim \Op(\veps)$. 
Furthermore, we have defined
\begin{align}
\kappa_1 &:= 2 N_f - N + 4\,,  \label{eq:def-Kappa1} \\
\kappa_2 &:= 2(4-N)-\kappa_1 + \sqrt{12 \left(N^2-16\right)+\kappa_1 (\kappa_1 +12 (N+4))}\,.
\label{eq:def-Kappa2}
\end{align}
Note that all the dependency on $N_f$ is contained in $\kappa_1$.

\subsection{Feynman rules}

We collect in this section the Feynman rules associated to the action \eqref{eq:GNY-lag}. The free propagators in $d$ dimensions are given by
\begin{align}
 \PropagatorScalar
 \;\, :=& \;
 \langle \phi_a(x_1) \phi_b(x_2) \rangle_{\lambda=g=0}
 \, = \;
\delta^{ab}\, I_{12}\,,
 \label{eq:free-prop} \\
 \PropagatorFermion
 \;\, :=& \;
\slashed{\partial}_1 I_{12}\,,
 \label{eq:free-prop-f}
\end{align}
where we have defined the scalar propagator function in $d=4-\veps$ dimensions:
\begin{equation}
I_{ij} := \frac{\Gamma(1-\veps/2)}{4 \pi^{2-\veps/2}_{\phantom{ij}} x_{ij}^{2 ( 1-\veps/2 )}}\,,
\label{eq:PropagatorFunction}
\end{equation}
with $x_{ij} := x_i - x_j$. For $d=4$ we have
\begin{equation}
I_{12} = \frac{1}{4\pi^2 x_{12}^2}\,.
\end{equation}
The scalar propagator satisfies the Green's equation
\begin{equation}
\partial_i^2 I_{ij} = - \delta^{(d)} (x_{ij})\,,
\label{eq:GreensEq}
\end{equation}
where $\delta^{(d)} (x)$ refers to the $d$-dimensional Dirac delta function.

The interaction terms yield the following vertices in position space:
\begin{align}
 \VertexFourScalars
 \;\; &:= \; 
 - \lambda_0 \int d^d x_5\, I_{15} I_{25} I_{35} I_{45} \, , \\
 \VertexYukawa
 \;\; &:= \; 
 - g_0 \int d^d x_4\, \slashed{\partial}_1 I_{14} \Sigma^a \slashed{\partial}_4 I_{34} I_{24} \, .
\end{align}
Note that one has to add a factor $1/n!$ when $n$ vertices of the same kind are being inserted, and that symmetry factors have to be taken into account.

\subsection{Bulk renormalization}

The couplings, as well as the (bulk) scalars $\phi^a$ and fermions $\Psib,\Psi^i$ get renormalized. We can define the \textit{bare} couplings and fields as
\begin{align}
  \lambda_0 = \mu^{\veps} \lambda Z_\lambda\:, \quad
  g_0 = \mu^{\frac{\veps}{2}} g Z_{g}\:, \quad \phi_0 = Z_{\phi} \phi\:,\quad   \Psi_0 = Z_{\Psi} \Psi\:,
\end{align}
where we have introduced rescaled couplings $g \to \mu^{\frac{\veps}{2}}g, \lambda \to \mu^{\veps}\lambda$ to ensure that the couplings in the renormalized Lagrangian are dimensionless. 
The expressions for the renormalization factors $Z_i$ up to $\Om(\veps^2)$ can be found in Appendix \ref{app:betafcn}.

The renormalization factors allow us to obtain the anomalous dimensions $\gamma_{\phi}, \gamma_{\Psi}$ for the scalar and fermionic fields, which are given here to first order in the couplings:
\begin{align}
  \gamma_{\phi} &= \frac{d \log Z_{\phi}}{d \log \mu} = \frac{2 g^2 N_f}{(4 \pi )^2} + \Om(\lambda^2, g^4, \lambda g^2)
  \:,\\
  \gamma_{\Psi} &= \frac{d \log Z_{\Psi}}{d \log \mu} = \frac{g^2 N}{2(4 \pi )^2} + \Om(\lambda^2,g^4,\lambda g^2) 
  \:.
\end{align}
This leads to the following values for the conformal dimensions evaluated at the WFY fixed point defined in Eq.~\eqref{eq:WFY-FP}:
\begin{align}
  \Delta_{\phi} &= 1 - \frac{\veps}{2} + \gamma_{\phi} = 1- \veps \frac{4-N}{2 \kappa_1}
  + \Om(\veps^2)\:,  \\
  \Delta_{\Psi} &= \frac{3}{2} - \frac{\veps}{2} + \gamma_{\Psi} = \frac{3}{2}-\frac{\veps}{4} \left(2 - \frac{N}{\kappa_1} \right) 
  + \Om(\veps^2)\:.
\end{align}
Furthermore, we need the normalization of their two-point functions, which are given by 
\begin{equation}
 \langle \phi^a (x_1) \phi^b (x_2) \rangle = \frac{\delta^{ab}\mathcal{N}_{\phi}^2}{x_{12}^{2\Delta_{\phi}}}\:, \quad \langle \bar{\Psi} (x_1) \Psi (x_2) \rangle = \frac{\mathcal{N}_{\Psi}^2 \bar{s}_1 \slashed{x}_{12} s_2}{x_{12}^{2(\Delta_{\Psi}+ 1/2)}}\:,
\end{equation}
with
\begin{equation}
 \mathcal{N}_{\phi} = \sqrt{\frac{\Gamma \left(\frac{d}{2}\right)}{2 (d-2) \pi ^{d/2}}}-\veps \frac{ (\kappa_1 + N - 4) (1 + \aleph )}{8 \pi\, \kappa_1} + \Om(\veps^2)\:, \quad  \mathcal{N}_{\Psi} = \frac{1}{\pi} + \Om(\veps)\:,
\end{equation}
where we have defined the following combination:
\begin{equation}
\aleph := 1 + \log \pi + \gamma_\text{E}\,,
\label{eq:def-aleph}
\end{equation}
with $\gamma_\text{E} = 0.57722 \ldots$ the Euler-Mascheroni constant.

\subsection{Defect fixed point}

One can define a defect CFT by adding a scalar line to the action \eqref{eq:GNY-lag}, in the same way as in the $\mathrm{O}(N)$ model \cite{Allais:2014fqa}.
This was shown in \cite{Giombi:2022vnz} for the GNY model, and generalized to the NJLY and chiral Heisenberg models in \cite{Pannell:2023pwz}. More precisely,
\begin{equation}
S_\text{defect} := S_0 + h_0 \int_{-\infty}^\infty d\tau\, \phi^1 (\tau)\,.
\end{equation}
Here $h_0$ is the bare coupling of the defect, which extends in the Euclidean time direction $\tau$, and $S_0$ is the bulk action in terms of the bare couplings $\lambda_0$ and $g_0$. 
The defect introduces a new vertex
\begin{align}
\VertexDefect
 \;\; \equiv \;
 - h_0 \int_{-\infty}^\infty d \tau_2\, I_{12} \, ,
\end{align}
with $\tau_2$ the point on the line, and where one should note that only $\phi^1$ and not $\phi^{\hat{a}}, \hat{a}=2, \ldots, N$ nor the fermions $\Psib, \Psi$ couple to the defect.  As for the bulk Feynman rules, one should add a factor $1/n!$ when $n$ vertices are inserted, and symmetry factors have to be accounted for.

We renormalize the defect coupling in a similar way to the bulk couplings.
We define the bare coupling $h_0$ in terms of the renormalized coupling $h$ as
\begin{equation}
 h_0 = \mu^{\frac{\veps}{2}} \, h \, Z_h\:,
\end{equation}
where $Z_h$ is given in Appendix~\ref{app:betafcn}, and can be computed by extracting the divergences from the one-point function of the renormalized scalar $\phi^a$ and requiring that it is finite:
\begin{equation}
 \vvev{\phi^a (x)} = \text{finite}\:.
\end{equation}

Note that the one-point function of a single fermion $\Psi$ is zero.  
The Feynman diagrams that contribute to the one-point function of $\phi^a$ up to $\Om(\veps^2)$ are given in Fig.~\ref{fig:one-pt-diags}. 
It is important to keep in mind that we are working perturbatively in the two bulk couplings $\lambda$ and $g$, since they are $\Om(\veps)$ and $\Om(\sqrt{\veps})$ respectively at the WFY fixed point, but we need to keep diagrams up to all orders in the defect coupling $h$
since it is of order $\Om(1)$.
There are however only a finite number of possible diagrams per order in $\lambda$ and $g$.
The diagrams in Fig.~\ref{fig:one-pt-diags} involving only scalar four-point couplings $\lambda$ (black dots) and defect couplings $h$ (blue dots) are the same as for the line defect in the $\mathrm{O}(N)$ model and were already computed in \cite{Allais:2014fqa}. 
The diagrams in Fig.~\ref{fig:one-pt-diags} that include the Yukawa coupling $g$ (red dots) were recently computed in \cite{Pannell:2023pwz}. 
Here we give the corresponding $\beta-$function for $h$ up to $\Om(\veps^2)$, which match the ones in \cite{Pannell:2023pwz}.
Some of the diagrams in Fig.~\ref{fig:one-pt-diags} are completely cancelled by the wavefunction renormalizations of $\phi^a$ and $\Psi$, while others do contribute to the defect counterterms.

\begin{figure}
\begin{subfigure}{\textwidth}
\centering
\OnePointOne
\OnePointTwo
\OnePointThree
\OnePointFour
\OnePointFive
\OnePointSix
\end{subfigure}
\begin{subfigure}{\textwidth}
\centering
\OnePointSeven
\OnePointEight
\OnePointNine
\OnePointTen
\OnePointEleven
\end{subfigure}
\caption{Diagrams contributing to the one-point function $\vvev{\phi^a}$ up to $\Om(\veps^2)$. The defect is denoted by a solid line, scalars by a dotted line, and fermions by solid arrowed lines. Bulk scalar couplings $\lambda_0$ are represented by a black dot, bulk Yukawa couplings $g_0$ by a red dot and defect couplings $h_0$ by a blue dot.}
\label{fig:one-pt-diags}
\end{figure}

We compute the $\beta-$function $\beta_h$ from the divergent part of the diagrams and we obtain:
\begin{align}\label{eq:betafcn-h}
\beta_h =& -\frac{\veps h}{2} + \frac{1}{(4\pi)^2} \left(\frac{\lambda h^3}{6}\right) + \frac{1}{(4\pi)^4} \Bigg\{\lambda^2 h \left(\frac{(2 + N)}{36} - \frac{h^2 (N+8)}{35} - \frac{h^4}{12}\right) - \lambda g^2 h^3 N_f \notag \\
 & + g^4 h \left(- \frac{(N+4)N_f}{2} + h^2 4 N_f \left(1 - \frac{\pi^2}{6}\right)\right)\Bigg\} + \Om(\lambda^3, g^6, \lambda^2 g^2, \lambda g^4)\:.
\end{align}
Using the values for $\lambda$ and $g$ at the WFY fixed point in Eq.~\eqref{eq:WFY-FP}, we find the corresponding defect fixed point
\begin{equation}
h^{2}_\star = -\frac{2 (N-4) (N+8)}{\kappa_2} + \Om(\veps)\:,
\end{equation}
where the $\Om(\veps)$ term is given in Appendix \ref{app:betafcn} for $N=1,2,3$.
If we include the finite part of the one-point function, we can extract the one-point function coefficient $a_{\phi}$ :
\begin{equation}
 \vvev{\phi^a (x)} = \frac{\delta^{a1} a_{\phi}}{|x^{\bot}|^{1 + \gamma_\phi}}\:, \quad a_{\phi}^2 = -\frac{(N-4) (N+8)}{2 \kappa_2} + \Om(\veps)\:.
 \label{eq:aphi}
\end{equation}
The $\Om(\veps)$ term is lengthy and given in the attached \textsc{Mathematica} notebook.

\section{Correlators of defect operators}
\label{sec:defcorr}

The bulk operators give rise to a plethora of defect operators.
In this section, we consider correlation functions between the lowest-lying defect operators.
The lowest-lying scalars are the first scalars appearing in the bulk-to-defect expansion of $\phi^a$, and are labelled in the following as $\phih^1$ and $t^{\hat{a}}$, with $\hat{a}=2\,, \ldots\,, N$.
These correspond to the two scalar operators of length $1$ that arise due to the breaking of $\mathrm{O}(N)$ symmetry in the bulk to $\mathrm{O}(N-1)$ symmetry on the defect, namely $\phi^1$ couples to the defect while $\phi^{\hat{a}}$ does not.
The conformal dimension of $ \phih^1$ was computed in \cite{Giombi:2022vnz} for the GNY model up to $\Om (\veps)$. 
It can be extracted from the $\beta-$function of the defect coupling at the fixed point:
\begin{equation}
 \Delta_{ \phih^1} = 1 + \frac{\partial \beta_h}{\partial h}|_{h = h_{\star}} = 1 + \frac{(4-N) \veps}{\kappa_1} + \Om(\veps^2)\:,
\label{eq:dim-phi-hat}
\end{equation}
which agrees with \cite{Giombi:2022vnz} for the case $N=1$ corresponding to the pure GNY model.
In this section, we extend their results to general $N$, as well as compute additional defect correlators.
The operator $t^{\hat{a}}$ (the \textit{tilt} operator) has protected conformal dimension
\begin{equation}
 \Delta_t = 1\:.
\label{eq:dim-t}
\end{equation}
Note that for the GNY model, there is no tilt operator, but only $ \phih^1 \equiv  \phih$ on the defect.

Besides the tilt there is another
scalar
defect operator with protected conformal dimension, namely the displacement operator $\Disp$. 
It is related to the bulk stress-energy tensor through the Ward identity
\begin{equation}
 \partial_{\mu} T^{\mu \nu} = \delta^{d-1}(x^{\bot}) \Dispb{\nuh}\:, \quad \nuh = 1\,, \ldots\,, d-1\:, 
\end{equation}
and has transverse spin $s = 1$ and conformal dimension
\begin{equation}
 \Delta_{\Disp} = 2\:.
\end{equation}

The expansion of the bulk fermion on the defect gives us the defect fermions $ \hat{\Psi},  \hat{\bar{\Psi}}$ with conformal dimension
\begin{equation}
 \Delta_{\hat{\Psi}} = \frac{3-\veps}{2} + \gamma_{\hat{\Psi}}\:.
\end{equation}
The anomalous dimension $\gamma_{\Psih}$ can be extracted from the two-point function.

Below we compute correlation functions between these operators and extract the corresponding defect CFT data.

\subsection{Two-point functions}
\label{sec:def-2pt}

We start by computing the two-point functions between the defect operators to obtain their anomalous dimensions and normalization constants.

\subsubsection{Two-point functions of scalars}
\label{sec:def-2pt-scal}

We consider first the two-point functions $\vev{ \phih^1 (\tau_1)  \phih^1 (\tau_2)}$ and $\vev{t^{\hat{a}} (\tau_1) t^{\hat{b}} (\tau_2)}$.
The two-point function of arbitrary (defect) scalars $ \phih^a$ takes the general form
\begin{equation}
 \langle  \phih ^a (\tau_1)  \phih^b (\tau_2) \rangle = \NormPhiHat^2\, \frac{\delta^{ab}}{\tau_{12}^{\smash{2 \Delta_{ \phih}}}}\:,
\end{equation}
with $\tau_{12}:= \tau_1 - \tau_2$, and where $\NormPhiHat$ and $\Delta_{ \phih}$ correspond respectively to the normalization constant and to the scaling dimensions given in \eqref{eq:dim-phi-hat} and \eqref{eq:dim-t}.

In terms of Feynman diagrams, this two-point function can be expressed as
\begin{align}\label{eq:2pt-def-phi}
 \langle \,  \phih^a(\tau_1) \,  \phih^b(\tau_2) 
         \, \rangle
 \;\; = \;\;
 \TwoPointScalarsOne
  \;\; + \;\; 
 \TwoPointScalarsTwo
 \;\; + \;\;
\TwoPointScalarsThree
 \; \; + \Om (\veps^2)\:,
 \end{align}
In the first diagram, the two external operators are connected through a single tree-level propagator. The second diagram corresponds to the bulk self-energy and consists of an internal fermion loop and two bulk Yukawa vertices (represented by red dots), while the third one is special to the defect theory and involves two integrals along the line (represented by blue dots) as well as a bulk four-scalar vertex (the black dot).
 
The fermion loop diagram is easy to compute and reads
\begin{align}
\TwoPointScalarsTwo
\;\; &= \; 
g_{0}^2\, \tr\, \Sigma^a \Sigma^b  B_{12} \notag \\
&= \;
- \frac{g_{0}^2 N_f}{4 \pi^2} \delta^{ab} I_{12}
\left( \frac{1}{\veps} + \aleph + \log \tau_{12}^2 + \Op(\veps^2) \right)\,,
\end{align}
where we have made use of the rule given in \eqref{eq:Sigma-identity1}.
The integral $B_{12}$ is defined in \eqref{eq:def-Bintegral} and solved in \eqref{eq:sol-Bintegral}, while the function $I_{12}$ corresponds to the scalar propagator and is defined in \eqref{eq:PropagatorFunction}. Finally, the constant $\aleph$ arising from dimensional regularization is defined in \eqref{eq:def-aleph}.  
The expressions for the two other diagrams can be found in \cite{Gimenez-Grau:2022czc}. 

Requiring that the sum of the diagrams is finite allows us to compute the renormalization factors for $ \phih^1$ and $t^{\hat{a}}$:
\begin{equation}
 \vev{ \phih^a (\tau_1)  \hat{\phi}^b (\tau_2)} = \frac{1}{Z_{ \phih}^2} \vev{ \phih^a_0 (\tau_1)  \phih^b_0 (\tau_2)} =\, \text{finite}\:,
\end{equation}
and leads to
\begin{align}
 Z_{ \phih^1 } &=
 1
 - \frac{1}{\veps} \frac{\lambda h^2 + 8 g^2 N_f}{64 \pi^2}
 + \Om\left(\veps^{-2}\right)\:, \\
 Z_{t} &=
 1
 - \frac{1}{\veps} \frac{\lambda h^2 - 24 g^2 N_f}{192 \pi^2}
 + \Om\left(\veps^{-2}\right)\:.
\end{align}
As a sanity check, we can read the scaling dimensions from the renormalization factors:
\begin{align}
\Delta_{\smash{ \phih^1}} &= \mu \frac{\partial \log Z_{ \phih^1}}{\partial \mu} = 1 + \veps \frac{4 - N}{\kappa_1} + \Op(\veps^2)\,, \\
\Delta_{t} &= \mu \frac{\partial \log Z_t}{\partial \mu} = 1 + \Op(\veps^2)\,,
\end{align}
which agree with the results given in \eqref{eq:dim-phi-hat} and \eqref{eq:dim-t}.

The normalization constants can now be extracted from the finite two-point functions, and we find for the two scalars
\begin{align}
\NormPhiHatIndex{1}^2 &= \frac{1}{4\pi^2} \left\lbrace 1 - \frac{\veps}{2} \left( 2 + \frac{(N - 4)(1 - 2 \aleph)}{\kappa_1} \right)  + \Op(\veps^2) \right\rbrace\,,  \label{eq:NormScalars1} \\
\NormTilt^2 &= \frac{1}{4\pi^2} \left\lbrace 1 - \frac{\veps}{2} \left( 2 + \frac{N - 4}{\kappa_1} \right)  + \Op(\veps^2) \right\rbrace\,,
\label{eq:NormScalarst}
\end{align}
where $\kappa_1$ depends on $N_f$ and $N$ and is defined in \eqref{eq:def-Kappa1}.

\subsubsection{Two-point function of the displacement}

We continue with the two-point function of the displacement.
The displacement has transverse spin $s = 1$ and can be constructed by taking a transverse derivative of the field $ \phih^1$:
\begin{equation}\label{eq:dispDef}
 \Dispb{\muh} \sim \partial_{\muh}  \phih^1\:,
\end{equation}
while there exist additional operators $\partial_{\muh} t^{\hat{a}}$ that correspond to taking the transverse derivative of the tilt.
The latter will not be considered here for brevity, but its correlators can be computed in a similar way as the displacement correlators. 

Because we can write the displacement as in Eq.~\eqref{eq:dispDef}, the diagrams that contribute to the two-point function are the same as for $ \phih_a$ and are given in Eq.~\eqref{eq:2pt-def-phi}.
For the evaluation of the diagrams, we need to first take the derivatives with respect to $x^{\bot}_1, x^{\bot}_2$ and then send $x^{\bot}_1, x^{\bot}_2 \to 0$.
This leads to the following expressions for the diagrams:
\begin{align}
\TwoPointScalarsTwo
\;\; &= \; %\;
2 g_{0}^2 \delta^{\hat{\mu} \hat{\nu}}\, \tr \Sigma^a \Sigma^b B_{12} \notag \\
\; \; &= \;
- \frac{g_{0}^2 \delta^{\hat{\mu} \hat{\nu}} N_f}{2 \pi^2} I_{12} \delta^{ab} \left( \frac{1}{\veps} + \aleph + \log \tau_{12}^2 + \Op(\veps^2) \right)\,.
\end{align}
The other diagrams were computed in \cite{Gimenez-Grau:2022czc}.

We compute the renormalization factor for the displacement in the usual way, by requiring that the two-point function is finite.
This results in
\begin{align}
 Z_\Disp
 = 1 
 - \frac{1}{(4\pi)^2 \veps} \left(\frac{\lambda h^2}{12} + 2 g^2 N_f \right) + \Om\left(\veps^{-2}\right)\, .
\end{align}
As a check, we compute the anomalous dimension of $\Disp$ and find
\begin{align}
\Delta_{\Disp} &= \mu \frac{\partial \log Z_{\Disp}}{\partial \mu} = 2 + \Op(\veps^2)\,,
\end{align}
where the $\Op(\veps)$ contributions cancel as expected.
We can extract the proper normalization from the finite parts of the diagrams and obtain
\begin{align}
\NormDisp^2 = \frac{1}{2 \pi ^2} \left\lbrace
1 -\veps \left(1 - \frac{N-4}{6 \kappa_1 } \right) + \Op(\veps^2) \right\rbrace\,.
\end{align}

\subsubsection{Two-point functions of fermions}\label{sec:def-2pt-fer}

Let us now turn our attention to the fermions. 
In $1d$, the two-point function of fermions takes the form
\begin{equation}
\vev{ \Psihb (\tau_1)  \Psih (\tau_2)} = \NormPsiHat^2\, \frac{\bar{s}_1 \gamma^0 s_2}{\tau_{12}^{\smash{2 \Delta_\Psih}}}\:,
\end{equation}
where we have use the polarization spinors $\bar{s}_1$, $s_2$ as defined in \eqref{eq:def-PolarizationSpinors} in order to avoid cluttering. 

The diagrams involved are 
\begin{align}
 \vev{ \Psihb (\tau_1)  \Psih (\tau_2)}
 \;\; &= \;\;
 \TwoPointFermionsOne
\;\; + \;\;
\TwoPointFermionsTwo
\notag \\
&+ \; \;
 \TwoPointFermionsThree
 \; \; + \; \; 
\TwoPointFermionsFour
\; \; + \Om (\veps^2)\:.
 \end{align}

As noted in \cite{Giombi:2022vnz}, the second diagram is zero at order $\Om(\veps)$.  This can be easily checked in the following way:
\begin{align}
\TwoPointFermionsTwo
\;\; &= \;\;
g_0 h_0\, \bar{s}_1 \int d\tau_3 \int d^d x_4\, \slashed{\partial}_4 I_{14}\, \Sigma^1 \slashed{\partial}_4 I_{24} I_{34}\, s_2 \notag \\
&\sim g_0 h_0 \int d\tau_3\, \tau_{13} \tau_{23}\, I_{13} I_{23}
\notag \\
&= \;\;
0 + \Om ( \veps^{3/2} )\,,
\end{align}
where in the second line we have used the $4d$ fermionic star-triangle relationship given in \eqref{eq:id-StarTriangle}, to which the corrections towards $d=4-\veps$ are of order $\Op(\veps)$ while $g \sim \Op(\sqrt{\eps})$.

The third diagram can be computed as follows:
\begin{align}
 \TwoPointFermionsThree
\;\; &= \;\;
g_{0}^2 \bar{s}_1 \int d^d x_3 \int d^d x_4\, \slashed{\partial}_1 I_{13} \, \Sigma^a\, \slashed{\partial}_3 I_{34}\, \Sigma^a\, \slashed{\partial}_4 I_{24} I_{34}\, s_2 \notag \\
&= \;\;
2 g_{0}^2 N \bar{s}_1 \slashed{\partial}_2 B_{12}\, s_2 \notag \\
&= \;\;
\frac{g_{0}^2 N}{32\pi^4} \frac{ \bar{s}_1 \gamma^0 s_2}{\tau_{12}^3} \left\lbrace \frac{1}{\veps} + \aleph -1 + \log \tau_{12}^2 + \Op(\veps) \right\rbrace\,.
\end{align}
Here we have made use of the rule given in \eqref{eq:Sigma-identity2} in order to be able to rewrite the integral as a derivative of $B_{12}$.

The fourth diagram is more involved and reads 
\begin{equation}
\TwoPointFermionsFour = g_{0}^2 h_{0}^2\, \bar{s}_1 \int d\tau_3 \int d\tau_4 \int d^d x_5 \int d^d x_6\, \slashed{\partial}_1 I_{15}\, \Sigma^1\, \slashed{\partial}_5 I_{56}\, \Sigma^1\, \slashed{\partial}_6 I_{26} I_{35} I_{46}\, s_2\,,
\end{equation}
with $\Sigma^1 = \mathds{1}$ for all our models of interest. The easiest way to compute it is to apply another slashed derivative on the integral and compare the result to an ansatz. We define
\begin{equation}
J_{12} := \int d\tau_3 \int d\tau_4 \int d^d x_5 \int d^d x_6\, \slashed{\partial}_1 I_{15} \slashed{\partial}_5 I_{56} \slashed{\partial}_6 I_{26} I_{35} I_{46}\,,
\end{equation}
and assume that
\begin{equation}
J_{12} = \frac{\gamma^0}{\tau_{12}^2} \left\lbrace \frac{A}{\veps} + B + C \log \tau_{12}^2 \right\rbrace\,.
\end{equation}
We then compute
\begin{equation}
\slashed{\partial}_1 J_{12} = - \int d\tau_3 \int d\tau_4\, I_{13}\, \slashed{\partial}_1 \slashed{\partial}_2 Y_{124}\,,
\end{equation}
where we have used $\slashed{\partial}_1 \slashed{\partial}_1 = \mathds{1} \partial_1^2$ and $\partial_1^2 I_{15} = - \delta^{(d)} (x_{15})$. After applying the identity \eqref{eq:id-StarTriangle}, we find
\begin{equation}
\slashed{\partial}_1 J_{12} = \frac{3}{16 \pi^6} \mathds{1} + \text{(quadratic divergences)}\,,
\end{equation}
from which we can read
\begin{equation}
A=C=0\,, \qquad B= - \frac{1}{16 \pi^6}\,.
\end{equation}
We see that this diagram is finite (after dropping the quadratic divergences) and so it contributes only to the normalization constant.

In the same way as for the scalars, we define a renormalization factor $Z_{\Psih}$ such that
\begin{equation}
\vev{ \Psihb (\tau_1)  \Psih (\tau_2)} = \frac{1}{Z_{ \Psih }^2} \vev{ \Psihb_0 (\tau_1)  \Psih_0 (\tau_2)} =\, \text{finite}\:,
\end{equation}
for which we find
\begin{equation}
Z_{ \Psih } = 1 - \frac{g^2 N}{32 \pi^2 \veps} + \Om\left(\veps^{-2}\right)\,,
\end{equation}
which agrees with the renormalization factor for the bulk given in Eq.~\eqref{eq:renorm_bulk}.

The scaling dimension is given by
\begin{equation}
\Delta_{\smash{ \Psih}} = \frac{3}{2} - \frac{\veps}{4} \left(2 - \frac{N}{\kappa_1} \right) + \Op(\veps^2)\,,
\end{equation}
while the normalization constant reads
\begin{equation}
\NormPsiHat^2 = - \frac{1}{2\pi^2} \left\lbrace 1 - \frac{\veps}{2 \kappa_1} \left( 2 \kappa_1 - \aleph \left( 1 - \frac{N}{2 \kappa_1} \right) + \frac{4}{\pi^2} \frac{(N-4)(N+8)}{\kappa_1 \kappa_2}  \right) + \Op(\veps^2) \right\rbrace\,.
\label{eq:NormFermions}
\end{equation}
Note that the renormalization factor as well the scaling dimension agree with the bulk computation, as the diagrams contributing to these results are the same.\footnote{This agreement is expected to be lifted at higher orders of $\veps$.}

\subsection{Three-point functions}\label{sec:def-threept}

We now compute three-point functions between the lowest-lying operators.
This gives us various defect OPE coefficients, which can be compared with the OPE coefficients coming from the conformal block expansion of the four-point function.

\subsubsection{Three-point functions of scalars}

The three-point function between three defect scalars $ \phih^{a,b,c}$, where $ \phih^a = \{  \phih^1, t^{\hat{a}} \}$, is given by a single Feynman diagram up to $\Om(\veps)$:
\begin{align}
 \vev{ \phih^a(\tau_1) \,  \phih^b(\tau_2) 
         \,  \phih^c(\tau_3)}
 \;\; &= \;\;
 \ThreePointScalars
  \; \; + \Om (\veps^2) \notag \\
&= \;\;
\NormPhiHatIndex{a} \NormPhiHatIndex{b} \NormPhiHatIndex{c} \frac{\lambda_{\smash{ \phih^a  \phih^b  \phih^c}}}{\tau_{12}^{2 \Delta_{abc}} \tau_{23}^{2 \Delta_{bca}} \tau_{13}^{2 \Delta_{cab}}}\:,
\label{eq:3pt-structure}
 \end{align}
where we have defined $\Delta_{abc} := \frac{1}{2} (\Delta_{\smash{ \phih^a}} + \Delta_{\smash{ \phih^b}} - \Delta_{\smash{ \phih^c}})$.

This diagram was already evaluated in \cite{Gimenez-Grau:2022czc}, and results in the following OPE coefficients:
\begin{align}
\lambda_{\smash{ \phih^1  \phih^1  \phih^1}} &= \frac{3 \pi\veps}{8}\frac{(4 \kappa_1 - N_f) \sqrt{2 (4-N)(N+8)\kappa_2}}{\kappa_1^2 (N+8)} + \Om (\veps^2) \:,  \\
\lambda_{\smash{t t  \phih^1}} &= \frac{\lambda_{\smash{ \phih^1  \phih^1  \phih^1}}}{3} + \Om (\veps^2)\:.
\end{align}

Since the OPE coefficients start at $\Om(\veps)$, they only appear at order $\Om(\veps^2)$ in the four-point function of scalars.

\subsubsection{Three-point functions involving $ \phih^2$}\label{sec:def-3pt-phi2}

The first scalar operators that appear in the OPE $ \phih^a \times  \phih^a$, $ \phih^a = \{ \phih^1, t^{\hat{a}}\}$, after $ \phih^1$ itself, are the degenerate operators $s_{\pm}$.
These operators have dimension close to 2, and can be constructed from $( \phih^1)^2$ and $( \phih^{\hat{a}})^2$.
In order to find the correct anomalous dimensions we need to require that the three-point functions involving $ \phih$ and $s_{\pm}$ are finite.
The diagrams that contribute up to $\Op(\veps)$ are
\begin{align}
 \langle \,  \phih^a(\tau_1) \,  \phih^b(\tau_2) \, 
             \phih^c  \phih^c(\tau_3) \, \rangle
 = 
 &
 \ThreePointScalarsSquareOne
  + 
 \ThreePointScalarsSquareTwo
 + 
 \ThreePointScalarsSquareThree
 + 
 \ThreePointScalarsSquareFour
 + \Om(\veps^2) \notag\\
&= \NormPhiHatIndex{a}^2 \Norms \frac{\lambda_{ \phih^a  \phih^a s_{\pm}}}{\tau_{12}^{2 \Delta_{a b c^2}} \tau_{23}^{2 \Delta_{b c^2 a }} \tau_{13}^{2 \Delta_{a c^2 b}}}\:.
 \label{eq:three-pt-PPP2}
\end{align}
The first three diagrams have been computed in \cite{Gimenez-Grau:2022czc}, while the last one is the wavefunction renormalization.
Requiring that this three-point function is finite in $\veps$ gives a renormalization matrix $Z_{s}$ that has a lengthy expression and is given in a \textsc{Mathematica} notebook.
The anomalous dimension can be computed by diagonalizing this matrix and taking the derivative:
\begin{align}
  \gamma_{s_{\pm}} =& \frac{\veps (-4 (N+8) (-\kappa_{1}+N-4)+\kappa_{2} (N+4) \pm \kappa_{3})}{4 \kappa_{1} (N+8)} + \Om(\veps^2)\,, \label{eq:defcorr-gamma-spm}
\end{align}
where we have defined
\begin{equation}
 \kappa_3 := \sqrt{\kappa_{2}^2 N^2+8 \kappa_{2} (N-4) (N-2) (N+8)+16 (N-4)^2 (N+8)^2}\:.
\end{equation}
We obtain the conformal dimension as 
$\Delta_{s_{\pm}} = 2 - \veps + \gamma_{s_{\pm}}$.

To complete the computation of the OPE coefficients, we also need the normalization of the two-point functions $\langle s_{\pm} (\tau_1) s_{\pm} (\tau_2)\rangle$.
From the two-point function, we get
\begin{align}
 &\Norms^2 = \pm \frac{(N-1) \left(\kappa_{2} (N-2)+4 (N-4) (N+8) \pm \kappa_3 \right)}{16 \pi ^4 \kappa_3} + \Om(\veps) \:.
\end{align}
%
%We have only displayed the $\Om(1)$ term here, while the $\Om(\veps)$ term is once again long and given in the \textsc{Mathematica} notebook.
We checked that the operators are now properly normalized, such that $\langle s_{+} (\tau_1) s_{-} (\tau_2) \rangle = 0$.

Putting everything together, we can now extract the OPE coefficients, for which we find:
\begin{align}
 \lambda_{ \phih^1  \phih^1 s_{\pm}} &= \pm \frac{2 \kappa_{2} \sqrt{N-1}}{\sqrt{\kappa_{3}^2\pm (\kappa_{2} (N-2)+4 (N-4) (N+8)) \kappa_3}} + \Om(\veps)\:,  \\
 \lambda_{t^{\hat{a}} t^{\hat{b}} s_{\pm}} &=  \delta^{\hat{a}\hat{b}}\frac{\sqrt{\kappa_3 \pm (\kappa_{2} (N-2)+4 (N-4) (N+8))}}{\sqrt{\kappa_3}\sqrt{N-1}} + \Om(\veps)\:. \label{eq:def-OPE-phi2}
\end{align}
While the diagrams of the three-point functions are computed to $\mathcal{O}(\varepsilon)$, the OPE coefficients can only be determined to $\mathcal{O}(1)$. The operators $s_{\pm}$ are linear combinations of $(\hat{\phi}^1)^2$ and $(\hat{\phi}^{\hat{a}})^2$, and their mixing will receive a correction at $\mathcal{O}(\varepsilon)$ that contributes to the OPE coefficients and the normalization of their two-point functions. Unfortunately, computing the correction to mixing would require knowing the anomalous dimensions of $s_{\pm}$ to $\mathcal{O}(\varepsilon^2)$, or alternatively computing additional four-point correlation functions of $\hat{\phi}$ and $s_{\pm}$ to unmix the operators, as was done recently in \cite{Belton:2025hbu}.
%The $\Om(\veps)$ terms can be found in the attached \textsc{Mathematica} notebook.

\subsubsection{Three-point functions of two fermions and one scalar}

An example of a mixed correlator is the three-point function $\vev{\Psihb \Psih \phih^a}$, which is given at leading order by
\begin{align}
\vev{ \Psihb (\tau_1)  \Psih (\tau_2) \phih^a (\tau_3)} &= \; \; 
\ThreePointScalarsTwoFermions
 \; \; + \Om(\veps) \notag \\
&= \NormPsiHat^2 \NormPhiHatIndex{a} \frac{ (\bar{s}_1 \gamma^0 \Sigma^a s_2)\lambda_{ \Psihb  \Psih  \phih^a}}{\tau_{12}^{2 \Delta_{\Psi \Psi a}} \tau_{23}^{2 \Delta_{\Psi a \Psi}} \tau_{13}^{2 \Delta_{a \Psi \Psi}}} + \mathcal{O}(\veps)\,,
\end{align}
with $\Delta_{ijk}$ following the same convention as given below \eqref{eq:3pt-structure}, where $a$ refers to $\phi^a$.

At order $\Op(\sqrt{\veps})$ we have a single diagram contributing.
It is easy to evaluate this diagram using the usual commutation rules for $\Sigma^a$ as well as the fermionic star-triangle identity given in \eqref{eq:id-StarTriangle}:
\begin{equation}
\ThreePointScalarsTwoFermions
 \; \;  =  \; \;  \NormPsiHat^2 \NormPhiHatIndex{a} g_0\, \left( \bar{s}_1 \gamma^0 \Sigma^a s_2 \right) \frac{1}{64 \pi^4 \tau_{12}^2 \tau^{\phantom{2}}_{23} \tau^{\phantom{2}}_{31}}\,.
\end{equation}
After inserting the normalization constants derived in Section \ref{sec:def-2pt} we find that the OPE coefficient is
\begin{equation}\label{eq:def-3pt-fer-scal}
\lambda_{ \Psihb  \Psih  \phih^a} =  \frac{\sqrt{\veps}}{4 \sqrt{2 \kappa_1}} + \Om(\veps)\,.
\end{equation}

\subsection{Four-point functions}

Let us now turn our attention to the four-point functions, which are the first correlators in our list to have non-trivial kinematics.
These correlators can be expanded in $1d$ conformal blocks to obtain defect CFT data, which we can compare with the OPE coefficients computed in the previous section.  We start by considering correlators of purely scalar operators, before moving on to fermions and concluding with an example of a mixed correlator including both scalars and fermions.

\subsubsection{Four-point functions of scalars}

We start this section by the four-point functions of scalars. Such correlators depend kinematically on a single cross-ratio $\chi$, which we define as\footnote{In higher $d$, four-point functions depend on \textit{two} cross-ratios $\chi$ and $\bar{\chi}$. In $1d$ the second cross-ratio is not independent of $\chi$ and becomes $\bar{\chi} = 1-\chi$.}
\begin{equation}
\chi := \frac{\tau_{12} \tau_{34}}{\tau_{13} \tau_{24}}\,.
\label{eq:def-CrossRatio}
\end{equation}
Four-point functions of scalars take the following form:
\begin{equation}
\vev{ \phih^a(\tau_1) \,  \phih^b(\tau_2) 
         \,  \phih^c(\tau_3) \,  \phih^d(\tau_4)} = \NormPhiHatIndex{a} \NormPhiHatIndex{b} \NormPhiHatIndex{c} \NormPhiHatIndex{d}\, \Km_4\,  f^{abcd} (\chi)\,.
\end{equation}
with the conformal prefactor
\begin{equation}
\Km_4 := \frac{1}{\tau_{12}^{\smash{\Delta_{ \phih^a} + \Delta_{ \phih^b}}} \tau_{34}^{\smash{\Delta_{ \phih^c}+\Delta_{ \phih^d}}}} \left( \frac{\tau_{24}}{\tau_{14}} \right)^{\smash{\Delta_{ \phih^a} - \Delta_{ \phih^b}}} \left( \frac{\tau_{14}}{\tau_{13}} \right)^{\smash{\Delta_{ \phih^c} - \Delta_{ \phih^d}}}\,.
\end{equation}
In terms of Feynman diagrams, the four-point function of arbitrary scalars is given up to $\Om(\veps)$ by 
\begin{align}
 \vev{ \phih^a(\tau_1) \,  \phih^b(\tau_2) 
         \,  \phih^c(\tau_3) \,  \phih^d(\tau_4)}
  = & \; \;
 \FourPointScalarsOne
\; \; + \; \;  
 \FourPointScalarsTwo
\notag \\
& + \; \; 
\FourPointScalarsThree
+
\FourPointScalarsFour
+ \Om(\veps^2)\:,
\label{eq:4pt-scal}
 \end{align}
where the first diagrams are products of two-point functions, and the last one was computed in \cite{Gimenez-Grau:2022czc}.

Adding all diagrams in Eq.~\eqref{eq:4pt-scal}, we obtain the following unit-normalized results:
\begin{align}
 f^{1111} (\chi) =&
 1 + \chi^{2 \Delta_{ \phih^1}} + \left(\frac{\chi}{1-\chi}\right)^{2 \Delta_{ \phih^1}}
 +\frac{3 \veps \kappa_2}{\kappa_1 (N+8)} \left(\chi \log (1-\chi) + \frac{\chi^2}{1 - \chi} \log \chi \right)  + \Op(\veps^2)\:,\\
 f^{\smash{1\hat{a}1\hat{b}}} (\chi) =&
\delta^{\smash{\hat{a}\hat{b}}} \chi^{\Delta_{ \phih^1} + \Delta_t} 
+ \veps \delta^{\smash{\hat{a}\hat{b}}} \frac{\kappa_2}{\kappa_1 (N+8)} \left(\chi \log(1-\chi) + \frac{\chi^2}{1-\chi} \log \chi \right)  + \Op(\veps^2)\:,\\
f^{\smash{\hat{a}\hat{b}\hat{c}\hat{d}}} (\chi) =&
\delta^{\smash{\hat{a}\hat{b}}} \delta^{\smash{\hat{c}\hat{d}}}
+\delta^{\smash{\hat{a}\hat{c}}}\delta^{\smash{\hat{b}\hat{d}}} \chi ^2
+\delta^{\smash{\hat{a}\hat{d}}} \delta^{\smash{\hat{b}\hat{c}}} \frac{\chi ^2}{(1-\chi)^2}
\notag \\
& + \veps (\delta^{\smash{\hat{a}\hat{b}}} \delta^{\smash{\hat{c}\hat{d}}} +\delta^{\smash{\hat{a}\hat{c}}} \delta^{\smash{\hat{b}\hat{d}}} +\delta^{\smash{\hat{a}\hat{d}}} \delta^{\smash{\hat{b}\hat{c}}}) \frac{\kappa_2}{\kappa_1 (N+8)}  \left(\chi \log (1-\chi)
+ \frac{\chi^2}{1-\chi} \log \chi \right) + \Op(\veps^2)\:,
\end{align}
where $\kappa_1$ and $\kappa_2$ are defined in \eqref{eq:def-Kappa1} and \eqref{eq:def-Kappa2}. Other orderings of $ \phih^1$ and $t^{\hat{a}}$ are not given here but can be computed in the same way straightforwardly from \eqref{eq:4pt-scal}.
The last correlator $f^{\smash{\hat{a}\hat{b}\hat{c}\hat{d}}} (\chi)$ can be decomposed into a scalar ($S$), an antisymmetric ($A$), and a traceless symmetric ($T$) contribution:
\begin{align}
  & f^{\smash{\hat{a}\hat{b}\hat{c}\hat{d}}}_S (\chi) = \frac{N}{N-1} \frac{\chi^2}{(1-\chi)^2} (2 + \chi(\chi-2)) \notag \\
  & \phantom{f^{\smash{\hat{a}\hat{b}\hat{c}\hat{d}}}_S (\chi) = } + \veps \frac{\kappa_2 (N+1)}{\kappa_1 (N-1)(N+8)} \frac{\chi}{1-\chi} (\chi \log \chi + (1-\chi) \log (1-\chi)) + \Om(\veps^2)\,, \\
 & f^{\smash{\hat{a}\hat{b}\hat{c}\hat{d}}}_T (\chi) = \frac{\chi^2}{(1-\chi)^2} (2 + \chi (\chi-2)) \notag \\
  & \phantom{f^{\smash{\hat{a}\hat{b}\hat{c}\hat{d}}}_T (\chi) = } + \veps \frac{2 \kappa_2}{\kappa_1 (N+8)} \frac{\chi}{1-\chi} (\chi \log \chi + (1-\chi) \log (1-\chi)) + \Om(\veps^2)\,, \\
  & f^{\smash{\hat{a}\hat{b}\hat{c}\hat{d}}}_A (\chi) = \frac{(2-\chi) \chi ^3}{2 (1-\chi)^2} + \Om(\veps^2)\,.
\end{align}

We can expand these four-point functions in the $1d$ conformal blocks
\begin{equation}\label{eq:1d-block}
 g_{\Delta}^{\Delta_{ij},\Delta_{kl}} (\chi) = \chi^{\Delta} \,_2 F_1 \left(\Delta - \Delta_{ij}, \Delta + \Delta_{kl}; 2 \Delta ; \chi \right)\:, \quad g_{\Delta} := g_{\Delta}^{0,0}\:,
\end{equation}
where $\Delta_{ij} = \Delta_i - \Delta_j$ are the conformal dimensions of the external operators. 

The first operators in the $ \phih^1 \times  \phih^1$ OPE are the degenerate operators $s_{\pm}$.
They can be unmixed, which we have done in Section~\ref{sec:def-3pt-phi2} in order to obtain the anomalous dimensions $\gamma_{s_{\pm}}$ and the OPE coefficients $\lambda_{ \phih^a  \phih^a s_{\pm}}$.
However, in the conformal block expansion we only see the \textit{average} of the conformal data for this operator, and we find
\begin{align}
 &f^{1111} (\chi) = 1 + \left(2-\frac{3 \kappa_{2}\veps }{\kappa_{1}(N+8)} \right) g_{2} (\chi) + \veps \left(\frac{3 \kappa_{2} + 4 (N+8) (4-N)}{ \kappa_{1}(N+8)} \right) \partial_{\Delta} g_{2} (\chi) + \ldots \:,  \\
 &(\Delta_{s_+} -2)\lambda_{ \phih^1  \phih^1 s_+}^2 + (\Delta_{s_{-}} - 2) \lambda_{ \phih^1  \phih^1 s_{-}}^2 = \veps \frac{3 \kappa_{2} + 4 (N+8) (4-N)}{ \kappa_{1}(N+8)} + \Om(\veps^2)\:,\\
 &\lambda_{ \phih^1  \phih^1 s_{+}}^2 + \lambda_{ \phih^1  \phih^1 s_{-}}^2 = 2-\frac{3 \kappa_{2}\veps }{\kappa_{1}(N+8)} + \Om(\veps^2)\:.
\end{align}
This is also the case for the correlators $f^{\smash{\hat{a}\hat{b}\hat{c}\hat{d}}}_S (\chi)$ and $f^{\smash{11\hat{a}\hat{b}}} (\chi)$, which contain in addition information on the OPE coefficients $\lambda_{tts_{\pm}}$.
The OPE coefficients given in Eq.~\eqref{eq:def-OPE-phi2} neatly obey these relations.

From the other correlators we can obtain the OPE coefficients and anomalous dimensions of $\hat{V}^{\hat{a}}$ appearing in $ \phih^1 \times t^{\hat{a}}$, and  $\hat{T}^{\hat{a}\hat{b}}$ and $\hat{A}^{\hat{a} \hat{b}}$, which are respectively a traceless symmetric and antisymmetric operator appearing in $t^{\hat{a}} \times t^{\hat{b}}$.
\begin{alignat}{2}
  \Delta_{\hat{V}} &= 2 + \veps \frac{2 ( \kappa_{2} + (4-N) (N+8))}{(N+8) \kappa_{1}} + \Om(\veps^2)\:,  \quad &&\lambda_{t \hat\phi^1 \hat{V}}^2 = 1- \veps \frac{\kappa_{2}}{\kappa_{1} (N+8)} + \Om(\veps^2)\:, \label{eq:vdef}\\
  \Delta_{\hat{T}} &= 2 + \veps \frac{\kappa_2}{\kappa_1 (8 + N)} + \Om(\veps^2) \:,  &&\lambda_{t t \hat{T}}^2 = 2- \veps\frac{2 \kappa_{2}}{\kappa_{1} (N+8)} + \Om(\veps^2)\:,\\
  \Delta_{\hat{A}} &= 3 + \Om(\veps^2) \:,  &&\lambda_{t t \hat{A}} = 1 + \Om(\veps^2)\:.\label{eq:adef}
\end{alignat}
Setting $N_f \to 0$, we obtain the results for the $\mathrm{O}(N)$ model found in \cite{Cuomo:2021kfm,Gimenez-Grau:2022czc} which provides a final check for our results.

\subsubsection{Four-point functions of the displacement}

We consider now the four-point function of the displacement:
\begin{align}
 \langle \Disp_{\muh} (\tau_1) \Disp_{\nuh} (\tau_2) \Disp_{\rhoh} (\tau_3) \Disp_{\sigmah} (\tau_4) \rangle = \Nm_{\Disp}^4 \, \Km_{4} \, f^{\smash{\muh \nuh \rhoh \sigmah}} (\chi)\,.
\end{align}
Similarly to the two-point function, the diagrams contributing to this four-point function are the same as for the four-point function of scalars, and are shown in Eq.~\eqref{eq:4pt-scal}.
Again, we take the derivative with respect to the transverse coordinates $\partial^{\muh}_{j}, j = 1, \ldots, 4$, and then set $x^{\bot}_{j} \to 0$.
The diagrams not involving fermions were already computed for the $\mathrm{O}(N)$ model in \cite{Gimenez-Grau:2022czc}, while the fermionic diagram is the renormalization of the wavefunction.
Adding all the diagrams and using the proper renormalization, we find the correlators up to $\Om(\veps)$:
\begin{align}
  f^{\smash{\muh \nuh \rhoh \sigmah}} (\chi) =\ &
\delta_{\muh \nuh} \delta_{\rhoh \sigmah}
+ \delta_{\muh \rhoh} \delta_{\nuh \sigmah} \chi^4
+ \delta_{\muh \sigmah} \delta_{\nuh \rhoh} \frac{\chi ^4}{(1-\chi)^4}  \notag \\
&+ \veps
(
\delta_{\muh \nuh} \delta_{\rhoh \sigmah}
+\delta_{\muh \rhoh} \delta_{\nuh \sigmah}
+\delta_{\muh \sigmah} \delta_{\nuh \rhoh}
)
\frac{\kappa_2}{10 \kappa_1 (N+8)} \frac{\chi}{(1-\chi)^3} \notag \\
& \phantom{+} \times
\left( 2 \chi (1-\chi) (\chi(1-\chi)-1)
+ \chi^3 (\chi (5 - 2\chi) - 5) \log \right. \chi  \notag \\
& \left. \phantom{+ \times (}
- (1-\chi)^3 (2\chi^2 + \chi + 2) \log (1-\chi)
\right) \notag \\
&
+\Op(\veps^2)\,.
\end{align}

\subsubsection{Four-point functions of fermions}

We now turn our attention to correlators involving four elementary fermions, identical up to their flavor index. This correlator is given by
\begin{align}
\vev{ \Psihb (\tau_1)  \Psih (\tau_2)  \Psihb (\tau_3)  \Psih (\tau_4) } =& \; \; 
\FourPointFermionsOne
\; \; + \; \;  
\FourPointFermionsTwo
\; \; + \; \; 
\FourPointFermionsThree
 \notag \\
 &+ \; \; 
 \FourPointFermionsFour
\; \; + \; \;  
\FourPointFermionsFive
\; \; + \Om(\veps^{3/2}) \notag \\ 
=& \frac{\NormPsiHat^4}{\tau_{12}^{\smash{2\Delta_{ \Psih}}} \tau_{34}^{\smash{2\Delta_{ \Psih}}}} \left( f_{12,34} (\chi) - \frac{\chi^3}{(1-\chi)^3} f_{14,32} (1-\chi) \right)\,,
\end{align}
where the second term follows by crossing symmetry.
The subscripts indicate the dependency on the polarization spinors $\bar{s}_1, s_2, \bar{s}_3, s_4$.
The disconnected part of the correlator is easy to compute and give
\begin{align}
\vev{\Psihb (\tau_1)  \Psih (\tau_2)  \Psihb (\tau_3)  \Psih (\tau_4)}_{\text{disc.}} =\ & \vev{ \Psihb (\tau_1)  \Psih (\tau_2)} \vev{ \Psihb (\tau_3)  \Psih (\tau_4)} \notag \\
&+ \vev{ \Psihb (\tau_1)  \Psih (\tau_4)} \vev{ \Psihb (\tau_3)  \Psih (\tau_2)} \notag \\
=\ & \frac{1}{\tau_{12}^{\smash{2\Delta_{ \Psih}}} \tau_{34}^{\smash{2\Delta_{ \Psih}}}} \left\lbrace (\bar{s}_1 \gamma^0 s_2) (\bar{s}_3 \gamma^0 s_4) \right. \notag \\
& \left. - \frac{\chi^3}{(1-\chi)^3} ( \lbrace s_2\,, j\,, \chi \rbrace \leftrightarrow \lbrace s_4\,,l\,, 1-\chi \rbrace) \right\rbrace\,.
\end{align}

The connected part consists of two diagrams:
\begin{equation}
\vev{\Psihb (\tau_1)  \Psih (\tau_2)  \Psihb (\tau_3)  \Psih (\tau_4)}_{\text{conn.}} = 
\FourPointFermionsFour
\; \; + \; \;  
\FourPointFermionsFive
 \:.
\end{equation}
These diagrams belong to a new class that we have not encountered yet and that we name \textit{$H$-diagrams}. They can be expressed as
\begin{equation}
\FourPointFermionsFour = g_{0}^2 (\bar{s}_1 \slashed{\partial}_1 \Sigma^a \slashed{\partial}_2 s_2) (\bar{s}_3 \slashed{\partial}_3 \Sigma^a \slashed{\partial}_4 s_4) H_{12,34}\,,
\label{eq:Hintegral-4F}
\end{equation}
where the integral $H_{12,34}$ is defined in \eqref{eq:BasicDefs} and has not been solved analytically yet.
It is however possible to solve the integral thanks to the derivatives in front, as shown in \eqref{eq:sol-G1234a}-\eqref{eq:sol-G1234}.
The second diagram can be calculated analogously, and we obtain the following unit-normalized correlator:
\begin{align}
f_{12,34} (\chi) =\ & (\bar{s}_1 \gamma^0 s_2)(\bar{s}_3 \gamma^0 s_4) \notag \\
&+ \frac{\veps}{64 \kappa_1} (\bar{s}_1 \Sigma^a \gamma^0 s_2)(\bar{s}_3 \Sigma^a \gamma^0 s_4) \notag \\
& \phantom{-\ } \times \frac{\chi}{(1-\chi)^2} \left( (1-\chi)(2-\chi) + \chi^2 (2-\chi) \log \chi + \chi (1-\chi)^2 \log (1-\chi) \right) \notag \\
&+ \Op(\veps^2)\,.
\end{align}
We can extract new defect CFT data from this correlator by expanding it in the $1d$ blocks of Eq.~\eqref{eq:1d-block}.
Since we have $N_f$ fermions, there is a $U(N_f)$ flavor symmetry and we need to decompose the fermions in the singlet ($S$) and adjoint ($\text{Adj}$) representations.
For this purpose, we reinstate the flavor indices $i=1, \ldots, N_f$, for which the decomposed correlator becomes:\footnote{This decomposition is more intricate for the cH model, and the results below are valid for the GNY and NJLY models.}
\begin{align}
 f^{ijkl}_{12,34}(\chi) =\ & \delta^{ij} \delta^{kl} f^{S}_{12,34} (\chi) + \left(\delta^{il} \delta^{jk} - \frac{\delta^{ij}\delta^{kl}}{N_f}\right) f^{\text{Adj}}_{12,34} (\chi)\:,\\
 f^{S}_{12,34} (\chi) =\ & \frac{4 + (\kappa_{1}+N) (\chi -1)^3 -2 \chi  ((\chi -6) \chi +6)}{(\chi -1)^3 (\kappa_{1}+N-4)} + \frac{\kappa_{1}+N}{(\kappa_{1}+N-4)}\notag \\
 &+\frac{\chi  \veps}{64 \kappa_{1} (\chi -1)^2 (\kappa_{1}+N-4)} \Bigg\{2 (\kappa_{1} -4 + N) -\chi  (14 + 3\kappa_{1}+3N)\notag \\
 &+\chi ^2 (\kappa_{1}+N-2) -\chi ^2 (\chi  (\kappa_{1}+N-2)-2 (\kappa_{1}+N-3)) \log \chi \notag \\
 &+(\chi -1)^2 (\chi  (\kappa_{1}+N-2)+2) \log (1-\chi ) \Bigg\}\:, \\
 f^{\text{Adj}}_{12,34} (\chi) =\ & \frac{\chi ^3}{(\chi -1)^3}+\frac{ \veps  \chi}{64 \kappa_{1} (\chi -1)^2}\Bigg\{\chi  (1 + \chi)+ (1+\chi) (1-\chi)^2 \log (1-\chi ) \notag \\
 &+(1-\chi) \chi^2  \log \chi\Bigg\}\:.
\end{align}
Let us decompose the singlet sector in the conformal blocks.
For the first few operators, we find
\begin{align}\label{eq:def-block-exp-4fer}
 f^{S}_{12,34} (\chi) =\ & g_{0} (\chi) + \frac{\veps}{32 \kappa_1} g_{1} (\chi) - \left( \frac{1}{N_f} - \frac{\veps (17 - 2 N - 2\kappa_1)}{384 \kappa_1 N_f} \right) g_{3} (\chi) \notag \\
 &-  \frac{\veps (3 - N - \kappa_1)}{64 \kappa_1 N_f} \partial_{\Delta} g_{3} (\chi) + \ldots\:.
\end{align}
The absence of a conformal block $g_{\Delta = 2} (\chi) $ indicates that 
\begin{equation}
 \lambda_{ \Psihb  \Psih s_{\pm}} = \Om(\veps)\:,
\end{equation}
such that the squared OPE coefficients only contribute at $\Om(\veps^2)$.
We can read off $\lambda^2_{ \Psihb  \Psih  \phih^1}$ as the coefficient in front of the block $g_{\Delta = 1} (\chi)$, which matches the expression in Eq.~\eqref{eq:def-3pt-fer-scal}.

\subsubsection{Four-point functions of fermions and scalars}

In this section, we compute the mixed correlator with two fermions $ \Psihb$, $\Psih$ and two elementary scalars $\phi^a$, $\phi^b$. The correlator takes the following form: 
\begin{equation}
\vev{ \Psihb (\tau_1)  \Psih (\tau_2)  \phih^a (\tau_3)  \phih^b (\tau_4)} = \frac{\NormPsiHat^2 \NormPhiHat^2}{\tau_{12}^{\smash{2\Delta_{ \Psih}}} \tau_{34}^{\smash{2\Delta_{ \phih}}}} f^{ab}_{12} (\chi)\,,
\end{equation}
with the kinematical cross-ratio $\chi$ defined in \eqref{eq:def-CrossRatio}, and where the $\mathrm{O}(N)$ tensor structure, as well as the dependence on the polarization spinors $\bar{s}_1, s_2$, are encoded in $f^{ab}_{12}$. As before, the disconnected part of the correlator is easy to obtain and consists of only one non-zero term:
\begin{align}
\vev{ \Psihb (\tau_1)  \Psih (\tau_2)  \phih^a (\tau_3)  \phih^b (\tau_4)}_{\text{disc.}} &= \vev{ \Psihb (\tau_1)  \Psih (\tau_2)} \vev{ \phih^a (\tau_3)  \phih^b (\tau_4)} \notag \\
&= \frac{(\bar{s}_1 \gamma^0 s_2)}{\tau_{12}^{\smash{2\Delta_{ \Psih}}} \tau_{34}^{\smash{2\Delta_{ \phih}}}} \delta^{ab}\,.
\end{align}
The connected part consists of two fermion-scalar $H$-diagrams:
\begin{equation}
\vev{ \Psihb (\tau_1)  \Psih (\tau_2)  \phih^a (\tau_3)  \phih^b (\tau_4)}_{\text{conn.}} = \; \;  
\FourPointFermionsScalarsOne
\; \; + \; \;  
\FourPointFermionsScalarsTwo
\,.
\end{equation}
After using the rules for the $\Sigma$-matrices, we find that the first diagram gives
\begin{equation}
\FourPointFermionsScalarsOne = \pm g_{0}^2 (\bar{s}_1\, \Sigma^a \Sigma^b F_{13,24}\, s_2)\,,
\end{equation}
with $+$ if $\phi^a$ is a scalar and $-$ if it is a pseudoscalar,\footnote{In this formulation, the index $b$ can be kept arbitrary since we have to commute $\Sigma^b$ with $\gamma$-matrices \textit{twice}.} and where the integral $F_{13,24}$ is defined in \eqref{eq:def-F1324} and solved in \eqref{eq:sol-F1324}. Putting everything together, the unit-normalized correlator reads
\begin{align}
f^{ab}_{12}(\chi) =\ & \delta^{ab} (\bar{s}_1 \cdot s_2) \notag \\
& \pm (\bar{s}_1 \Sigma^a \Sigma^b s_2) \frac{\veps}{8 \kappa_1} 
\frac{\chi}{(1-\chi)^2} \left(
\chi^3 \log \chi - (1-\chi)^2(2+\chi) \log (1-\chi)
\right) \notag \\
&+ \Op(\veps^2)\,.
\end{align}

We expand this correlator in the $1d$ blocks of Eq.~\eqref{eq:1d-block} for the case of equal external scalars, such that 
\begin{equation}
 \Sigma^a = \Sigma^b\:, \quad \Sigma^a \Sigma^b = \mathds{1}_{N_f} \mathds{1}_{4}\:.
\end{equation}
We find:
\begin{align}
  f^{aa}_{12} (\chi) &=  g_{0}(\chi) +  \frac{\veps}{4 \kappa_1} g_{2} (\chi) + \frac{19\veps}{240 \kappa_1} g_{4} (\chi) - \frac{\veps}{8 \kappa_1} \partial_{\Delta} g_{4} (\chi) + \ldots \:,
\end{align}
where we emphasize that no sum is implied by the repetition of indices on the left-hand side. Since the correlator and the block expansion have the same expression for $ \phih^1$ and $t^{\hat{a}}$, we find the same relations for the OPE coefficients $\lambda_{ \phih^1  \phih^1 \Om}$ and $\lambda_{t t \Om}$, which we denote as $\lambda_{ \phih^a  \phih^a \Om}$ for brevity.

From the block expansion we see that for $s_{\pm}$, which has dimension $\Delta_{s_{\pm}} \sim 2$, we obtain
\begin{align}
  &\lambda_{ \Psihb  \Psih s_{+}} \lambda_{ \phih^a  \phih^a s_{+}} + \lambda_{ \Psihb  \Psih s_{-}} \lambda_{ \phih^a  \phih^a s_{-}}  = \frac{\veps}{4 \kappa_1}\:,\\
  &(\Delta_{s_{+}} - 2)\lambda_{ \Psihb  \Psih s_{+}} \lambda_{ \phih^a  \phih^a s_{+}} + (\Delta_{s_{-}} - 2)\lambda_{ \Psihb  \Psih s_{-}} \lambda_{ \phih^a  \phih^a s_{-}} = 0\:.
\end{align}
Using the expressions for $\Delta_{s_{\pm}}$ and $\lambda_{ \phih^1  \phih^1 s_{\pm}}, \lambda_{t t s_{\pm}}$ in eqs.~\eqref{eq:defcorr-gamma-spm} and \eqref{eq:def-OPE-phi2}, we can extract the OPE coefficients involving the fermions:
\begin{align}
 \lambda_{ \Psihb  \Psih s_{\pm}} =\ &\veps \frac{(N\kappa_2 - 4(N-4)(N+8) \pm \kappa_{3}) \sqrt{\kappa_3 \pm 4 (N-4)(N+8) \pm (N-2)\kappa_2}}{16 \kappa_{1} \kappa_{2} \sqrt{\kappa_{3}} \sqrt{N-1}} \notag \\
 &+ \Om(\veps^2) \:.
\end{align}
As expected from the conformal block expansion in Eq.~\eqref{eq:def-block-exp-4fer}, the OPE coefficients start at $\Om(\veps)$.

\section{Correlators of bulk operators with a defect}\label{sec:bulkcorr}
The three- and four-point functions of scalars and fermions on the defect provided us with important defect data. In addition, we can also study bulk operators in the presence of the line defect, and obtain new data such as bulk-to-defect OPE coefficients.
In this section we study two-point functions of bulk and defect operators, as well as two-point functions of bulk scalars.
At the end of the section we give a short outlook on how to generalize our analysis to the case of fermionic operators.

\subsection{One-point functions}\label{sec:one-pt}

\paragraph{Squared scalar.}
We computed the one-point function of $\phi^a$ in Section~\ref{sec:setup} to extract the $\beta$-function of the defect coupling.
The coefficient of this one-point function, $a_{\phi}$, appears in the bulk channel expansion of the two-point function of $\phi^a$ in the presence of the line defect.
One-point function coefficients of other operators appear as well, the first one being the one-point function of $\phi^2$ and the traceless symmetric tensor $T^{ab} = \phi^a \phi^b - \frac{\delta^{ab}}{N} (\phi^{c})^2$.\footnote{Note that $T^{ab}$ does not appear for the GNY model, where $N=1$.}
These observables were computed for the $\mathrm{O}(N)$ model in \cite{Cuomo:2021kfm}.
At $\Om(\veps)$ there are four diagrams that contribute:
\begin{align}
 \langle \langle \phi^a \phi^b (x) \rangle \rangle = 
 \OnePointSquaredScalarOne
  \, + \, 
\OnePointSquaredScalarTwo
 \, + \, 
\OnePointSquaredScalarThree
 \, + \, 
\OnePointSquaredScalarFour
 \,  + \Om(\veps^2)  \:.
\end{align}
The diagrams not including any fermionic contributions were computed in \cite{Cuomo:2021kfm,Gimenez-Grau:2022ebb,Bianchi:2022sbz}, while the diagram with the fermionic loop cancels the wavefunction renormalization of $\phi^a$.
In order to compute the one-point function coefficient, we need the renormalization factor, anomalous dimension, and normalization factor of $\phi^2$ and $T^{ab}$.
The anomalous dimensions for the GNY and NJLY models can be found in \cite{Fei:2016sgs}, while the others can be obtained from computing the corrections to the propagator $\langle \phi^a \phi^b  (x_1) \phi^c \phi^d (x_2)\rangle$.
Generalized for $N = 1,2,3$, we find for $\phi^2$ up to $\Om(\veps)$ \cite{Fei:2016sgs,Zerf:2017zqi}
\begin{align}
&Z_{\phi^2} = 1 - \frac{1}{\veps (4\pi)^2} \left(\frac{\lambda (N+2)}{3} + 4 g^2 N_f \right) +\Om\left(\veps^{-2}\right)\:,  \\
&\gamma_{\phi^2} = \frac{\lambda (N+2)}{3 (4\pi)^2} + \frac{4 g^2 N_f}{(4\pi)^2} +\Om(\lambda^2,g^4,\lambda g^2)\:,  \\
&\NormPhiSquared = \frac{\Gamma \left(\frac{d}{2}\right) \sqrt{2 N}}{2 \pi ^{\frac{d}{2}} (d-2)} \Bigg\{1- \veps \Bigg(\frac{2 (\aleph+1) (N+8) (\kappa_1+N-4)+\kappa_2 \aleph (N+2)}{4 \kappa_1 (N+8)}\Bigg) +\Om(\veps^2)\Bigg\}\:.
\end{align}
For $T^{ab}$, we obtain the following results:
\begin{align}
&Z_{T} = 1 - \frac{1}{\veps (4\pi)^2} \left(\frac{2 \lambda}{3} + 4 g^2 N_f \right) +\Om\left(\veps^{-2}\right)\:, \\
&\gamma_{T} = \frac{2 \lambda}{3 (4\pi)^2} + \frac{4 g^2 N_f}{(4\pi)^2} +\Om(\lambda^2,g^4,\lambda g^2)\:,  \\
&\NormPhiSquared = \frac{\Gamma \left(\frac{d}{2}\right) }{\sqrt{2} \pi ^{\frac{d}{2}} (d-2)} \Bigg\{1- \veps \Bigg( \frac{(\aleph+1) (N+8) (\kappa_1+N-4)+\kappa_2 \aleph}{2 \kappa_1 (N+8)} \Bigg)  +\Om(\veps^2)\Bigg\}\:,
\end{align}
from which we can extract the one-point function coefficients $a_{\phi^2}$ and $a_{T}$:
\begin{align}
  a_{\phi^2} &= \frac{(4 - N)(N+8)}{2 \kappa_2 \sqrt{2 N} } + \Om(\veps)\:, \quad a_{T} = \frac{(4 - N)(N+8)}{2 \kappa_2 \sqrt{2} } + \Om(\veps)\:.
  \label{eq:one-pt-phi2T}
\end{align}
The $\Om(\veps)$-terms are lengthy and can be found in the attached \textsc{Mathematica} notebook.

\paragraph{Fermion bilinear.}

Another interesting one-point function is $\vvev{\Psib \Psi (x)}$, which appears in the two-point function $\vvev{\Psib (x_1) \Psi (x_2)}$.
In this case, $\Psib \Psi$ is not a conformal primary, but rather a conformal descendant of $\phi$.
This can be seen from its conformal dimension being $\Delta_{\phi} + 2 + \Om(\veps^2)$ \cite{Fei:2016sgs}.
The one-point function can be computed through Feynman diagrams, and receives a contribution at $\Om(\sqrt{\veps})$.
We will only consider the GNY and NJLY model here, for which the one-point function is given by:
\begin{equation}
\vvev{\Psib \Psi (x)} = 
 \OnePointBilinear
  \, + \Om(\veps)\:.
\end{equation}
This egg-shaped diagram is new and has the following expression: 
\begin{equation}
\OnePointBilinear = \NormBilinear \frac{g_0 h_0 N_f}{16 \pi^3 |x^\perp|^3} \, \text{tr}\, \Sigma^1\:.
\end{equation}
Hence, the one-point function coefficient can be written as
\begin{equation}
\vvev{\Psib \Psi (x)} = \frac{a_{\Psib \Psi}}{|x^\perp|^{\Delta_{\Psib\Psi}}}\:, \quad a_{\Psib\Psi} = - \frac{\sqrt{(4-N) (N+8)}}{\sqrt{\kappa_1 \kappa_2} } \frac{8 \pi^2 N_f}{(4 \pi)^2 } \sqrt{\veps} + \Om(\veps)\:.
\end{equation}

\subsection{Bulk-to-defect two-point functions}

The correlators of a bulk and a defect operator give us the OPE coefficients of the bulk-to-defect OPE.
The fundamental bulk scalar $\phi^a$ can be decomposed into $\phih^1$ and $t^{\hat{a}}$, and the two-point function between these operators is given by the following diagrams:
\begin{align}
  \vvev{ \phi^a (x_1) \phih^b (\tau_2) } = 
  \BulkToDefectTwoPointOne
   \, + \,
\BulkToDefectTwoPointTwo
 \, + \,
\BulkToDefectTwoPointThree
 \, + \Om(\veps^2) \:.
\end{align}

The first diagram does not involve any integration, while the second is a self-energy correction.
The third diagram is less trivial, and was computed in \cite{Gimenez-Grau:2022ebb}.
Adding all diagrams and the proper renormalization terms, we obtain the following bulk-to-defect OPE coefficients:
\begin{align}
  &\vvev{\phi^a (x_1) \phih^1 (\tau_2)} = \frac{\delta^{a1} \NormPhi \NormPhiHatIndex{1} \hat{b}_{\phi \phih^1}}{(\hat{x}_{1\hat{2}}^2)^{\hat{\Delta}_{\phih^1}} |x^{\bot}_1|^{\Delta_{\phi} - \hat{\Delta}_{\phih^1}}}\:, \quad &&\hat{b}_{\phi \phih^1} = 1 + \veps \frac{3 (4 - N)(\log 2 - 1)}{2 \kappa_1} + \Om(\veps^2)\:, \\
  &\vvev{\phi^a (x_1) t^{\hat{b}} (\tau_2)} = \frac{\delta^{a\hat{b}} \NormPhi \NormTilt \hat{b}_{\phi t}}{(\hat{x}_{1\hat{2}}^2)^{\hat{\Delta}_{t}} |x^{\bot}_1|^{\Delta_{\phi} - \hat{\Delta}_{t}}}\:, \quad &&\hat{b}_{\phi t} = 1 + \veps \frac{(4 - N)(\log 2 - 1)}{2 \kappa_1} + \Om(\veps^2)\:.
 \label{eq:bulkdefOPE-phiphi2}
\end{align}

\subsection{Two-point functions of bulk scalars}

In the presence of a defect, the two-point function of bulk operators is no longer fixed by kinematics.
Instead, it depends on two \textit{defect cross-ratios} determined by the distance to the defect and the distance between the bulk operators.
The scalar two-point function then takes the form
\begin{equation}
 \vvev{ \phi^a (x_1) \phi^b (x_2)} = \frac{\NormPhi^2 \Fm^{ab} (r, w)}{|x_{1}^{\bot}|^{ \Delta_{\phi}} |x_{2}^{\bot}|^{\Delta_{\phi}}}\:,
\end{equation}
where $\Fm^{ab} (r,w)$ is a function of the cross-ratios
\begin{equation}
 r + \frac{1}{r} = \frac{\tau^{2}_{12} + (x^{\bot}_{1})^2 + (x^{\bot}_{2})^2}{|x^{\bot}_1| |x^{\bot}_2|} \:, \quad w + \frac{1}{w} = \frac{2 x^{\bot}_1 \cdot x^{\bot}_2}{|x^{\bot}_1| |x^{\bot}_2|} \:.
\end{equation}
It is sometimes convenient to switch to different cross-ratios $z,\bar{z}$, which are related to $r,w$ as
\begin{equation}\label{eq:zzb-coord}
 z = rw \:, \quad \bar{z} = \frac{r}{w}\:.
\end{equation}
The diagrams contributing to this two-point function, up to $\Om(\veps)$, are shown in Fig.~\ref{fig:2pt-bulk-scal-diags}.
They consist of diagrams we already encountered when computing the one-point function of $\phi^a$, of diagrams coming from the wavefunction renormalization in the bulk, and one non-trivial one. 

\begin{figure}
 \begin{subfigure}{\textwidth}
 \centering
 \TwoPointBulkScalarsOne
 \TwoPointBulkScalarsTwo
 \TwoPointBulkScalarsThree
\end{subfigure}
\begin{subfigure}{\textwidth}
\centering
\TwoPointBulkScalarsFour
 \TwoPointBulkScalarsFive
 \TwoPointBulkScalarsSix
 \end{subfigure}
 \caption{Contributions to the two-point function $\vvev{ \phi^a(x_1) \phi^b(x_2)}$ up to $\Om(\veps)$. The defect is denoted by a solid line, scalars by a dotted line, and fermions by solid arrowed lines. Bulk scalar couplings $\lambda_0$ are represented by a black dot, bulk Yukawa couplings $g_0$ by a red dot and defect couplings $h_0$ by a blue dot.}
 \label{fig:2pt-bulk-scal-diags}
\end{figure}

The non-trivial diagram is the X-shaped diagram, which was computed for a line defect in the $\mathrm{O}(N)$ model in \cite{Gimenez-Grau:2022ebb,Bianchi:2022sbz}.
Evaluating it in $d = 4$ gives:
\begin{align}
 \scalebox{.75}{\raisebox{2.5ex}{\TwoPointBulkScalarsSix}}
  \; \; &= 
-\frac{\lambda_0 h_{0}^2\, \Gamma^4 \left(\frac{d}{2}\right)}{32 \pi ^{2 d} (d-2)^4} \int d\tau_3\, d\tau_4\, X_{1234} \notag\\
&= \frac{3 \lambda_0 h_{0}^2  H(r,w )}{768 \pi ^4 |x_{1}^{\bot}| |x_{2}^{\bot}|} + \Om(\veps^2)\:,
\end{align}
with $X_{1234}$ defined in \eqref{eq:BasicDefs}, and where $H(r,w)$ contains one unevaluated integral over a Schwinger parameter $\alpha$ \cite{Gimenez-Grau:2022ebb}:
\begin{equation}
\label{eq:H_integral}
 H(r,w) = - \int_{0}^{\infty} d\alpha \sqrt{\frac{z\bar{z}}{(\alpha+1)(\alpha + z \bar{z}) (\alpha + z)(\alpha + \bar{z})}} \tanh^{-1} \sqrt{\frac{(\alpha + z)(\alpha + \bar{z})}{(\alpha + 1)(\alpha + z \bar{z})}}\:.
\end{equation}
Even though the integral is unevaluated, the series expansions in the bulk and defect channels are known. 

Adding all diagrams in Fig.~\ref{fig:2pt-bulk-scal-diags} and properly renormalizing them, we obtain
\begin{equation}\label{eq:phiphi-corr}
 \Fm^{ab} (r,w) = \delta^{ab} \xi^{-\Delta_{\phi}}+ \delta^{a1} \delta^{b1} a_{\phi}^2  + \veps (\delta^{ab} + 2 \delta^{a1} \delta^{b1}) \frac{3(4-N)}{4 \kappa_1} H(r,w ) + \Om(\veps^2)\:.
\end{equation}
Here, $a_{\phi}$ is the one-point function coefficient given in Eq.~\eqref{eq:aphi}.
We see that the general form is the same as in \cite{Gimenez-Grau:2022ebb,Bianchi:2022sbz}, except for additional fermionic contributions to the coefficient in front of $H(r,w)$.
We can expand this expression in bulk and defect conformal blocks to extract CFT data, and check with the explicit calculations in the previous sections.

\subsubsection{Defect channel}

In the defect channel, the correlator $\Fm^{ab} (r,w)$ contains two types of operators: $\mathrm{O}(N)-$singlets $\hat{\Om}^{S}_{s,n}$ and $\mathrm{O}(N)-$vectors $\hat{\Om}^{V}_{s,n}$.
Their conformal dimensions are given by
\begin{equation}
 \Delta_{\hat{\Om}^{S,V}_{s,n}} = \Delta_{\phi} + s + n + \gamma_{\hat{\Om}^{S,V}_{s,n}} \:.
\end{equation}
These operators are in general degenerate, except for $n=0$, where they can be expressed as derivatives of $\phih^1$ and $t$:
\begin{equation}
 \hat{\Om}_{s,0}^{S} \sim \partial^{\bot}_{i_1} \ldots \partial^{\bot}_{i_s} \phih_1\:, \quad \hat{\Om}_{s,0}^{S} \sim \partial^{\bot}_{i_1} \ldots \partial^{\bot}_{i_s} t_{\hat{a}}\:,
\end{equation}
where $i_n = 1, \ldots , d-1$ are the directions transverse to the defect.
For higher $n$, one needs to solve a mixing problem.
This has been done in \cite{Cuomo:2021kfm,Gimenez-Grau:2022czc} for the $\mathrm{O}(N)$ model and repeated in Section~\ref{sec:def-threept} for the case $n=1, s=0$ to obtain the anomalous dimension $\gamma_{s_{\pm}}$.
For general $n>0,s$ we give the averaged CFT data.

The correlator $\Fm^{ab} (r,w)$ can be decomposed in the two symmetry channels $S$ (singlet) and $V$ (vector):
\begin{align}
  &\Fm^{ab} (r,w) = \delta^{a1} \delta^{b1} \hat{\Fm}_{S} (r,w) + (\delta^{ab} - \delta^{a1} \delta^{b1}) \, \hat{\Fm}_{V} (r,w)\:, \\
  &\hat{\Fm}_S (r,w) = a_{\phi}^2 + \frac{1}{\xi^{\Delta_{\phi}}} + \veps \frac{3(4-N)}{4 \kappa_1} H(r,w)\:, \, \, \hat{\Fm}_V (r,w) = \frac{1}{\xi^{\Delta_\phi}} + \veps \frac{(4-N)}{4 \kappa_1} H(r,w)\:,
  \label{eq:bulk2pt-def-symm-exp}
\end{align}
with
\begin{equation}
\xi := \frac{(1-rw)(w-r)}{rw} = \frac{x_{12}^2}{|x_1^\perp| |x_2^\perp|}\,.
\end{equation}
Each of the channels can be composed in defect conformal blocks, which are known in closed form \cite{Billo:2016cpy}:
\begin{align}
  &\Fm (z,\bar{z}) = \sum_{\hat{\Om}} 2^{-s} \, \hat{b}^{2}_{\Om \hat{\Om}} \hat{f}_{\hat{\Delta},s} (z,\bar{z})\:, \\
  &\hat{f}_{\hat{\Delta},s} (z,\bar{z}) = (z \bar{z})^{\frac{\hat{\Delta}}{2}} \left(\frac{\bar{z}}{z}\right)^{\frac{s}{2}} \,_2 F_1 \left(\frac{p}{2}, \hat{\Delta}; \hat{\Delta} + 1 - \frac{p}{2};z \bar{z}\right) \,_2 F_1 \left(-s, \frac{q}{2} - 1;2 - \frac{q}{2} - s;\frac{z}{\bar{z}}\right) \:,
\end{align}
where $p = 1$ is the dimension of the defect, $q = d-1$ the codimension, $s$ is the transverse spin and we have switched variables from $r,w$ to $z, \bar{z}$ using the definition in Eq.~\eqref{eq:zzb-coord}.
The factor of $2^{-s}$ ensures that the blocks have a convenient normalization.

To expand the correlator in terms of these blocks, we need to know how to decompose the function $H(r,w)$.
It turns out there is an elegant expression found in \cite{Gimenez-Grau:2022ebb,Bianchi:2022sbz}:
\begin{equation}\label{eq:Hdef-exp}
 H(r,w) = \sum_{s=0}^{\infty} \left(\frac{H_s  - H_{s - \frac{1}{2}}}{s + \frac{1}{2}} - \frac{1}{(s + \frac{1}{2})^2} + \frac{1}{s + \frac{1}{2}} \partial_{\hat{\Delta}} \right) \hat{f}_{s+1,s} (r,w)\:.
\end{equation}
The derivative of the block gives us the anomalous dimension of the corresponding operator, and hence Eq.~\eqref{eq:Hdef-exp} provides a straightforward way to extract defect CFT data.

The constant terms in Eq.~\eqref{eq:bulk2pt-def-symm-exp} correspond to the defect identity given by $\hat{f}_{0,0} (r,w)$.
This leaves us with the factors $\xi^{- \Delta_{\phi}}$, whose expansion in defect blocks is well known \cite{Lemos:2017vnx}.

Combining all the pieces together we are ready to extract the CFT data.
Let us start with the singlet channel.
Expanding in conformal blocks gives us
\begin{equation}\label{eq:FSdef-block-exp}
 \hat{\Fm}_S (r,w) = a_{\phi}^2 \, \hat{f}_{0,0} (r,w) + \sum_{s = 0}^{\infty} 2^{-s} \, \hat{b}^{2}_{\phi \hat{\Om}^S_{s,0}} \hat{f}_{\hat{\Delta}_{\hat{\Om}^{S}_{s,0}} ,s} (r,w) + \Om(\veps^2)\:,
\end{equation}
where up to $\Om(\veps)$ only operators with $n=0$ appear.
Combining Eq.~\eqref{eq:FSdef-block-exp} and Eq.~\eqref{eq:bulk2pt-def-symm-exp}, and using the expression in Eq.~\eqref{eq:Hdef-exp}, we obtain the following OPE coefficients:
\begin{align}
 \hat{b}_{\phi \hat{\Om}^S_{s,0}} = &2^{\frac{s}{2}} \Bigg\{1 + \frac{\veps}{4\kappa_1} \Bigg(\frac{6 (N-4)}{(2 s+1)^2}-\frac{(\kappa_1 (2s + 1)+3 (N-4))}{(2 s+1)} H_s \notag \\
 &+ 3 (N-4) H_{s-\frac{1}{2}}\Bigg) + \Om(\veps^2)\Bigg\}\:.
\end{align}
For $s = 0$, we see that this matches exactly the bulk-to-defect OPE coefficient $\hat{b}_{\phi \phih_1}$ given in Eq.~\eqref{eq:bulkdefOPE-phiphi2}.
As stated above, to extract the anomalous dimension we only have to look at the derivative term in Eq.~\eqref{eq:Hdef-exp}.
This results in
\begin{align}
 \Delta_{\hat{\Om}^S_{s,0}} &= \Delta_{\phi} + s + n + \frac{3 (4-N) \veps }{4 \kappa_1 \left(s+\frac{1}{2}\right)} + \Om(\veps^2) \notag \\
 &= 1 + s + \frac{(N-4) (s-1) \veps}{\kappa_1 (2 s+1)} + \Om(\veps^2)\:.
\end{align}
For $s = 0$, this matches with $\Delta_{\phih_1}$ given in Eq.~\eqref{eq:dim-phi-hat}, while for $s = 1$ this should give us the dimension of the displacement $\Delta_\Disp = 2$.
Indeed, we see that for $s=1$, the $\Om(\veps)$ correction is zero and the dimension is protected and equal to 2.

We are ready to move on to the vector channel.
The expansion in conformal blocks results in
\begin{equation}
 \hat{\Fm}_{V} (r,w) = \sum_{s= 0}^{\infty} 2^{-s} \, \hat{b}^{2}_{\phi \hat{\Om}^{V}_{s,0}} \hat{f}_{\hat{\Delta}_{\hat{\Om}^{V}_{s,0}},s} (r,w) + \Om(\veps^2)\:,
\end{equation}
where we see that also here up to $\Om(\veps)$, only the $n=0$ family of operators appears. 
The defect identity is not present in this case.
We follow the same procedure as for the singlet channel, and extract the bulk-to-defect OPE coefficients

\begin{align}
 \hat{b}_{\phi \hat{\Om}^{V}_{s,0}} = 2^{\frac{s}{2}} \left\{1 + \frac{\veps (N-4)}{4 \kappa_1 } \Bigg(\frac{2s + 3}{(2 s+1)^2} +  \frac{2 s H_s+H_{s-\frac{1}{2}}}{(2 s+1)} \Bigg) + \Om(\veps^2)\right\}\:.
\end{align}
We can now compare this for $s = 0$ with $\hat{b}_{\phi t}$ in Eq.~\eqref{eq:bulkdefOPE-phiphi2} and find a perfect match.
The anomalous dimensions are once again read off from the derivative term in the expansion of $H(r,w)$, and result in the following conformal dimensions:
\begin{equation}
 \hat{\Delta}_{\hat{\Om}^{V}_{s,0}} = \Delta_{\phi} + s + n + \frac{\veps  (4-N) }{4 \kappa_1 \left(s+\frac{1}{2}\right)} + \Om(\veps^2) = 1 + s + \frac{\veps (N-4) s}{\kappa_1 (2 s+1)}+\Om(\veps^2)\:.
\end{equation}
As a check, we see that for $s = 0$ the $\Om(\veps)$ term disappears and we find the protected dimension of the tilt $\Delta_t = 1$.

\subsubsection{Bulk channel}

In the bulk channel, the operators that appear in the $\phi^a \times \phi^b$ OPE are $\mathrm{O}(N)$ singlets $\Om^{S}_{\ell,n}$, where the first one is $\phi^2$, and traceless symmetric representations $\Om^{T}_{\ell,n}$, the first one of which is $T^{ab}$.
The operators in the lowest twist family after $\phi^2$ and $T^{ab}$ can be written as 
\begin{equation}
 \Om^{S}_{\ell,0} \sim \partial_{\mu_1} \ldots \partial_{\mu_\ell} (\phi^{a})^2\:, \quad \Om^{T}_{\ell,0} \sim \partial_{\mu_1} \ldots \partial_{\mu_\ell} \left( \phi^a \phi^b  - \frac{\delta^{ab}}{N} (\phi^c)^2 \right)\:,
\end{equation}
where $\ell \geq 2$.
In the free theory, they are the higher-spin currents and hence their conformal dimension and OPE coefficients are protected up to $\Om(\veps)$ and given by the conformal dimension of $\phi$ and their spin.
The CFT data is given by 
\begin{align}
 &\Delta_{\Om^{S,T}_{\ell,0}} = 2 \Delta_{\phi} + \ell + \Om(\veps^2)\:, \\
 &\lambda^{2}_{\phi \phi \Om^{S}_{\ell,0}}  = \frac{2^{\ell + 1} (\Delta_{\phi})_{\ell}^2}{N \ell! (2 \Delta_{\phi} + \ell - 1)_{\ell}} + \Om(\veps^2)\:, \\
 & \lambda^{2}_{\phi \phi \Om^{T}_{\ell,0}} = N \lambda^{2}_{\phi \phi \Om^{S}_{\ell,0}} + \Om(\veps^2)\:.
 \label{eq:bulkn0-CFT-data}
\end{align}
Operators with $n>0$ are not protected up to this order, and are also degenerate.
The correlator can be decomposed in the two symmetry channels as:
\begin{align}
  &\Fm^{ab} (r,w) = \delta^{ab} \Fm_{S} (r,w) + \left(\delta^{a1} \delta^{b1} - \frac{\delta^{ab}}{N}\right) \Fm_{T} (r,w)\:,  \\
  &\Fm_{S} (r,w) = \frac{1}{\xi^{\Delta_{\phi}}} + \frac{a_{\phi}^2}{N} + \frac{\veps(4-N)(N-2)}{4 N \kappa_1} H(r,w)\:,  \\\
  &\Fm_{T} (r,w) = a_{\phi}^2 + \frac{\veps(4-N)}{2\kappa_1} H(r,w)\:.
  \label{eq:bulk2pt-bulk-exp-symm}
\end{align}
The decomposition in bulk channel blocks is more difficult, since they are not known in closed form.
One should also keep in mind that the correlator gets multiplied by a factor of $\xi^{\Delta_{\phi}}$ coming from the prefactor.
However, as pointed out, Eq.~\eqref{eq:phiphi-corr} has a similar form to the correlator $\langle \langle \phi^a \phi^b \rangle \rangle$ computed for the $\mathrm{O}(N)$ model in \cite{Gimenez-Grau:2022ebb,Bianchi:2022sbz}, and we can reuse known results.
In particular, they found an expression for $H(r,w)$ in terms of bulk blocks as well:
\begin{equation}\label{eq:Hbulk-exp}
 \xi H(r,w) = (\partial_{\Delta} - 1 - \log 2) f^{0}_{2,0}(r,w) + \Om(\eps)\:,
\end{equation}
where $f^{\Delta_{12}}_{\Delta,\ell}(r,w)$ are the bulk channel conformal blocks, which are known as a double sum \cite{Billo:2016cpy,Isachenkov:2018pef}.

From Eq.~\eqref{eq:Hbulk-exp} we see that $H(r,w)$ only corrects $\phi^2$, hence, for the CFT data of the other operators we can directly use the results from \cite{Gimenez-Grau:2022ebb,Bianchi:2022sbz}.
The other terms in Eq.~\eqref{eq:bulk2pt-bulk-exp-symm}, after multiplication with $\xi^{\Delta_{\phi}}$, are a constant term that corresponds to the bulk identity $f^{0}_{0,0} (r,w)$, and a term proportional to $\xi^{\Delta_{\phi}}$, whose expansion in bulk blocks is given in equation (167) of \cite{Gimenez-Grau:2022ebb}.
Putting eveything together, the expansion of $\Fm_{S,T}$ in blocks can be written as follows:
\begin{align}
  \xi^{\Delta_{\phi}} \Fm_{S} (r,w) = \, &1 + \lambda_{\phi \phi \phi^2} \, a_{\phi^2} f_{\Delta_{\phi^2},0} + \sum_{\ell = 2, 4, \ldots}^{\infty} 2^{-\ell} \, \lambda_{\phi \phi \Om^{S}_{\ell,0}} a_{\Om^{S}_{\ell,0}} f^{0}_{2 \Delta_{\phi} + \ell,\ell}\notag \\
  &+ \sum_{\ell = 0,2, \ldots}^{\infty} 2^{-\ell} \,  \overline{\lambda_{\phi \phi \Om^{S}_{\ell,1}} a_{\Om^{S}_{\ell,1}}} f^{0}_{2 \Delta_{\phi} + \ell + 2,\ell} + \Om(\veps^2)\:,  \\
  \xi^{\Delta_{\phi}} \Fm_{T} (r,w) = \, &\lambda_{\phi \phi T} \, a_{T} f_{\Delta_{T},0} + \sum_{\ell = 2, 4, \ldots}^{\infty} 2^{-\ell}\,  \lambda_{\phi \phi \Om^{T}_{\ell,0}} a_{\Om^{T}_{\ell,0}} f^{0}_{2 \Delta_{\phi} + \ell,\ell} \notag \\
  &+ \sum_{\ell = 0,2, \ldots}^{\infty} 2^{-\ell} \, \overline{\lambda_{\phi \phi \Om^{T}_{\ell,1}} a_{\Om^{T}_{\ell,1}}} f^{0}_{2 \Delta_{\phi} + \ell + 2,\ell} + \Om(\veps^2)\:,
  \label{eq:bulk2pt-bulk-block-exp}
\end{align}
where the bar indicates an average over CFT data since mixing needs to be solved before one is able to extract the individual OPE and one-point coefficients.
We can now extract the CFT data of all operators except $\phi^2$ and $T$.
For the $\mathrm{O}(N)$ singlets, using Eq.~\eqref{eq:bulkn0-CFT-data}, we extract the following one-point functions of twist-two operators:
\begin{align}
  a_{\Om^{S}_{\ell,0}} =\ & \frac{(4-N) (N+8) \Gamma \left(\frac{\ell+1}{2}\right)^2 \sqrt{\frac{2^{1-\ell} \Gamma (\ell+1)}{N \Gamma \left(\ell+\frac{1}{2}\right)}}}{\pi^{\frac{3}{4}} \kappa_2 \, \ell^2 \Gamma\left(\frac{\ell}{2}\right)^2} \Bigg\{ 1 + \veps \Bigg(-\frac{2 \kappa_{2} a_{\phi}^{(1)}}{(N-4) (N+8)} \notag \\
  &+ \frac{(N-4)}{2 \kappa_{1}} \left( 2 H_{\frac{\ell-1}{2}} + H_{2 \ell} - 2 H_\ell -  H_{\ell-\frac{1}{2}} + 2 \log 2\right)  \Bigg) + \Om(\veps^2)\Bigg\} \:,
\end{align}
where $a_{\phi}^{(1)}$ is the $\Om(\veps)$ correction to $a_{\phi}$, which can be found in the attached \textsc{Mathematica} notebook.
The averaged CFT data for the higher-twist $n=1$ operators is given by
\begin{align}
  \overline{\lambda_{\phi \phi \Om^{S}_{\ell,1}} a_{\Om^{S}_{\ell,1}}} = \veps\frac{(\ell+1)^2 (4-N) (N+8)  \Gamma \left(\frac{\ell+1}{2}\right)^3}{64 \pi  \kappa_2 N \Gamma \left(\frac{\ell}{2}+2\right) \Gamma \left(\ell+\frac{3}{2}\right)} + \Om(\veps^2)\:.
\end{align}
For the traceless symmetric operators, we find the following conformal dimensions and OPE coefficients:

\begin{align}
  a_{\Om^{T}_{\ell,0}} =\ & \frac{(4 - N) (N+8) \Gamma \left(\frac{\ell+1}{2}\right)^2 \sqrt{\frac{2^{1-\ell} \Gamma (\ell+1)}{\Gamma \left(\ell+\frac{1}{2}\right)}}}{\pi^{\frac{3}{4}} \kappa_{2} \, \ell^2 \, \Gamma \left(\frac{\ell}{2}\right)^2} \Bigg\{1+ \veps \Bigg( - \frac{2 \kappa_{2} a_{\phi}^{(1)} }{(N-4) (N+8)} \notag \\
  &+ \frac{(N-4) \left(2 H_{\frac{\ell-1}{2}}-2 H_\ell+H_{2 \ell}-H_{\ell-\frac{1}{2}}+\log (4)\right)}{2 \kappa_{1}} \Bigg) + \Om(\veps^2) \Bigg\}\: \\
  \overline{\lambda_{\phi \phi \Om^{T}_{\ell,1}} a_{\Om^{T}_{\ell,1}}} =\ & \veps\frac{(\ell+1)^2 (N+8)  \Gamma \left(\frac{\ell+1}{2}\right)^3}{128 \pi  \Gamma \left(\frac{\ell}{2}+2\right) \Gamma \left(\ell+\frac{3}{2}\right)} + \Om(\veps^2)\:.
\end{align}
To compare the expansion in Eq.~\eqref{eq:bulk2pt-bulk-block-exp} with the one-point functions of $\phi^2$ and $T_{ab}$ computed in Section~\ref{sec:one-pt}, we need to know the bulk OPE coefficients $\lambda_{\phi \phi \phi^2}$ and $\lambda_{\phi \phi T}$. These can be easily calculated and are given by 
\begin{align}
\lambda_{\phi \phi \phi^2} 
&= \delta^{ab} \left(\sqrt{\frac{2}{N}}-\frac{\veps \kappa_2 (N+2)  }{2 \sqrt{2 N} \kappa_1 (N+8)} + \Om(\veps^2) \right)\,, \\
\lambda_{\phi \phi T} &= \sqrt{2}-\frac{\veps \kappa_2}{\sqrt{2} \kappa_1 (N+8)} + \Om(\veps^2)\,.
\end{align}
With these OPE coefficients and the one-point functions in Eq.~\eqref{eq:one-pt-phi2T}, we can check the block expansion in Eq.~\eqref{eq:bulk2pt-bulk-block-exp} and see that it indeed reproduces the desired conformal data of $\phi^2$ and $T^{ab}$.                        

\subsection{Towards two-point functions of bulk fermions}

We conclude this section by commenting on how to generalize the two-point function analysis when the external operators are fermions. 
This is an interesting problem as the $\veps$-expansion was originally designed to capture physics in three dimensions, however four-dimensional fermions are very different objects compared to three-dimensional fermions. 
In order to understand how fermionic correlators in $d=3$ are encoded in the $\veps$-expansion, we can start by calculating them in perturbation theory.
We do not bring this calculation to completion in this section, as the diagrams involved are a lot more challenging than the ones we have studied so far. 
Nevertheless we sketch out the computation, and in the conclusion we discuss in more detail possible future directions. 
Just as for the one-point function of the fermion bilinear, here we will also restrict ourselves to the GNY and NJLY models.

The 
disconnected part of the correlator corresponds to the wavefunction renormalization of the bulk fermion, and the diagrams are the first and third ones shown in \ref{fig:2pt-bulk-fer-diags}.
The Feynman diagrams contributing to the
connected part of the
fermionic two-point function $\vvev{\Psib (x_1) \Psi (x_2)}$ up to $\Om(\veps)$ are given in figure~\ref{fig:2pt-bulk-fer-diags} (second and fourth diagrams).

\begin{figure}
 \begin{subfigure}{\textwidth}
 \centering
 \TwoPointBulkFermionsOne
 \TwoPointBulkFermionsTwo
 \TwoPointBulkFermionsThree
 \TwoPointBulkFermionsFour
 \end{subfigure}
 \caption{Contributions to the two-point function $\vvev{\Psib (x_1) \Psi (x_2)}$ up to $\Om(\veps)$. The defect is denoted by a solid line, scalars by a dotted line, and fermions by solid arrowed lines. 
 Bulk Yukawa couplings $g_0$ are represented by a red dot and defect couplings $h_0$ by a blue dot. The first and third diagrams correspond to the \textit{disconnected} part of the correlator, while the second and fourth are \textit{connected} and are the diagrams which make the correlator different to a defectless two-point function.}
 \label{fig:2pt-bulk-fer-diags}
\end{figure}

The $Y$-diagram in figure~\ref{fig:2pt-bulk-fer-diags} is the first connected diagram at $\Om(g) \sim \Om(\sqrt{\veps})$.
Setting for convenience $\tau_1 = \tau_2 = 0$, it is given by
\begin{align}
\TwoPointBulkFermionsTwo \,  &= \NormPsi^2 g_0 h_0 \bar{s}_1 \int d\tau_3 \int d^4 x_4\, \slashed{\partial}_1 I_{14}\, \Sigma^1\, \slashed{\partial}_2 I_{24} I_{34}\, s_2 \notag \\
&= - \frac{\pi\, g_0 h_0}{8 (|x_1^\perp| + |x_2^\perp|)} \bar{s}_1 \left(
\frac{\slashed{x}_1 \slashed{x}_2}{|x_1^\perp| |x_2^\perp|} + \mathds{1}
\right) s_2\,,
\end{align}
where we have used the fermionic star-triangle identity given in \eqref{eq:id-StarTriangle}. The remaining one-dimensional integral is trivial to compute.

At $\Op(\veps)$ we have one $H$-diagram that connects scalars insertions on the line to the external fermions through two Yukawa vertices.
This diagram contains a challenging ten-dimensional (finite) integral that we only solve partially for now.
We provide however a solution for the $4d$ bulk integral, i.e., before performing the $\tau_3\,, \tau_4$ integrals.
After Wick contractions the diagram gives
\begin{equation}
\TwoPointBulkFermionsFour  = - \NormPsi^2\, g_{0}^2 h_{0}^2\, \bar{s}_1 \int d\tau_3 \int d\tau_4 \int d^4 x_5 \int d^4 x_6\, \slashed{\partial}_1 I_{15} \Sigma^1 \slashed{\partial}_5 I_{56} \Sigma^1 \slashed{\partial}_6 I_{26} I_{35} I_{46}\, s_2\,.
\end{equation}
Using the fact that $\Sigma^1 = \mathds{1}$, and that one four-dimensional integral can be lifted by using the fermionic star-triangle identity given in \eqref{eq:id-StarTriangle}, we then have
\begin{equation}
\TwoPointBulkFermionsFour  = \NormPsi^2 \pi^2\, g_{0}^2 h_{0}^2\, \bar{s}_1\,  \slashed{\partial}_1 \int d\tau_3 \int d\tau_4\, I_{24} \left( \int d^4 x_5\, \slashed{x}_{54}\, I_{15} I_{25} I_{35} I_{45} \right) \slashed{x}_{24}\, s_2\,.
\end{equation}
The tensor integral between the brackets can be solved by applying tensor decomposition. There exists many automated tools to perform this step, and here we use the package \textsc{X} \cite{Patel:2016fam}. We find
\begin{equation}
J_{123;4} := \int d^4 x_5\, \slashed{x}_{54}\, I_{15} I_{25} I_{35} I_{45} = \frac{2}{\phi_K} j_{123;4}\,,
\end{equation}
with $\phi_K$ the Kibble function defined as
\begin{alignat}{2}
&\phi_K :=&& \Phi_{1234} + \Phi_{1324} + \Phi_{1423} + \Psi_{123} + \Psi_{124} + \Psi_{134} + \Psi_{234}\,, \\
&\Phi_{1234} =&& - \frac{1}{64 \pi^6 I_{12} I_{34}} \left( \frac{1}{I_{12}} + \frac{1}{I_{34}} - \frac{1}{I_{13}} - \frac{1}{I_{14}} - \frac{1}{I_{23}} - \frac{1}{I_{24}} \right)\,, \\
&\Psi_{123} =&& - \frac{1}{64 \pi^6 I_{12} I_{13} I_{23}}\,,
\end{alignat}
and with
\begin{equation}
j_{123;4} := \slashed{f}_{1234}\, X_{1234} + \slashed{g}_{123;4} Y_{123} + \slashed{g}_{124;3} Y_{124} + \slashed{g}_{134;2} Y_{134} + \slashed{g}_{234;1} Y_{234}\,.
\end{equation}
The $X$- and $Y$-integrals are defined in \eqref{eq:BasicDefs} and solved in \eqref{eq:X-sol} and \eqref{eq:Y-sol}.  The prefactor function $\slashed{f}_{1234}$ can be expressed in terms of propagators and read
\begin{equation}
\slashed{f}_{1234} = a_{1234}\, \slashed{x}_1 + a_{2341}\, \slashed{x}_2 + a_{3412}\, \slashed{x}_3 + (a_{4123}\, - 1) \slashed{x}_4\,,
\end{equation}
with
\begin{align}
a_{1234} := - \frac{1}{I_{23} I_{24} I_{34}} & \left( 2
+ \frac{I_{24} I_{34}}{I_{14} I_{23}}
+ \frac{I_{23} I_{34}}{I_{13} I_{24}}
+ \frac{I_{23} I_{24}}{I_{12} I_{34}} \right. \notag \\
& \left. - I_{34} \left( \frac{1}{I_{13}} + \frac{1}{I_{14}} \right)
- I_{24} \left( \frac{1}{I_{12}} + \frac{1}{I_{14}} \right)
- I_{23} \left( \frac{1}{I_{12}} + \frac{1}{I_{13}} \right)
 \right)\,.
\end{align}
The function $\slashed{g}_{123;4}$ can also be expressed in an elegant way as
\begin{equation}
\slashed{g}_{123;4} := b_{123;4} \slashed{x}_1
+ b_{231;4} \slashed{x}_2
+ b_{312;4} \slashed{x}_3
+ c_{123} \slashed{x}_4\,,
\end{equation}
with
\begin{align}
b_{123;4} &:= \frac{1}{I_{23}} \left( \frac{1}{I_{12}}
+ \frac{1}{I_{13}}
- \frac{1}{I_{23}}
+ \frac{1}{I_{24}}
+ \frac{1}{I_{34}}
- \frac{2}{I_{14}} \right)
- \left( \frac{1}{I_{12}} - \frac{1}{I_{13}} \right) \left( \frac{1}{I_{24}} - \frac{1}{I_{23}} \right)\,, \\
c_{123} &:= \frac{1}{I_{12}^2} + \frac{1}{I_{13}^2} + \frac{1}{I_{23}^2}
-2 \left( \frac{1}{I_{12} I_{13}} + \frac{1}{I_{12} I_{23}} + \frac{1}{I_{13} I_{23}} \right)\,.
\end{align}
This is as far as we can go for now and we are left with a difficult two-dimensional integral as well as a slashed derivative with respect to $x_1$. We note however that this integral can efficiently be computed numerically. 

There is another path that one can take in order to try and solve this integral. Instead of computing the bulk integrals, one could start with the \textit{defect} integral over $\tau_3\,, \tau_4$ and use, e.g., Schwinger parametrization for computing the remaining eight-dimensional integral. This approach was indeed useful for the computation of the $X$-diagram in the scalar two-point case, however here it is not clear at present how these $8$ integrals could be solved efficiently.

\section{Conclusions}
\label{sec:conclusion}

In this work we studied defect correlators for line defects in fermionic models using the $\veps$-expansion. Our setup is a natural generalization of line defects previously considered in $\mathrm{O}(N)$ models. Indeed, the definition of the defect as the integral of a scalar along a line is identical to the magnetic line defect studied in \cite{Cuomo:2021kfm,Gimenez-Grau:2022czc}. The main difference is the presence of fermions in the bulk, which induce new fermionic excitations on the $1d$ defect.

We calculated a host of $1d$ correlators, putting special emphasis on the new fermionic excitations.  Closed-form expressions for four-point functions on the line were obtained in terms of the unique $1d$ cross-ratio. These correlators can be used to easily extract CFT data by means of a conformal block expansion, and can also be used as input in the numerical bootstrap. The numerical bootstrap for magnetic line defects was initiated in \cite{Gimenez-Grau:2022czc}. The numerical bootstrap plots should accommodate the models studied in this paper, where the numbers of fermions $N_f$ is a free parameter. One can also use the data calculated here to steer the numerics, and hopefully solve particular models of interest. In Table \ref{tab:nums} it is shown how the presence of fermions affects the CFT data for a low number of fermions, and for $N=3$, which is the particular case considered in \cite{Gimenez-Grau:2022czc}. One remarkable observation is that the scaling dimension of the lowest-lying scalar $\hat\phi^1$ decreases for higher values of $N_f$. Because of this, the magnetic line defect in fermionic models seems to explain the numerics of \cite{Gimenez-Grau:2022czc} better than the same line defect in the $\mathrm{O}(3)$ model. It would be interesting to see if the inclusion of fermionic correlators to the numerics will improve the bounds.

\begin{table}
\centering
 \begin{tabular}{|c|c|c|c|c|}
 \hline
  & $N_f = 0$ & $N_f = 1$ & $N_f = 2$ & $N_f \to \infty$\\
  \hline
  \hline
  $\Delta_t$ & 1 & 1 & 1 & 1\\
  $\Delta_{\hat{\phi}^1}$ & $1 + \veps$ & $1 + \frac{2 \veps}{3}$ & $1+\frac{\veps}{5}$ &1\\
  $\Delta_{s_{-}}$ & $2 + 0.35502$ & $2 + 0.78832$ & $2 + 0.75055 $ &2\\
  $\Delta_{T}$ & $2 + 0.\bar{18} \veps$ & $2 + 0.433 \veps$ & $2 + 0.29\bar{09} \veps$ &2\\
  $\Delta_A$ & 3 & 3 & 3 & 3\\
  $\Delta_V$ & $2 + 1.\bar{18}\veps$ & $2 + 1.100 \veps$ & $2 + 0.49\bar{09} \veps$ &2\\
  \hline
  \hline
  $\lambda_{tt \hat\phi^1}$& $0.947226 \veps$ & $2.1893 \veps$ & $1.6075 \veps$ &0\\
   $\lambda_{\hat\phi^1 \hat\phi^1 \hat\phi^1}$ & $2.84168 \veps$ & $6.56789 \veps$ & $4.82249 \veps$ &0 \\
  \hline
 \end{tabular}
 \caption{Values of the conformal dimensions and OPE coefficients of the lowest-lying operators in the $\hat\phi^1 \times \hat\phi^1, t^{\hat{a}} \times t^{\hat{b}}$ and $\hat\phi^1 \times t^{\hat{a}}$ OPEs at first order in $\veps$, see ~\eqref{eq:vdef} to~\eqref{eq:adef}. The values are given for a pinning line defect in the $\mathrm{O}(3)$ model without fermions, and a pinning line defect in the chiral Heisenberg model with $N_f = 1,2,\infty$.}
 \label{tab:nums}
\end{table}

In addition to correlators constrained to the line, we also studied how excitations in the bulk are modified by the presence of the defect. We focused on two-point functions, which have non-trivial kinematics and depend on two conformal invariants. For $\mathrm{O}(N)$ models this analysis had been done recently in \cite{Gimenez-Grau:2022ebb,Bianchi:2022sbz}. Due to the similarity of the Feynman diagram calculation, we could recycle several of their results, in particular the non-trivial integral presented in \eqref{eq:H_integral}.

Having understood two-point functions of scalars, the next step is to study two-point functions of fermions. This analysis comes with several conceptual and technical challenges. The question of how to analytically continue fermions across dimensions has not been studied systematically, and a naive counting of tensor structures already shows disagreement between three and four dimensions. This problem opens several avenues for future research. On the one hand there is the explicit perturbative calculation, which we sketched at the end of section \ref{sec:bulkcorr}. Regardless if one knows how to interpolate fermions across dimensions, the Yukawa models considered here are well-defined pertubartive CFTs, and correlators involving fermions exist and can be calculated. On the other hand, one can also investigate the kinematics of fermion correlators at a more fundamental level, understanding for example the structure of conformal blocks and how they depend on the spacetime dimension $d$. We should point out that, even though in this paper we focused on a line defect, the questions raised above are relevant for standard CFTs without defects. For example, the following \textit{bulk} four-point function $\langle \bar\Psi \Psi \phi \phi \rangle$ already exhibits interesting non-trivial behavior across dimensions \cite{Buric:202X}. To our knowledge this type of correlator has never been studied using the $\veps$-expansion, and would form an excellent starting point for the study of fermions across dimensions.

%%%%%%%%%%%%%%%%%%%%%%%%%%%%%%%%%%%%%
\section*{Acknowledgements }
%%%%%%%%%%%%%%%%%%%%%%%%%%%%%%%%%%%%%

We thank A.~Antunes, I.~Buric, A.~Gimenez-Grau, E.~Lauria, J.~Plefka, V.~Schomerus, and A.~Stergiou for discussions and comments. 
PL and PvV acknowledge support from the DFG through the Emmy Noether research group `The Conformal Bootstrap Program' project number 400570283, and through the German-Israeli Project Cooperation (DIP) grant `Holography and the Swampland'.
JB is funded by the Deutsche Forschungsgemeinschaft (DFG, German Research Foundation) – Projektnummer 417533893/GRK2575 “Rethinking Quantum Field Theory”.

%%%%%%%%%%%%%%%%%%%%%%%%%%%%%%%%%%%%%
\appendix
\section{Spinor conventions}\label{app:spinors}

In this appendix we describe our conventions for the spinor fields for the cases $N=1,2$, which can be straightforwardly generalized to the case $N=3$.
In the action given in \eqref{eq:GNY-lag}, the fermions $\Psi$ are presented as vectors of Dirac fields $\psi^i$ ($i$ is the flavor index), which can be decomposed into two basic Weyl spinors as follows:
\begin{equation}
\psi^A = \binom{\chi_\alpha}{\xi^{\dagger \dot{\alpha}}}\,, \quad \bar{\psi}^A = \left( \xi^\alpha\ \chi^{\dagger}_{\dot{\alpha}} \right)\,,
\label{eq:defchixi}
\end{equation}
with $A=1,2,3,4$, and $\alpha, \dot{\alpha} = 1,2$. The Weyl spinors are two-component vectors defined as
\begin{equation}
\chi = \binom{\chi_1}{\chi_2}\,, \quad \xi^\dagger = \left( \xi_1\ \xi_2 \right)\,.
\end{equation}
Spinors with an undotted index $\alpha$ transform as left-handed spinors $(1,0)$, while right-handed spinors $(0,1)$ are complex conjugates of the $(1,0)$ representation and carry a dotted index $\dot{\alpha}$. The dot is here to indicate the transformation property, i.e.,
\begin{equation}
\chi^\dagger_{\smash{\dot{\alpha}}} = (\chi_\alpha)^\dagger\,.
\end{equation}
Indices can be raised and lowered in the following way:
\begin{equation}
\chi^\alpha = \eps^{\alpha\beta} \chi_\beta = - \eps^{\beta\alpha} \chi_\beta\,,
\end{equation}
which implies
\begin{equation}
\chi^\alpha \xi_\alpha = - \chi_\alpha \xi^\alpha\,.
\end{equation}
Here the tensor $\eps^{\alpha\beta}$ is defined as
\begin{equation}
\eps^{12} = - \eps^{21} = \eps_{21} = - \eps_{12} = +1\,,
\end{equation}
and a similar definition can be formulated for dotted indices:
\begin{equation}
\eps_{\smash{\dot{\alpha} \dot{\beta}}} = \eps_{\alpha\beta}\,, \quad \eps^{\smash{\dot{\alpha} \dot{\beta}}} = \eps^{\alpha\beta}\,.
\end{equation}

For external operators it is convenient to use polarization spinors $s^A$, $\bar{s}^{A}$ in order to avoid cluttering of the indices.
We define
\begin{equation}
\psi^i (s,\tau) := s^A \psi^{i, A} (\tau)\,, \quad \bar{\psi}^i (s,\tau) := \bar{s}^A \bar{\psi}^{i, A} (\tau)\,,
\label{eq:def-PolarizationSpinors}
\end{equation}
and a similar definition holds for the Weyl fermions as well.

The four-dimensional (Euclidean) $\gamma$-matrices are defined in the chiral representation as
\begin{equation}
(\gamma^{\mu})^{AB} := \begin{pmatrix}
0 & (\sigma^\mu)_{\alpha \dot{\beta}} \\
(\bar{\sigma}^\mu)^{\dot{\alpha} \beta} & 0
\end{pmatrix}\,,
\end{equation}
where we have introduced
\begin{equation}
(\sigma^\mu)_{\alpha\dot{\beta}} := \left( \sigma^0\,, i \sigma^i \right)\,, \quad (\bar{\sigma}^\mu)^{\dot{\alpha}\beta} := \left( \sigma^0\,, - i \sigma^i \right)\,.
\end{equation}
The Pauli matrices $\sigma^0\,, \sigma^i$ are defined as
\begin{equation}
\sigma^0 = \mathds{1}_2\,, \quad
\sigma^1 =  \renewcommand{\arraystretch}{.85}\begin{pmatrix} 0 & 1 \\ 1 & 0 \end{pmatrix}\,, \quad
\sigma^2 = \begin{pmatrix} 0 & -i \\ i & 0 \end{pmatrix}\,, \quad
\sigma^3 = \begin{pmatrix} 1 & 0 \\ 0 & -1 \end{pmatrix}\,.
\end{equation}
The $\gamma$-matrices satisfy the Euclidean Clifford algebra
\begin{equation}
\lbrace \gamma^\mu, \gamma^\nu \rbrace^{AB} = 2 \delta^{\mu\nu} \mathds{1}^{AB}\,,
\end{equation}
and we can define an additional $\gamma$-matrix as
\begin{equation}
(\gamma^5)^{AB} := \begin{pmatrix}
\mathds{1}_\alpha^{\phantom{\alpha}\beta} & 0 \\
0 & - \mathds{1}^{\smash{\dot{\alpha}}}_{\phantom{\alpha}\smash{\dot{\beta}}}
\end{pmatrix}\,.
\label{eq:gamma5}
\end{equation}
This definition ensures that $\gamma^5$ satisfies the following properties:
\begin{equation}
\lbrace \gamma^5\,, \gamma^\mu \rbrace = 0\,, \quad (\gamma^5)^\dagger = \gamma^5\,, \quad (\gamma^5)^2 = \mathds{1}\,.
\end{equation}
As mentioned above, it is easy to generalize these conventions to higher-dimensional $\gamma$-matrices, keeping the representation arbitrary and relying on \eqref{eq:Sigma-identity1} and \eqref{eq:Sigma-identity2}.
\section{\texorpdfstring{More details on the $\beta-$functions}{More details on the beta functions}}\label{app:betafcn}

The general $\beta-$function for the bulk coupling constants $\lambda^{abcd},\Sigma^a$, as well as the anomalous dimensions, are given in the appendix of \cite{Fei:2016sgs} up to $\Om(\veps^2)$.
The renormalization constants up to $\Om(\veps^2)$ are:
\begin{align}
  Z_\lambda = 1 &+ \frac{1}{(4\pi)^2 \veps}\Bigg(\frac{(N + 8)\lambda}{3}  + 8 N_f g^2 -\frac{4 (N+8) N_f g^2 \lambda}{6 (4\pi)^2}- \frac{12 N_f g^4}{\lambda}+\frac{12 N_f g^6}{\pi^2 \lambda}\notag \\
  &+\frac{4 (12 - 5 N) N_f g^4}{2 (4 \pi)^2}-\frac{(14 + 3N)\lambda^2}{6 (4\pi)^2} \Bigg) \notag \\
  &+ \frac{1}{(4\pi)^4 \veps^2}\Bigg(-96 N_f \left(4 + 4 N_f - N\right) \frac{g^6}{\lambda}+ 4(N+8)N_f g^2 \lambda \notag \\
  &+ 12 N_f \left(4 N_f - 2 (N + 4) \right) g^4 +\frac{(N+8)^2}{9} \lambda^2\Bigg) + \Om(\lambda^3, g^6, \lambda^2 g^2, \lambda g^4)
  \:, \\
  Z_{g} = 1 &+ \frac{1}{(4\pi)^2 \veps}\left(\kappa_1 g^2-\frac{N+2}{3 (4\pi)^2}g^2 \lambda-\frac{9N^2  - 40 N - 32 + 24 \kappa_1}{8 (4\pi)^2}g^4+\frac{N+ 2}{72 (4\pi)^2}\lambda^2\right) \notag \\
  &+ \frac{1}{(4\pi)^4 \veps^2}\left(\frac{(N+2)(5 \kappa_1 - 32)}{72 (4\pi)^2}g^2 \lambda^2 +\frac{3 \kappa_{1}^2}{2} g^4\right) \notag \\
  &+ \Om(\lambda^3, g^6, \lambda^2 g^2, \lambda g^4) 
  \:,  \\
  Z_{\phi} = 1 &+ \frac{1}{(4\pi)^2 \veps }\left(-2 N_f g^2 + \frac{4 (N+4) N_f }{8 (4 \pi)^2} g^4-\frac{N+2}{72 (4\pi)^2 }\lambda^2 \right)+\frac{1}{(4\pi)^4 \veps ^2} \left(\frac{(N-4)^2 - \kappa_{1}^2}{2} g^4\right) \notag \\
  &+ \Om(\lambda^3, g^6, \lambda^2 g^2, \lambda g^4)
  \:,  \\
  Z_{\Psi} = 1 &+\frac{1}{(4\pi)^2 \veps}\left(- \frac{N}{2}g^2+ \frac{N (7 N + 6(\kappa_1 - 4))}{16 (4 \pi )^2} g^4\right) - \frac{1}{(4\pi)^4 \veps^2}\left(\frac{N(N - 4 \kappa_1) }{8} g^4\right) \notag \\
  &+ \Om(\lambda^3, g^6, \lambda^2 g^2, \lambda g^4)
  \:.
  \label{eq:renorm_bulk}
\end{align}

The $\beta-$function for the defect coupling was computed up to $\Om(\veps^2)$ in \cite{Pannell:2023pwz}.
The corresponding renormalization factor is given by
\begin{align}
 Z_{h} &= 1 +\frac{1}{(4\pi)^2 \veps} \Bigg\{\frac{\lambda h^2}{12}-\frac{g^2 \lambda h^2 N_f}{3 (4\pi)^2}-\frac{\left(\pi ^2-6\right) g^4 h^2 2 N_f}{9 (4\pi)^2} \notag \\
 &-\frac{g^4 (N+4) N_f}{4 (4\pi)^2}+g^2 2 N_f +\lambda^2 \left(-\frac{h^2 (N+8)}{108 (4\pi )^2}-\frac{h^4}{48 (4\pi)^2}+\frac{N+2}{72 (4\pi)^2}\right)\Bigg\}\notag \\
 &+ \frac{1}{(4\pi)^4\veps^2} \Bigg\{ \frac{g^2 \lambda h^2 N_f}{2} -\frac{g^4 h^2 4 N_f}{3} +g^4 \left(6 N_f+ 8 -2 NN\right)+\lambda^2 \left(\frac{h^2 (N+8)}{108}+\frac{h^4}{96}\right)\Bigg\} \notag\\
 &+ \Om(\lambda^3, g^6, \lambda^2 g^2, \lambda g^4)
  \:.
\end{align}
Let us look at each model individually.

\subsection{\texorpdfstring{Gross--Neveu--Yukawa model ($N = 1$)}{Gross--Neveu--Yukawa model (N = 1)}}

We start with considering the GNY model, which contains a single scalar field $\phi$ and $N_f$ fermions. Hence, the matrix $\Sigma^a = \Sigma^1$ in Eq.~\eqref{eq:GNY-lag}, which corresponds to $\Gamma_i$ in \cite{Fei:2016sgs}, is given by the identity matrix:
\begin{equation}
 \Sigma = \mathds{1}_{N_f} \mathds{1}_4\:.
\end{equation}
The $\beta-$functions up to $\Om(\veps^2)$ were computed in \cite{Karkkainen:1993ef} and we adopted the same convention as \cite{Fei:2016sgs}:
\begin{align}
  \beta^{\text{GNY}}_{\lambda} &= -\veps \lambda +\frac{1}{(4 \pi )^2} \left(8 g^2 \lambda N_f-48 g^4 N_f+3 \lambda^2\right) \notag \\
  &-\frac{1}{(4 \pi )^4}\left(-12 g^2 \lambda^2 N_f+28 g^4 \lambda N_f+384 g^6 N_f-\frac{17 \lambda^3}{3}\right)+ \Om(\lambda^3, g^6, \lambda^2 g^2, \lambda g^4)\:,\\
  \beta^{\text{GNY}}_{g} &= -g \frac{\veps }{2}+\frac{1}{(4 \pi )^2}\left(\frac{g^3 (4 N_f+6)}{2}\right) + \frac{1}{(4 \pi )^4}\left(-2 g^3 \lambda-\frac{3}{4} g^5 (16 N_f+3)+\frac{g \lambda^2}{12}\right) + \Om(g^6)\:.
\end{align}
The Wilson-Fisher-Yukawa (WFY) fixed point can be reached for the following values of the couplings at one loop in $\veps:=4-d$:
\begin{align}
 (g^{\text{GNY}}_\star)^{2} &= (4 \pi)^2 \left(\frac{\veps}{2 \kappa_{1}} + \frac{\veps^2 (2 \kappa_{1} (\kappa_{2}+288)+15 (4 \kappa_{2}-99))}{432 \kappa_{1}^3} + \Om(\veps^3)\right) \,,  \\
\lambda^{\text{GNY}}_{\star} &= (4 \pi )^2 \Bigg\{\frac{\veps \kappa_{2}}{6 \kappa_{1}} + \frac{\veps^2}{216 \kappa_{1}^3 (\kappa_{1}+\kappa_{2}-6)} \Bigg(-12 \kappa_{1}^3 (\kappa_{2}-36) \notag \\
&+2 \kappa_{1}^2 (97 \kappa_{2}+72)+15 \kappa_{1} (47 \kappa_{2}-1188)-2925 \kappa_{2}+40500\Bigg) + \Om(\veps^3) \Bigg\} \:.
\end{align}
The $\beta-$function for $h \equiv h_1$ is given by 
\begin{align}
  \beta^{\text{GNY}}_h &= - \frac{h \veps}{2} + \frac{1}{(4\pi)^2} \left(2 g^2 h N_f+\frac{\lambda h^3}{6}\right) \notag \\
  &+ \frac{1}{(4\pi)^4} \left(4 g^4 h^3 N_f - g^2 \lambda h^3 N_f -\frac{2}{3} \pi ^2 g^4 h^3 N_f-\frac{5}{2} g^4 h N_f-\frac{\lambda^2 h^5}{12} -\frac{\lambda^2 h^3}{4}+\frac{\lambda^2 h}{12}\right) \notag  \\
  &+ \Om(\lambda^3, g^6, \lambda^2 g^2, \lambda g^4)\:,
\end{align}
leading to the following fixed point at $\Om(\veps)$:
\begin{align}
  (h^{\text{GNY}}_{\star})^{2} &= \frac{54}{\kappa_{2}}+\frac{\veps}{2 \kappa_{1}^2 \kappa_{2}^2 (4 \kappa_{1} (\kappa_{2}-54)+(\kappa_{2}-24) \kappa_{2}+648)}\Bigg\{(\kappa_{1} (149 \kappa_{1}-816)+1125) \kappa_{2}^3 \notag \\
  &+3 \left(\kappa_{1} \left(-216 \pi ^2 (\kappa_{1}-3)+\kappa_{1} (105 \kappa_{1}-302)+1545\right)-4500\right) \kappa_{2}^2 \notag \\
  &-216 \left(\kappa_{1} \left(3 \pi ^2 (\kappa_{1}-6) (\kappa_{1}-3)+\kappa_{1} (69 \kappa_{1}-112)+171\right)-1125\right) \kappa_{2} \notag \\
  &+69984 (\kappa_{1}-6) (\kappa_{1}-3) \kappa_{1}\Bigg\} + \Om(\veps^2)\:.
\end{align}

\subsection{\texorpdfstring{Nambu--Jona-Lasinio--Yukawa model ($N = 2$)}{Nambu--Jona-Lasinio--Yukawa model (N = 2)}}

When we extend the number of scalars to a real scalar and a real pseudoscalar, or one complex scalar, and keep the number of fermions arbitrary at $N_f$, we obtain the NJLY model.
The matrix $\Sigma^a = \Sigma^1, \Sigma^2$ is now given by: $\Sigma^1 = \mathds{1}_{N_f} \mathds{1}_4, \Sigma^2 = \mathds{1}_{N_f} i \gamma_5$. 
The $\beta$-functions for $\lambda$ and $g$  are given up to two loop orders by \cite{Fei:2016sgs}:
\begin{align}
 \beta^{\text{NJLY}}_{\lambda} &= - \veps \lambda + \frac{1}{(4\pi)^2} \left( \frac{10}{3} \lambda^2 + 8 N_f \lambda g^2 - 48 N_f g^4\right) \notag \\
 &- \frac{1}{(4\pi)^4} \left(\frac{20}{3} \lambda^2 - 384 N_f g^6 - 8 N_f \lambda g^4 + \frac{40}{3} N_f g^2 \lambda^2\right) \notag\\
 &+ \Om(\lambda^3, g^6, \lambda^2 g^2, \lambda g^4)\:, \\
 \beta^{\text{NJLY}}_g &=  - \frac{\veps}{2} g  + \frac{1}{(4 \pi)^2} \left( 2 N_f + 2 \right) + \frac{1}{(4\pi)^4} \left( - \frac{8}{3} g^3 \lambda + \frac{1}{9} g \lambda^2 + (7 - 12 N_f) g^5 \right) + \Om(g^7)\:.
\end{align}
The zeros of these $\beta-$functions give us the fixed points $\lambda_{\star}, g_{\star}$ of the NJLY model:
\begin{align}
  \frac{\lambda_{\star}^{\text{NJLY}}}{(4\pi)^2} =\ & \frac{3 \kappa_{2} \veps }{20 \kappa_{1}}-\frac{9 \veps^2 \left(3 \kappa_{1}^3 (\kappa_{2}-40)+\kappa_{1}^2 (160-75 \kappa_{2})+\kappa_{1} (5000-219 \kappa_{2})+674 \kappa_{2}-9680\right)}{500 \kappa_{1}^3 (\kappa_{1}+\kappa_{2}-4)} \notag \\
  &+ \Om(\veps^3)\:,\\
  \frac{(g^{\text{NJLY}}_{\star})^2}{(4\pi)^2} =\ & \frac{\veps}{2 \kappa_{1}} +  \frac{\veps^2 (\kappa_{1} (\kappa_{2}+260)+36 \kappa_{2}-870)}{200 \kappa_{1}^3} + \Om(\veps^3)\:.
\end{align}
The defect $\beta$-function is given by
\begin{align}
  \beta_{h}^{\text{NJLY}} =\ & - \frac{\veps h }{2} + \frac{1}{(4\pi)^2}\left(2 g^2 h N_f+\frac{\lambda h^3}{6}\right) \notag \\
  &+ \frac{1}{(4\pi)^4} \Bigg( 4 g^4 h^3 N_f-3 g^4 h N_f- g^2 \lambda h^3 N_f-\frac{2}{3} \pi ^2 g^4 h^3 N_f-\frac{\lambda^2 h^5}{12} -\frac{5 \lambda^2 h^3}{18}+\frac{\lambda^2 h}{9} \Bigg) \notag \\
  &+ \Om(\lambda^3, g^6, \lambda^2 g^2, \lambda g^4)\:,
\end{align}
which has a fixed point for $g_{\star}, \lambda_{\star}$ and
\begin{align}
  (h^{\text{NJLY}}_{\star})^{2} =\ & \frac{40}{\kappa_{2}}+\frac{\veps}{15 \kappa_{1}^2 \kappa_{2}^2 (4 \kappa_{1} (\kappa_{2}-60)+(\kappa_{2}-16) \kappa_{2}+480)}\Bigg\{6 (\kappa_{1} (187 \kappa_{1}-1096)+1452) \kappa_{2}^3 \notag\\
  &+2 \left(\kappa_{1} \left(-2000 \pi ^2 (\kappa_{1}-2)+3 \kappa_{1} (353 \kappa_{1}+1620)+3708\right)-34848\right) \kappa_{2}^2 \notag\\
  &-160 \left(\kappa_{1} \left(25 \pi ^2 (\kappa_{1}-4) (\kappa_{1}-2)+3 \kappa_{1} (179 \kappa_{1}-72)+2820\right)-8712\right) \kappa_{2} \notag \\
  &+288000 (\kappa_{1}-4) (\kappa_{1}-2) \kappa_{1}\Bigg\} + \Om(\veps^2)\:. 
\end{align}

\subsection{\texorpdfstring{Chiral Heisenberg model ($N = 3$)}{Chiral Heisenberg model (N = 3)}}

The last model we consider is the chiral Heisenberg model. It contains three real scalars and the model is invariant under $\mathrm{O}(3)$ rotations.
The matrix $\Sigma^a = \Sigma^1, \Sigma^2, \Sigma^3$ is given by the Pauli matrices $\sigma_i$:
\begin{equation}
 \Sigma^a = \sigma_a \otimes \mathds{1}_{2 N_f}\:.
\end{equation}
The $\beta-$functions for $\lambda$ and $g$ up to $\Om(\veps^2)$ were calculated in \cite{Rosenstein:1993zf}, together with various critical exponents.
They are given by
\begin{align}
 \beta^{\chi\text{H}}_{\lambda} =\ & - \veps \lambda + \frac{1}{(4\pi)^2} \left( 8 g^2 \lambda N_f-48 g^4 N_f+\frac{11 \lambda^2}{3}\right) \notag \\
 &- \frac{1}{(4\pi)^4} \left(-\frac{44}{3} g^2 \lambda^2 N_f- 12 g^4 \lambda N_f+384 g^6 N_f-\frac{23 \lambda^3}{3}\right) + \Om(\lambda^3, g^6, \lambda^2 g^2, \lambda g^4)\:,\\
 \beta^{\chi\text{H}}_g =\ &  - \frac{\veps}{2} g  + \frac{1}{(4 \pi)^2} \left( 2 g^3 N_f+g^3 \right) + \frac{1}{(4\pi)^4} \left( -\frac{10 g^3 \lambda}{3}-12 g^5 N_f+\frac{47 g^5}{4}+\frac{5 g \lambda^2}{36}\right) + \Om(g^7)\:.
\end{align}
The corresponding WFY fixed points are
\begin{align}
  \frac{\lambda^{\chi\text{H}}_{\star}}{(4\pi)^2} =\ &\frac{3 \kappa_{2} \veps }{22 \kappa_{1}} + \frac{\veps^2}{10648 \kappa_{1}^3 (\kappa_{1}+\kappa_{2}-2)}\Bigg(-564 \kappa_{1}^3 (\kappa_{2}-44) \notag \\
  &+6 \kappa_{1}^2 (2951 \kappa_{2}-9064)+57 \kappa_{1} (1027 \kappa_{2}-16412)-74853 \kappa_{2}+965052\Bigg) + \Om(\veps^3)\:,\\
  \frac{(g^{\chi\text{H}}_{\star})^2}{(4\pi)^2} =\ & \frac{\veps }{2 \kappa_{1}} +\frac{\veps ^2 (2 \kappa_{1} (5 \kappa_{2}+1232)+420 \kappa_{2}-8151)}{1936 \kappa_{1}^3}+ \Om(\veps^3)\:.
\end{align}
The $\beta-$function of $h$ is given by
\begin{align}
  \beta_{h}^{\chi\text{H}} =\ & - \frac{\veps h }{2} + \frac{1}{(4\pi)^2}\left(2 g^2 h N_f+\frac{\lambda h^3}{6}\right) \notag \\
  &+ \frac{1}{(4\pi)^4} \Bigg(- g^2 \lambda h^3 N_f-\frac{2}{3} \pi ^2 g^4 h^3 N_f+4 g^4 h^3 N_f-\frac{7}{2} g^4 h N_f-\frac{\lambda^2 h^5}{12} -\frac{11 \lambda^2 h^3}{36}+\frac{5 \lambda^2 h}{36}\Bigg) \notag\\
  &+\Om(\lambda^3, g^6, \lambda^2 g^2, \lambda g^4)\:,
\end{align}
with the corresponding fixed point
\begin{align}
  (h^{\chi\text{H}}_{\star})^2 =\ & \frac{22}{\kappa_{2}}+\frac{\veps}{66 \kappa_{1}^2 \kappa_{2}^2 (4 \kappa_{1} (\kappa_{2}-66)+(\kappa_{2}-8) \kappa_{2}+264)}\Bigg\{\Big(\kappa_{1} \Big(7947 \kappa_{1}^2+98706 \kappa_{1} \notag \\
  &-10648 \pi ^2 (\kappa_{1}-1)-18717\Big)-87732\Big) \kappa_{2}^2+3 (\kappa_{1} (1627 \kappa_{1}-8928)+7311) \kappa_{2}^3 \notag\\
  &-88 \left(\kappa_{1} \left(121 \pi ^2 (\kappa_{1}-2) (\kappa_{1}-1)+3 \kappa_{1} (813 \kappa_{1}+280)+18753\right)-21933\right) \kappa_{2} \notag \\
  &+383328 (\kappa_{1}-2) (\kappa_{1}-1) \kappa_{1}\Bigg\} + \Om(\veps^2)\:.
\end{align}

\section{Integrals}
\label{app:integrals}

We gather in this appendix the integrals useful for the computations performed in this work. Integrals are computed using dimensional regularization with $d=4-\veps$. In our perturbative computations, we encounter the following integrals:
\begin{align}
Y_{123} &:= \int d^d x_4\, I_{14} I_{24} I_{34}\,, \\
X_{1234} &:= \int d^d x_5\, I_{15} I_{25} I_{35} I_{45}\,, \\
H_{12,34} &:= \int d^d x_5 \int d^d x_6\, I_{15} I_{25} I_{36} I_{46} I_{56} = \int d^d x_5\, I_{15} I_{25} Y_{345}\,,
\label{eq:BasicDefs}
\end{align}
where $I_{ij}$ corresponds to the scalar propagator in $d$ dimensions (see \eqref{eq:PropagatorFunction}). The three- and four-point massless integrals $X$ and $Y$ are finite in $d=4$ and have been solved analytically \cite{Usyukina:1994iw,Usyukina:1994eg}.  The $X$-integral is given by
\begin{equation}
X_{1234} = \frac{I_{12} I_{34}}{16 \pi^2} \chi \bar{\chi}\, D(\chi\,, \bar{\chi})\,,
\label{eq:X-sol}
\end{equation}
with the Bloch-Wigner function
\begin{equation}
D(\chi, \bar{\chi}) := \frac{1}{\chi - \bar{\chi}} \left( 2\text{Li}_2 (\chi) - 2\text{Li}_2 (\bar{\chi}) + \log \chi \bar{\chi} \log \frac{1-\chi}{1-\bar{\chi}} \right)\,,
\label{eq:BlochWigner}
\end{equation}
and where the variables $\chi, \bar{\chi}$ are defined via
\begin{equation}
\chi \bar{\chi} = \frac{I_{13} I_{24}}{I_{12} I_{34}}\,, \qquad (1-\chi) (1-\bar{\chi}) = \frac{I_{13} I_{24}}{I_{14} I_{23}}\,.
\end{equation}

In the case where all the external points are aligned (here in the $\tau$-direction), the $X$-integral can be expressed as a special limit of the result above:
\begin{align}
X_{1234} &= \frac{I_{12} I_{34}}{16 \pi^2} \chi^2\, D(\chi\,, \chi) \notag \\
& = - \frac{I_{12} I_{34}}{8\pi^2} \frac{\chi}{1-\chi} \left( \chi \log \chi + (1-\chi) \log (1-\chi) \right)\,.
\end{align}
Note that in $1d$, $X_{1234}$ is one degree of transcendentality lower than in higher $d$, and that although the prefactor in \eqref{eq:BlochWigner} implies a divergence in the limit $\bar{\chi} \to \chi$, it turns out to be compensated by the numerator.

The $Y$-integral can easily be obtained starting with $X_{1234}$ and sending one of the external points to $\infty$:
\begin{equation}
Y_{123} = \lim\limits_{x_4 \to \infty} 4\pi^2 x_4^2\, X_{1234}\,.
\label{eq:Y-sol}
\end{equation}
For the $1d$ limit mentioned above, this gives
\begin{equation}
Y_{123} = - \frac{I_{12}}{8\pi^2} \left( \frac{\tau_{12}}{\tau_{13}} \log \frac{\tau_{12}}{\tau_{13}} + \frac{\tau_{23}}{\tau_{13}} \log \frac{\tau_{23}}{\tau_{13}} \right)\,,
\end{equation}
with $\tau_{ij} := \tau_i - \tau_j$. It is also useful to consider derivatives of the $Y$-integral, e.g.,
\begin{align}
\partial_1^2 Y_{123} &= - I_{12} I_{13} \,,  \label{eq:id_Y1} \\
(\partial_1 \cdot \partial_2) Y_{123} &= \frac{1}{2} \left( I_{12} I_{13} + I_{12} I_{23} - I_{13} I_{23} \right) \,. \label{eq:id_Y2}
\end{align}

To the best of our knowledge there exists no analytical solution for the $H$-integral. However several identities relate derivatives of the $H$-integral to its $X$ and $Y$ siblings \cite{Eden:1998hh,Eden:1999kh}:
\begin{align}
\partial_1^2 H_{12,34} &= - I_{12} Y_{134} \,, \label{eq:id_H1} \\
(\partial_1 \cdot \partial_2) H_{12,34} &= \frac{1}{2} \left[ I_{12} (Y_{134} + Y_{234}) - X_{1234} \right]\,. \label{eq:id_H2}
\end{align}
Other combinations can be obtained by using
\begin{equation}
H_{12,34} = H_{21,34} = H_{12,43} = H_{34,12}\,.
\end{equation}
In our calculations we only encounter the $H$-integral in the following special "spinor" combinations:
\begin{align}
(F_{13,24})^{AB} := (\slashed{\partial}_1 (\slashed{\partial}_1 + \slashed{\partial}_3) \slashed{\partial}_2)^{AB} H_{13,24}\,,\label{eq:def-F1324} \\
(G_{12,34})^{ABCD} := (\slashed{\partial}_1 \slashed{\partial}_2)^{AB} (\slashed{\partial}_3 \slashed{\partial}_4)^{CD} H_{12,34}\,, \label{eq:def-G1234}
\end{align}
where we have written the matrix indices explicitly to avoid confusion.

The $F$-integral is finite and can be solved by using integration by parts, the fermionic star-triangle relation
\begin{equation}
\int d^d x_4\, \slashed{\partial}_4 I_{14} I_{24} \slashed{\partial}_4 I_{34} = - 4 \pi^2 \slashed{x}_{12} \slashed{x}_{23} I_{12} I_{13} I_{23}\,,
\label{eq:id-StarTriangle}
\end{equation}
and going to a conformal frame. For the case where all the external points are aligned, this gives
\begin{align}
F_{13,24} & \overset{\tau_4 \to \infty}{\sim} \frac{\gamma^0}{4} I_{34} \partial_{\tau_1} Y_{123} \notag \\
& \overset{\phantom{\tau_4 \to \infty}}{=} \frac{\gamma^0}{\tau_{12}^3 \tau_{34}^2} \frac{1}{512 \pi^6} \frac{\chi}{1-\chi} \left( \chi^2 \log \chi + (1+\chi)(1-\chi) \log (1-\chi) \right)\,,
\label{eq:sol-F1324}
\end{align}
where we have suppressed the indices for compactness.

The $G$-integral is also finite and can be solved by observing that the correlator given in \eqref{eq:Hintegral-4F} for the case $N=1$ needs to have the following structure in terms of spinor matrices:
\begin{equation}
G_{12,34}^{ABCD} =  \frac{(\gamma^0)^{AB} (\gamma^0)^{CD}}{\tau_{12}^3 \tau_{34}^2} g_{12,34}(\chi)\,.
\label{eq:sol-G1234a}
\end{equation}
This implies
\begin{equation}
g_{12,34} (\chi) = \frac{1}{4} \tau_{12}^3 \tau_{34}^3 (\partial_1 \cdot \partial_2) (\partial_3 \cdot \partial_4) H_{12,34}\,,
\end{equation}
which, after using the identities given in \eqref{eq:id_Y2} and \eqref{eq:id_H2}, turns into
\begin{equation}
g_{12,34} (\chi) = \frac{1}{2048 \pi^6} \frac{\chi}{(1-\chi)^2} \left( (1-\chi)(2-\chi) + \chi^2 (2-\chi) \log \chi + \chi (1-\chi)^2 \log (1-\chi) \right)\,.
\label{eq:sol-G1234}
\end{equation}

The integrals described above are $\log$-divergent in the limit where two external points coincide. In particular, we encounter repeatedly the integral $Y_{112}$ in self-energy diagrams, which reads:
\begin{equation}
Y_{112} = - \frac{1}{32 \pi^4 \tau_{12}^2} \left( \frac{1}{\varepsilon} + \aleph + \log \tau_{12}^2 + \Op(\varepsilon) \right)\,.
\end{equation}

Another divergent integral that appears in two-point fermion loops is the following:
\begin{equation}
B_{12} := \int d^d x_3 \int d^d x_4\, I_{13} I_{24} \slashed{\partial}_3 I_{34} \slashed{\partial}_3 I_{34}\,.
\label{eq:def-Bintegral}
\end{equation}
This integral is easy to relate to $Y_{112}$ by using $\gamma$-matrix identities and integration by parts:
\begin{align}
B_{12} &= \frac{1}{2} \int d^d x_3 \int d^d x_4\, I_{13} I_{24} \partial_3^2 I_{34}^2 \notag \\
&= \frac{1}{2} Y_{112}\,,
\label{eq:sol-Bintegral}
\end{align}
where in the final result there is a $4 \times 4$ (or $2N_f \times 2N_f$) identity matrix implied. Note that in the last line we have made use of Green's equation \eqref{eq:GreensEq}. In the course of the computation, a quadratic divergence dropped out as dimensional regularization is insensitive to it.

%%%%%%%%%%%%%%%%%%%%%%%%%%%%%%%%%%%%%

%%%%%%%%%%%%%%%%%%%%%%%%%%%%%%%%%%%%%
%\bibliography{bib}
%\bibliographystyle{auxi/JHEP}

\providecommand{\href}[2]{#2}\begingroup\raggedright\endgroup

\end{document}